\documentclass[sn-nature, pdflatex, iicol]{sn-jnl}
\newcommand{\onecolumnfig}{1.\linewidth}

\usepackage{graphicx}%
\usepackage{multirow}%
\usepackage{amsmath,amssymb,amsfonts}%
\usepackage{amsthm}%
\usepackage{mathrsfs}%
\usepackage[title]{appendix}%
\usepackage [dvipsnames]{xcolor}%
\usepackage{textcomp}%
\usepackage{manyfoot}%
\usepackage{booktabs}%
\usepackage{algorithm}%
\usepackage{algorithmicx}%
\usepackage{algpseudocode}%
\usepackage{listings}%
\usepackage{siunitx}
\usepackage{soul}
\usepackage{url}
\usepackage[skip=4pt]{caption}
\usepackage{multirow}
\usepackage{pifont}
\theoremstyle{thmstyleone}%

\theoremstyle{thmstyletwo}%

\theoremstyle{thmstylethree}%

\DeclareSIUnit{\calorie}{\text{cal}}
\DeclareSIUnit{\angstrom}{\text{Å}}
\DeclareSIUnit{\atm}{\text{atm}}
\DeclareSIUnit{\rpm}{\text{rpm}}
\DeclareSIUnit{\atom}{\text{atom}}

\newcommand{\red}[1]{{\color{red}{#1}}} %
\newcommand{\commentGT}[1]{{\color{Cerulean}{1}}} %

\usepackage{enumitem}                    %
\newlist{todolist}{itemize}{2}           %
\setlist[todolist]{label=$\square$}      %
\usepackage{pifont}                      %

\begin{document}

\title{Energy-GNoME: A Living Database of Selected Materials for Energy Applications}

\author[1]{\fnm{Paolo} \sur{De Angelis}}
\author[1]{\fnm{Giovanni} \sur{Trezza}}
\author[1]{\fnm{Giulio} \sur{Barletta}}
\author[1,2]{\fnm{Pietro} \sur{Asinari}}
\author*[1]{\fnm{Eliodoro} \sur{Chiavazzo}}\email{eliodoro.chiavazzo@polito.it}

\affil*[1]{\orgdiv{Department of Energy}, \orgname{Politecnico di Torino}, \orgaddress{\street{Corso Duca degli Abruzzi, 24}, \city{Torino}, \postcode{10129}, \country{Italy}}}
\affil[2]{\orgname{Istituto Nazionale di Ricerca Metrologica}, \orgaddress{\street{Strada delle Cacce, 91}, \city{Torino}, \postcode{10135}, \country{Italy}}}

\abstract{
Artificial Intelligence (AI) in materials science is driving significant advancements in the discovery of advanced materials for energy applications. 
The recent GNoME protocol identifies over 380,000 novel stable crystals. 
From this, we identify over 33,000 materials with potential as energy materials forming the Energy-GNoME database. 
Leveraging Machine Learning (ML) and Deep Learning (DL) tools, our protocol mitigates cross-domain data bias using feature spaces to identify potential candidates for thermoelectric materials, novel battery cathodes, and novel perovskites.
Classifiers with both structural and compositional features identify domains of applicability, where we expect enhanced accuracy of the regressors.
Such regressors are trained to predict key materials properties like, thermoelectric figure of merit ($zT$), band gap ($E_{\mathrm{g}}$), and cathode voltage ($\Delta V_{\mathrm{c}}$).
This method significantly narrows the pool of potential candidates, serving as an efficient guide for experimental and computational chemistry investigations and accelerating the discovery of materials suited for electricity generation, energy storage and conversion.
}


\maketitle

\section{Introduction}\label{sec:introduction}
The growing commitment to environmental sustainability and preservation has catalyzed a shift towards a green economy, emphasizing usage of Renewable Energy Sources (RES), decarbonization strategies, and sustainable resource management to ensure long-term ecological balance and economic resilience~\cite{Sun2019}.
In this context, energy-related materials play a central role in driving the transition to a new, eco-friendly industrial paradigm. 
Materials for renewable energy conversion --- such as perovskites for photovoltaic (PV) solar cells~\cite{kojima2009organometal,green2014emergence}, materials for efficient energy usage --- such as thermoelectric~\cite{snyder2008complex,Bell2008} materials, along with materials for energy storage devices --- like cathode materials for batteries~\cite{whittingham_electrical_1976, mizushima_lixcoo2_1980, lazzari_cyclable_1980} --- are the key to attenuate the intermittent nature of RES, unlocking the full potential of clean energy and achieving the overarching goal of minimizing our environmental footprint while shifting to a sustainable green economy~\cite{kittner2017energy,halkos2020reviewing,nadeem2018comparative}.

Advancements in these fields are strictly correlated to the discovery of novel materials with enhanced properties. 
Most of these physical and chemical properties can be accurately determined by first principles methods based on Density Functional Theory (DFT)~\cite{hohenberg1964inhomogeneous,kohn1965self} 
However, an exhaustive screening of such innovative materials with traditional strategies is to date impractical due to the high-dimensional composition spaces and the often unaffordable computational cost of DFT simulations. Furthermore, efforts in investigating hypothetical materials typically rely heavily on the intuition of the researcher for identifying promising candidates, as well as on heuristics with limited extrapolation capacities on unseen samples~\cite{Nandy2022, back2024accelerated, schrier2023pursuit}.

Despite these difficulties, the efforts required to address the task of materials discovery have been greatly reduced over the last years by the development of high-throughput platforms and of data-driven Machine Learning (ML) techniques.
Indeed, in recent years, ML techniques have greatly impacted the way research~\cite{zdeborova_new_2017, himanen_datadriven_2019} and industry~\cite{lee_cyber-physical_2015, casini_machine_2024, li_methods_2023} approach to several applications, including those in the energy field~\cite{casini_current_2024, koroteev_artificial_2021, liu_machine_2021}.
Coupled with the creation of extensive materials databases such as Materials Project (MP)~\cite{jain2013commentary}, the Inorganic Crystal Structure Database (ICSD)~\cite{hellenbrandt2004inorganic}, the Open Quantum Materials Database (OQMD)~\cite{saal2013materials}, NOMAD~\cite{draxl2019nomad}, and AFLOWLIB~\cite{curtarolo2012aflowlib}, these advanced tools have matured to unlock new potential in the materials discovery process~\cite{butler_machine_2018, tabor_accelerating_2018}. 

The combination of high-throughput computational methods and ML approaches has been successfully applied in recent research to predict novel materials and determine key properties, driving innovation in energy storage, generation, and conversion.
Fanourgakis et al.~\cite{Fanourgakis2020} applied ML methodologies to screen a wide virtual space of hypothetical Metal-Organic Frameworks (MOFs), introducing a universal strategy employing the ``atom types" as the only descriptors to predict the MOFs' adsorption capacities.
A similar work was carried out by Trezza et al.~\cite{Trezza_Bergamasco_Fasano_Chiavazzo_2022}: the authors exploited ML regressors trained to predict MOFs' adsorption capacities to establish a minimal set of important crystallographic features, and investigated the role of such ``genetic code" when using Sequential Learning (SL) algorithms.
Nandy et al.~\cite{Nandy2022} exploited Natural Language Processing (NLP) procedures to leverage the available MOF literature, obtaining stability measures and thermal decomposition temperatures for structurally characterized MOFs. Furthermore, the authors trained Artificial Neural Network (ANN) models to predict solvent removal stability and thermal stability.
Cerqueira et al.~\cite{cerqueira2024sampling} created a computational dataset for conventional, i.e. Bardeen-Cooper-Schrieffer, superconductors containing \emph{ab initio} electron-phonon calculations, which was also used for training a ML model to identify superconducting compounds with a critical temperature $T_{\mathrm{c}}$ greater than $\SI{5}{\kelvin}$, taking as input features the compositional, structural, and ground-state properties. 
Moses et al.~\cite{moses2021machine} leveraged data retrieved from the MP database to train deep Neural Network (DNN) models able to predict the change in volume and the average voltage of battery electrode materials during the charging and discharging processes. The authors also investigated the screening capabilities outside of the training dataset.
Rutt et al.~\cite{rutt_expanding_2022} applied a computational screening approach to identify promising multivalent cathodes, demonstrating the importance of evaluating both relative stability and ion mobility in materials not initially containing the working ion of interest. Here a high-throughput material exploration strategy was applied, in which the MP crystal candidates were iteratively defined in 4 steps, starting by screening the materials with relative stability above $\SI{0.2}{\electronvolt\per\atom}$, and selecting the crystals showing reducible potential with respect to Mg ions. The authors then proceeded with insertion site identification, and finally measured the migration path using approximate Nudged Elastic Band (ApproxNEB) algorithms~\cite{rong_efficient_2016}.
Wang et al.~\cite{Wang2023} developed a computational band gap database of single and double perovskites based on highly accurate DFT calculations, and used it to identify an accurate expression to predict the band gap. This model was then employed to screen Pb-free perovskites in the MP database, finding 14 unreported crystals potentially suitable for PV applications.
Kim et al.~\cite{kim2018machine} trained a Random Forest (RF) based ML model on the whole OQMD to identify novel quaternary Heusler compounds. The model was employed to screen 3.2 million possible structures, predicting their stability and identifying 303 promising compositions, of which 55 were confirmed to yield stable compounds through DFT calculations.
Kang et al.~\cite{kang2020machine} used ML to characterize the heat of explosion of potential candidates of energetic materials, based on the constituent elements-averaged cohesive energy and on the oxygen balance. The authors applied the model to perform a high-level screening of over 140 million molecules in the PubChem database~\cite{wang2009pubchem}, followed by a theoretical fine-level screening which eventually identified 262 molecular candidates with the required properties.
Rao et al.~\cite{rao2022machine} investigated the compositional design of high-entropy alloys, proposing an active learning framework which combines a generative model, regression ensemble, physics-driven learning, and experiments. The authors demonstrated the framework's capabilities in the design of high-entropy Invar alloys with low thermal expansion coefficient.

In addition to these previous screening material successes, the application of ML in materials science has unlocked a vast and largely untapped resource: the Graph Networks for Materials Exploration (GNoME) database~\cite{merchant_scaling_2023}. 
GNoME is an Artificial Intelligence (AI) driven platform designed to explore the vast chemical space through an iterative pipeline that combines active learning algorithms and Graph Neural Networks (GNNs).
This process generates and filters numerous candidate solid state materials using a GNN trained and validated with DFT to predict formation energies.
This active learning framework enables GNoME to continuously refine its predictions, culminating in the discovery of over 2.2 million stable materials. 
Remarkably, it has identified over 380,000 novel stable crystals, which reside on the updated convex hull of formation energies. 
To the best of our knowledge, this vast database of materials has not yet been screened to identify potential materials for disparate energy applications. 

Therefore, in this study, we aim to perform a preliminary screening of the GNoME database to identify potential materials for further numerical or experimental investigation within three relevant domains in the energy field: thermoelectric materials, perovskites, and batteries.
Specifically, we adopt specialized datasets available in the literature for training, validating and testing proper ML regressors towards the prediction of relevant properties of interest across the above four domains.
A straightforward practice would consist in using those models to directly predict such properties of interest over the GNoME materials.
However, the specialized datasets utilized for training represent only a localized subset of the entire materials space. 
As a consequence, the trained ML models -- being not extrapolative \cite{schrier2023pursuit} -- are able to reliably forecast the corresponding property of interest only for those GNoME samples that fall within the same localized subset as the training materials.
To take into account this biased nature of specialized datasets, we adopt the protocol recently proposed and validated by some of the authors of this work \cite{trezza_classification_2024}. 
Specifically, it consists of a set of binary classifier-based filters, trained over samples from the specialized dataset (class 1) and random subsets from a \emph{less biased} general-purpose database like MP (class 0).
By applying these classifiers to the GNoME materials, we can effectively rule out samples for which the regression models are likely to provide unreliable predictions, thereby we expect an enhanced reliability and accuracy of our screening process.

The protocol identified 7,530 thermoelectric candidate materials, 4,259 perovskite candidates for PV applications, and 21,243 cathode material candidates for lithium and eight \emph{post-lithium} kind batteries.

This article is structured as follows: Section~\ref{sec:results} includes a brief overview of the proposed protocol and workflow and presents the findings of our AI-driven screening process for various energy-related materials, including candidates for thermoelectric, perovskites, and cathodes applications. Section~\ref{sec:conclusions} discusses the significance and limitations of these results, their potential applications in the energy sector, and the further development of these methods. Section~\ref{sec:methods} details the computational approaches, data handling, and ML protocols used to predict material properties and identify promising candidates within the GNoME database.
\section{Results}
\label{sec:results}

The proposed method --- introduced in the next Subsection~\ref{sec:protocol} and detailed in Section~\ref{sec:methods} --- requires  three key components: a specialized energy material database containing experimental measurements or numerical predictions of desired properties, a significantly less biased general-purpose materials database, and a set of unexplored materials.
The selection of the last two components is straightforward. 
We utilize the Materials Project (MP) database, which, from its deployment in 2018 to today, reached a cardinality of over 150,000 materials (of which approximately 34,000 stable and 23,000 experimentally observed) characterized using DFT. 
For the unexplored materials set, we use the recent GNoME database, which includes materials that have yet to be experimentally synthesized or numerically simulated for specific energy applications of interest in this work.
Thus, the primary challenge is to obtain reliable and comprehensive energy material data for various applications.
In the following Subsections, we introduce the adopted protocol and present the results of the ML-based screening, along with the identified potential candidates, for three case-studies corresponding to three classes of materials with significant energy relevance: thermoelectric materials, perovskites, and cathode materials.
We also report the ML models predictions for the related properties: figure of merit for thermoelectric materials, band gap for perovskites, and reduction potential for cathode materials.
\subsection{Protocol Overview}\label{sec:protocol}

\begin{figure*}[ht!]
  \centering
  \includegraphics[width=1.\linewidth]{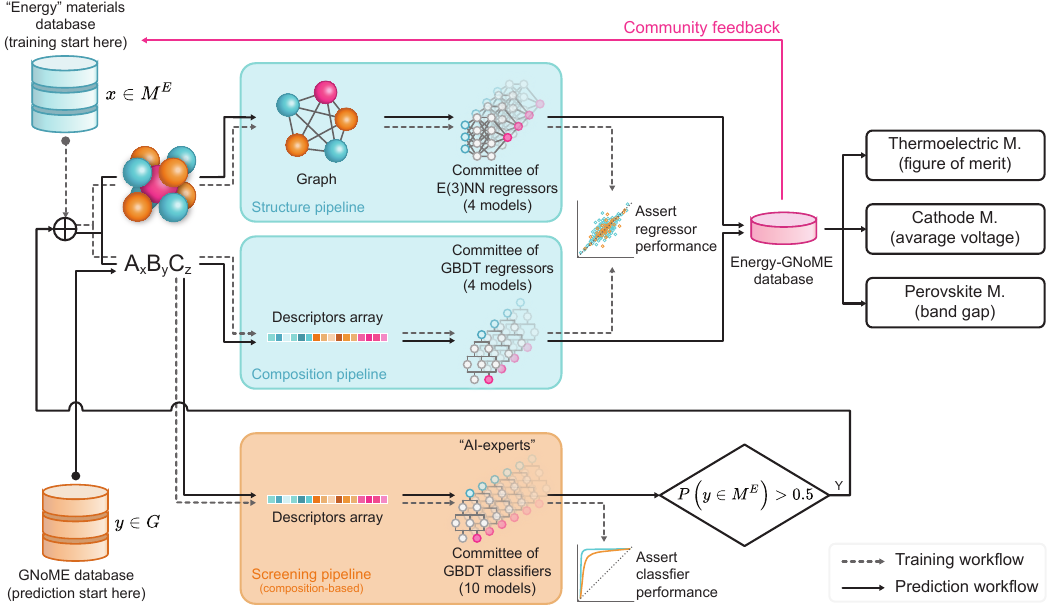}
\caption{
The schematic shows the protocol for creating the Energy-GNoME database, illustrating training (grey dashed line) and predictive (black solid line) phases. 
Training begins with the cyan database and ends with ML model evaluations (e.g., parity plots, ROC curves). 
Feature extraction depends on material storage in the ``Energy'' database $M^E$ and may use composition- or structure-based pipelines, as indicated by the OR switch symbol $\oplus$.
The \textit{structure pipeline} applies a graph representation, while the \textit{composition pipeline} uses chemical descriptors, each feeding a committee of E(3)NN or GBDT. 
Concurrently, the \textit{screening pipeline} (orange box) trains GBDT classifiers -- ``AI-experts'' -- to identify $M^E$-like materials. 
In prediction mode, screened GNoME materials ($y$) with over 50\% likelihood $\left(P(y\in M) > 0.5\right)$ of matching $M^E$ materials ($x$) biases enter the regressor pipeline to predict properties.
These candidates, with predicted properties, are added to the Energy-GNoME database, initiating a continuous active learning cycle (see magenta arrow).     
}
\label{fig:protocol}
\end{figure*}

We propose that there exists a region $E$ within the high-dimensional feature space of all materials, containing all materials suited for a given energy application.
The intersection between $E$ and a general-purpose database of known materials (e.g. MP) contains known materials for energy applications.
This ``energy-specific" subset $M^E~=~M \cap E$ and its complement $M \setminus E$ are leveraged by the AI-experts to delineate the boundary of $E$.
This approach allows for the identification of the intersection between the dataset of unexplored materials $G$ (namely the GNoME database) and $E$, $G^E = G \cap E$, i.e., the subset of crystals in $G$ that share properties with $M \cap E$.
Here, regression yields more reliable results than when applied to the entire set $G$.

For a detailed explanation of the protocol and methods, please refer to Section~\ref{sec:methods}.
Additionally, Figure~\ref{fig:method-concept} visually represents the relationships among these different sets within the high-dimensional material feature space aforementioned.

The discovery of energy materials within the GNoME database by identifying the $G^E$ subset is achieved through the designed protocol illustrated in Figure~\ref{fig:protocol}.
The process involves two distinct workflows: the training workflow and the prediction workflow, indicated in Figure~\ref{fig:protocol} with a grey dashed line and a black solid line, respectively.

The \textit{training workflow} goes through all the steps to instruct all the ML models, starting from the specialized ``energy'' material database $M^E$.
The training set data is used to qualify the AI-expert algorithms to classify and, therefore, hypothetically identify the boundary, $\partial E$, of the energy material region in the n-dimensional space and to train regression machine learning algorithms to predict the specific property of interest for each class of energy material.
In the proposed case study, we have chosen, the figure of merit ($zT$) for thermoelectric materials, the band gap ($E_g$) for perovskites, and the reduction potential ($\Delta V$) for cathode materials.
This workflow ends with assessing the performance of all the models.
The available data for materials in energy applications do not always include structural information, such as crystallographic files (e.g., CIF or XYZ files) or unique string identifiers like the International Chemical Identifier (InChI).
Consequently, for the regression, a conditional OR switch splits the workflow into two pipelines: one optimized for materials with complete structural information and another designed to handle cases with only compositional data.
If structural data is available, the workflow follows the ``structure pipeline'' (top blue box in Figure~\ref{fig:protocol}) that uses the graph representation of the material to train a committee of four E(3)NNs to predict the material property of interest. 
If only composition information is available, the dataflow goes through the ``composition pipeline'' (center blue box in Figure~\ref{fig:protocol}), which uses the descriptors array to train four Gradient Boosted Decision Trees (GBDT) models instead.
On the other hand, the AI-experts, which are composed of a committee of ten binary GBDT classifiers, are always trained using the chemical formula of the compounds, which can also be obtained from the crystal structure, to be able to identify the biases of the specialized energy material database ($M^E$).

Once all the models are trained and validated, the data flow transitions to the \textit{prediction workflow}, which begins from the GNoME database $G$.
All the data points $y \in G$ are processed through the ``screening pipeline'' (bottom orange box in Figure~\ref{fig:protocol}) where the AI-experts are consulted to compute the probability that the crystal shares the same biases as those in the specialized training set, i.e. $P(y \in M^E)$.
According to our hypothesis, this probability also coincides with the likelihood of falling inside the energy material region $E$ ($P(y \in M^E) = P(y \in E)$).
All the crystals that have passed the screening process -- i.e., those with an average probability from the AI-experts higher than 50\% -- continue to the regressors pipeline. 
Here, depending on the specialized databases used for training, the materials are featurized either by converting them into graphs if a committee of E(3)NNs was trained or into descriptor arrays if a committee of GBDTs was used. 
The trained regressors then predict the property related to the specialized energy material under investigation of only the screened material. 
In this way, we have the computational benefit of working with a significantly smaller dataset than the entire GNoME database, reducing computational costs and resources. 

Finally, the candidates with the predicted properties are stored in the Energy-GNoME database. 
The resulting database can then be evaluated, refined, and validated by both the computational and experimental community, expanding the initial specialized energy material database used to train the workflow. 
Consequently, the entire workflow can be rerun, thereby improving both the screening and prediction accuracy, initiating an iterative process that makes the Energy-GNoME a \textit{living database}. 
\subsection{Thermoelectrics (figure of merit $zT$)}\label{ssec:resutls-thermoelectrics}
When an electrical current is supplied, thermoelectric materials are able to generate a temperature gradient while also releasing Joule heat; vice versa, a temperature gradient can generate an open-circuit voltage, allowing these materials to play the role of electric generators \cite{sootsman2009new}.
As a result, thermoelectric-based devices may utilize a range of heat sources, such as solar radiation and industrial waste heat, making them an interesting asset for the advancement of sustainable and energy-efficient technologies. 
The effectiveness of a material in thermoelectric systems is determined by the dimensionless thermoelectric figure of merit $zT = (S^2\sigma/\kappa)T$, with $S$ denoting the Seebeck coefficient, $\sigma$ the electrical conductivity, $\kappa$ the thermal conductivity, all varying with the temperature $T$.
\begin{figure}[ht!]
  \centering
  \includegraphics[width=\onecolumnfig]{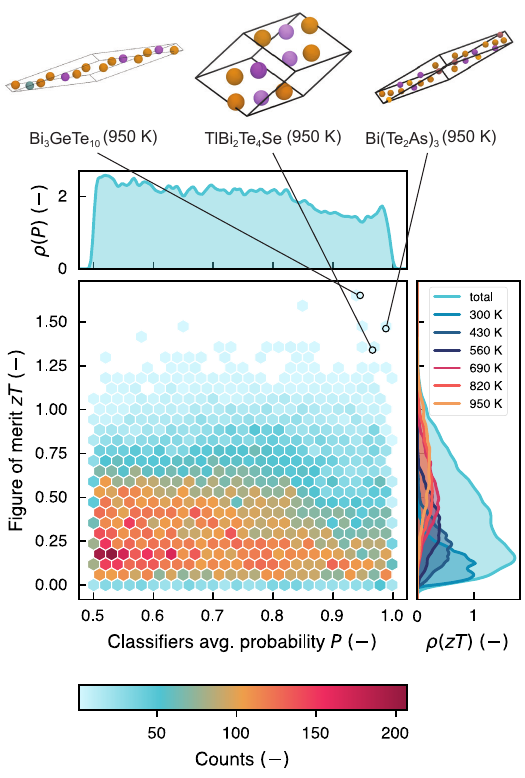}
\caption{
Hexagonal plot of thermolesctrics candidate materials in the Energy-GNoME database.
The hexagon colors represent material counts per region as indicated by the color bar.
Density distributions, $\rho$, are shown on the plot's top and right, calculated using Gaussian KDE for the average AI-expert probability, $P$, and predicted figure of merit, $zT$. 
The thermoelectric performance was assessed across six temperatures, with combined $zT$ values displayed in a color-coded distribution on the right. 
The crystal structures above show three notable candidates among the  top-ranked screened thermoelectric materials as determined by $R^{T}(y)$ (see Subsubsection~\ref{sssec:m-case-specific-thermoelectrics}). 
Atom colors follow the extended CPK~\cite{corey_molecular_1953} scheme by Jmol~\cite{noauthor_jmol}.
}
\label{fig:result-thermoelectrics}
\end{figure}

Here, we aim at finding new potential thermoelectric materials within the GNoME database following the same aforementioned protocol.
As data source, in this case, we use the Experimentally Synthesized Thermoelectric Materials (ESTM) database \cite{na2022public}, along with the experimental $zT$s of 869 materials as function of the temperature $T$. 
In this case only the composition is available and, as such, we extract a set of 145 composition-based features for those 869 brute formulae by means of Matminer~\cite{ward2018matminer}.
Specifically, as detailed by Ward et al.~\cite{ward2016general}, these descriptors include stoichiometric features, statistics on elemental properties, characteristics related to electronic structure, and specific attributes for ionic compounds.
Furthermore, since on average each material in ESTM comes with 6 distinct values of the temperature $T$ -- in general different across the database -- we make $T$ act as the 146$^{\mathrm{th}}$ feature.
However, to prevent potential unfairness that may arise from having the same material -- only with a different $T$ -- over various random splits across training/validation/testing sets, we ensure that all instances of the same material (coming with various $T$ values) are included in the same split.
In order to take into account the $146^{\mathrm{th}}$ feature for materials not coming from ESTM (i.e., MP), for each of them we create 6 replicas, each with one of 6 evenly spaced temperatures $T$, namely $\SI{300}{\kelvin}$,
$\SI{430}{\kelvin}$,
$\SI{560}{\kelvin}$,
$\SI{690}{\kelvin}$,
$\SI{820}{\kelvin}$,
$\SI{950}{\kelvin}$.
On average, the 4 regression models show reasonably high performance in test ($R^{2}\approx0.71$, see Table \ref{tab:thermoelectrics_summary}) and the 10 classifiers turn out to be highly skilled as well ($\mathrm{AUC} \lesssim 1$).
The metrics of the individual regression and classification models are reported in Table~\red{S1} and Table~\red{S2} respectively.
For further details about pre-processing and featurization, ML models training, and the whole protocol, refer to Section~\ref{sec:methods}.

By incorporating $T$ as an additional feature, we predict the average classification probabilities and the average $zT$s values for each of the energy material candidates within the GNoME database, across 6 replicas at the same 6 evenly spaced temperatures mentioned above.
As a result, we identify 7,530 unique GNoME compositions, corresponding to 30,664 $T$-based replicas samples, showing an average probability $P>0.5$ to fall within the materials space of the ESTM database.
Among those, at $T=\SI{950}{\kelvin}$, Bi$_3$GeTe$_{10}$ is predicted to exhibit an average $zT=1.65$ (over the 4 regressors) with an average classification probability (over the 10 classifiers) $P=0.95$; 
Bi(Te$_2$As)$_3$ is predicted to exhibit an average $zT=1.46$ (over the 4 regressors) with an average classification probability (over the 10 classifiers) $P=0.99$;
TlBi$_2$Te$_4$Se is predicted to exhibit an average $zT=1.34$ (over the 4 regressors) with an average classification probability (over the 10 classifiers) $P=0.97$ (see Figure~\ref{fig:result-thermoelectrics}).

\begin{table}[th!]
	\caption{
    The table reports the total number of thermoelectrics ($\left| M^E \right|$, with $|\cdot|$ denoting the cardinality) and a summary of regressor model testing performances (coefficient of determination $R^2$ and root mean square error RMSE) for the committee of 4 GBDT (ensemble). 
    The models predict the thermoelectric figure of merit $zT$.
    }
	\centering
	\label{tab:thermoelectrics_summary}
    \begin{tabular}{>{\centering\arraybackslash}m{26mm}>{\centering\arraybackslash}m{18mm}>{\centering\arraybackslash}m{18mm}}
        \toprule
        $\mathbf{\left| M^E \right|}$
        & $\mathbf{R^2}$ & $\mathbf{RMSE}$\\
        $\left(-\right)$ & $\left(-\right)$ & $\left(-\right)$  \\
       \midrule
       869 (5,061 replicas) & $0.730$ & $0.170$ \\
       \bottomrule
       \end{tabular}
\end{table}

\subsection{Perovskites (band gap $E_g$)}\label{ssec:resutls-perovskite}

Perovskite solar cells have gained extensive popularity due to their high absorption coefficient, high charge carrier mobility, controllable band gap, and ease and low cost of fabrication~\cite{roy_review_2020,nair_recent_2020}.
\begin{figure*}[ht!]
  \centering
  \includegraphics[width=1.\linewidth]{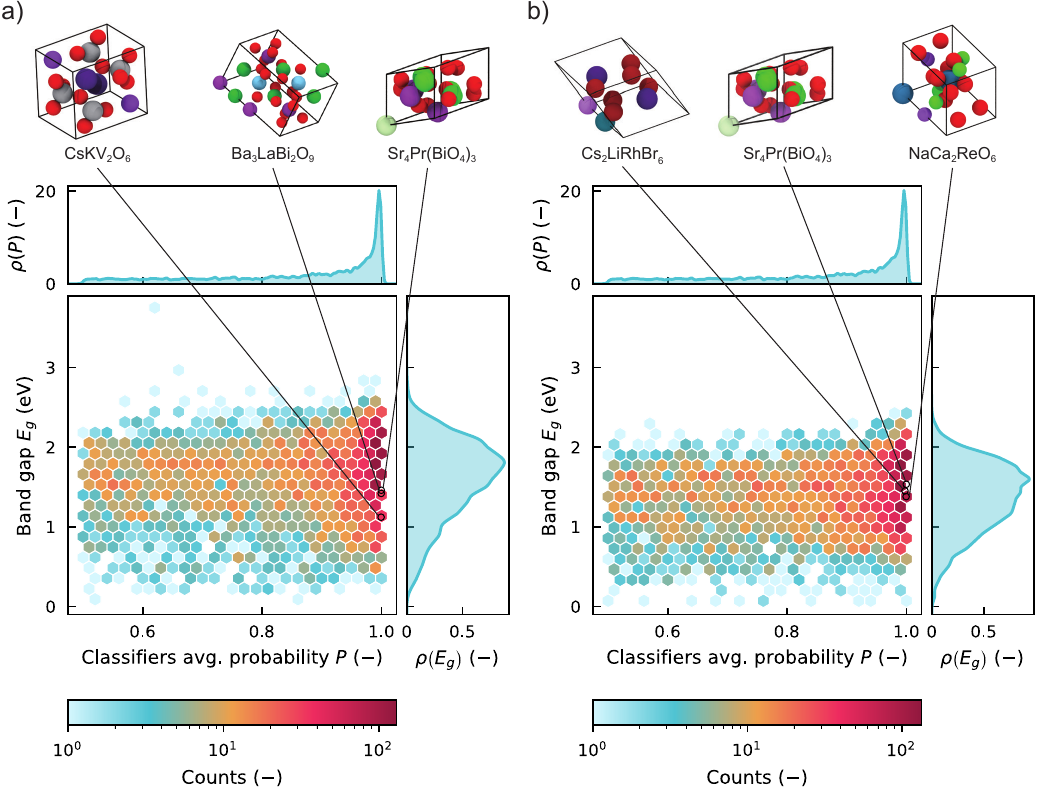}
\caption{
Hexagonal plot of perovskite candidates materials in the Energy-GNoME database. 
The hexagon colors represent material counts per region on a logarithmic scale, as indicated by the color bar. 
Density distributions, $\rho$, are shown on the plot's top and right, calculated using Gaussian KDE for the average AI-expert probability, $P$, and the predicted band gap, $E_g$. 
For $E_g$, results are displayed from regressors trained on (a) perovskite data alone and (b) an augmented dataset.
The crystal structures above show three notable candidates among the top-ranked perovskite materials, as determined by $R^{P}(y)$ (see Subsubsection~\ref{sssec:m-case-specific-perovskite}).
Atom colors follow the extended CPK~\cite{corey_molecular_1953} scheme by Jmol~\cite{noauthor_jmol}.
}
\label{fig:results-perovskite}
\end{figure*}
In fact, the suitability of a perovskite as photovoltaic material is most importantly determined by its band gap $E_{\mathrm{g}}$.
It is worth noting that the possibility of engineering synthetic perovskites gives rise to a vast compositional material space available for exploration, allowing researchers to fine-tune their properties for specific applications~\cite{osterrieder_autonomous_2023}.
Hence, a reliable and cheap methodology to determine potential interesting structures would be a strong tool to investigate such space.

In this case-study, we aim at finding new potential perovskites suitable for PV applications within the GNoME database following the same aforementioned protocol.
We thus leverage a valuable source of data, namely the MP database \cite{jain2013commentary}, which provides the structure along with both the computed and the experimentally measured properties of more than 150,000 solid-state materials.
Screening the MP database, we identify 648 perovskites with unique structures displaying an $E_{\mathrm{g}}$ suitable for PV applications, i.e. $\SI{0.0}{\electronvolt}<E_{\mathrm{g}}\leq\SI{2.5}{\electronvolt}$~\cite{hu_review_2019,horantner_potential_2017,liu_highly_2021}.
We train and test 4 regressors (hereafter referred to as the ``pure models") over the extracted database able to predict the $E_{\mathrm{g}}$.
Taking advantage of the MP database's detailed crystal structure information, we utilize an Euclidean Equivariant Neural Network (E(3)NN) architecture, following the structure-based prediction pipeline (Figure~\ref{fig:protocol}) described in the Methods (refer to Section~\ref{sec:methods}).
As the $E_{\mathrm{g}}$ is a property shared by all materials, we also randomly select 1,988 non-perovskite materials displaying an $E_{\mathrm{g}}$ in the same aforementioned range, and merge them with the 648 perovskites.
Again, we train and test 4 E(3)NN regression models (hereafter referred to as the ``mixed models") over the merged energy material database to predict the $E_{\mathrm{g}}$.
Furthermore, we identify a subset of the 648 perovskites containing 576 samples with unique chemical formula, and select 620 random non-perovskite materials, from which we extract a set of 694 composition-based and structure-based features by means of Matminer~\cite{ward2018matminer}.
For further details, refer to Subsection~\ref{ssec:m-materials-featurization}.
These features are used to train and test 10 classifiers (AI-experts) able to discriminate perovskites from other materials in the MP database. 
\begin{table}[th!]
	\caption{
    The table reports the total number of perovskites $\left( \left| M^E \right| \right)$ and a summary of regressor model testing performances (coefficient of determination $R^2$ and root mean square error RMSE) for the committee of 4 E(3)NNs (ensemble). 
    The models predict one key perovskite property for PV applications, namely the band gap $E_{\mathrm{g}}$.
    }
	\centering
	\label{tab:perovskite_summary}
    \begin{tabular}{>{\centering\arraybackslash}m{10mm}>{\centering\arraybackslash}m{11mm}>{\centering\arraybackslash}m{11mm}>{\centering\arraybackslash}m{11mm}>{\centering\arraybackslash}m{11mm}}
        \toprule
        \multirow{3}*{$\mathbf{\left| M^E \right|}$} & \multicolumn{2}{c}{$\mathbf{E_g}$ (pure)} & \multicolumn{2}{c}{$\mathbf{E_g}$ (mixed)} \\
        & $\mathbf{R^2}$ & $\mathbf{RMSE}$ & $\mathbf{R^2}$ & $\mathbf{RMSE}$ \\
        & $\left(-\right)$ & $\left(\si{\electronvolt}\right)$ & $\left(-\right)$ & $\left(\si{\electronvolt}\right)$ \\
       \midrule
       648 & $0.557$ & $0.422$ & $0.572$ & $0.415$ \\
       \bottomrule
       \end{tabular}
\end{table}

On average, both pure and mixed regressors in testing show reasonable predictability (coefficient of determination $R_{\mathrm{pure}}^2\approx0.56$ and $R_{\mathrm{mixed}}^2\approx0.57$, see Table~\ref{tab:perovskite_summary}), while AI-experts are highly skilled ($\mathrm{AUC}\approx0.98$). 
The metrics of the individual pure and mixed regressors are reported in Table~\red{S3} and Table~\red{S4}, respectively.
The metrics of the individual classification models are reported in Table~\red{S5}.
For further details about pre-processing and featurization, ML models training, and the whole protocol, refer to Section~\ref{sec:methods}.

As a result, we identify 4,259 GNoME materials showing an average probability $P>0.5$ to fall within the materials space of the perovskites included in the MP database.
Figure~\ref{fig:results-perovskite} shows the potential perovskite candidates in the Energy-GNoME database, along with the classifier committee's average probability and the average predictions of the individual $E_{\mathrm{g}}$ values, obtained through either the pure or the mixed regressor models.

Among the most promising candidates according to the ``pure models" predictions and the ranking function Eq.~\ref{eq:perovskite_ranking} (described in Subsubsection~\ref{sssec:m-case-specific-perovskite}), interesting materials are $\mathrm{CsKV_2O_6}$, $\mathrm{Ba_3LaBi_2O_9}$, and $\mathrm{Sr_4Pr(BiO_4)_3}$.
In particular, the regressor committee predicts $\mathrm{CsKV_2O_6}$ to exhibit an average $E_{\mathrm{g}} = \SI{1.12}{\electronvolt}$, $\mathrm{Ba_3LaBi_2O_9}$ to exhibit an average $E_{\mathrm{g}} = \SI{1.42}{\electronvolt}$, and $\mathrm{Sr_4Pr(BiO_4)_3}$ to exhibit an average $E_{\mathrm{g}} = \SI{1.46}{\electronvolt}$.
All three materials have an average classification probability (over the 10 AI-experts) $P=1.00$.

On the other hand, among the most promising candidates according to the ``mixed models" predictions and the ranking function Eq.~\ref{eq:perovskite_ranking}, interesting materials are $\mathrm{Cs_2LiRhBr_6}$, $\mathrm{Sr_4Pr(BiO_4)_3}$, and $\mathrm{NaCa_2ReO_6}$.
The regressor committee predicts $\mathrm{Cs_2LiRhBr_6}$ to exhibit an average $E_{\mathrm{g}} = \SI{1.45}{\electronvolt}$, $\mathrm{Sr_4Pr(BiO_4)_3}$ to exhibit an average $E_{\mathrm{g}} = \SI{1.54}{\electronvolt}$ (similar to that predicted by the ``pure models"), and $\mathrm{NaCa_2ReO_6}$ to exhibit an average $E_{\mathrm{g}} = \SI{1.38}{\electronvolt}$.
All three materials have an average classification probability (over the 10 AI-experts) $P=1.00$.

\subsection{Cathodes (average voltage $\Delta V_c$)}\label{ssec:resutls-batteries}
\begin{table*}[h!]
	\caption{
    The table reports the total number of materials per working ion $\left( \left| M^E \right| \right)$ and a summary of regressor model testing performances (coefficient of determination $R^2$ and root mean square error RMSE) for the committee of 4 E(3)NNs (ensemble). 
    The models predict four key cathode properties: average cathode potential ($\Delta V_c$), maximum relative volume difference ($\max \left(\Delta \text{Vol}\right)$), stability of the charged state ($\Delta E_{\mathrm{charge}}$), and stability of the discharged state ($\Delta E_{\mathrm{discharge}}$).
    }
	\centering
	\label{tab:cathodes_regressor}
    \begin{tabular}{>{\arraybackslash\let\newline}m{4mm}|>{\centering\arraybackslash\let\newline}m{11mm}cccccccc}
        \toprule
        & \multirow{3}*{$\mathbf{\left| M^E \right|}$} & \multicolumn{2}{c}{$\mathbf{\Delta V_c }$}  & \multicolumn{2}{c}{$ \mathbf{max}\mathbf{\left( \Delta Vol \right)}$ } & \multicolumn{2}{c}{$\mathbf{ \Delta E_{charge}}$} & \multicolumn{2}{c}{$\mathbf{\Delta E_{discharge}}$} \\
        & & $\mathbf{R^2}$ & $\mathbf{RMSE}$ & $\mathbf{R^2}$ & $\mathbf{RMSE}$  & $\mathbf{R^2}$ & $\mathbf{RMSE}$  & $\mathbf{R^2}$ & $\mathbf{RMSE}$ \\
         &   & $\left(-\right)$ & $\left(\si{\volt}\right)$ & $\left(-\right)$ & $\left(\si[per-mode = symbol]{\meter^3\per\meter^3}\right)$ & $\left(-\right)$ & $\left(\si[per-mode = symbol]{\electronvolt\per\atom}\right)$  & $\left(-\right)$ & $\left(\si[per-mode = symbol]{\electronvolt\per\atom}\right)$ \\
       \midrule
       Li &  2440 & $0.733$ & $0.563$ & $0.442$ & $0.030$ & $0.315$ & $0.041$  & $0.263$ & $0.034$  \\
       Na &  309 & $0.606$ & $0.744$ & $0.205$ & $0.055$  & $0.432$ & $0.049$  & $-0.112$ & $0.034$  \\
       Mg &  423 & $0.619$ & $0.876$ & $0.335$ & $0.040$  & $0.491$ & $0.069$  & $0.669$ & $0.091$  \\
       K &  107 & $-0.173$ & $1.023$ & $0.440$ & $0.109$  & $0.545$ & $0.028$  & $-0.115$ & $0.092$  \\
       Ca &  435 & $0.695$ & $0.654$ & $0.550$ & $0.043$  & $0.477$ & $0.076$  & $0.276$ & $0.072$  \\
       Cs &  33 & $0.594$ & $0.865$ & $-0.246$ & $0.094$  & $0.082$ & $0.017$  & $-0.199 $ & $0.033$  \\
       Al &  95 & $0.783$ & $0.571$ & $-0.186$ & $0.059$  & $0.335$ & $0.158$  & $0.596$ & $0.076$  \\
       Rb &  50 & $-0.944$ & $2.196$ & $0.040$ & $0.192$  & $-0.079$ & $0.065$  & $-0.335$ & $0.102$  \\
       Y &  93 & $0.365$ & $0.757$ & $0.488$ & $0.096$  & $0.493$ & $0.122$  & $0.600$ & $0.092$  \\
       \bottomrule
       \end{tabular}
\end{table*}

The search for new cathode materials is a critical focus in the electrochemical community as the demand for next-generation batteries continues to rise. 
Indeed, the widespread use of batteries in electronic devices, electric vehicles, and for energy storage during RES's surplus production is driving the demand for new battery technologies that are safer, more reliable, cost-effective, and sustainable, even moving beyond the conventional Lithium-ion Battery (LIB) technology~\cite{larcher_towards_2015, canepa_odyssey_2017, manthiram_reflection_2020}.

In this Subsection, we present the results of our ML-based screening protocol applied to the GNoME database using the ``Battery Explore'' specialized database from Materials Project~\cite{jain2013commentary}. 
This specialized database --- at the time of this study --- consisted of 3,985 batteries, each comprising a pair of charge and discharge cathode materials and intermediate stable phases used to compute the average voltage potential.
The database includes intercalation-type cathode materials designed for nine different monovalent and multivalent working ions: Li, Na, Mg, K, Ca, Cs, Al, Rb, and Y. 

This case study aims to identify possible new cathode material candidates for lithium and \emph{post-lithium} batteries within the GNoME database.
Critical parameters present in the MP database for batteries include the average voltage ($\Delta V_c$), maximum relative volume difference ($\max \left(\Delta \text{Vol}\right)$), stability of the charged state ($\Delta E_{\text{charge}}$), and stability of the discharged state ($\Delta E_{\text{discharge}}$).
Following the protocol we are presenting, we trained four regressors.
Similar to the perovskite case, we can also access the crystal structure of the materials by querying the MP database, therefore also here we use E(3)NN, following the structure-based pipeline of the deataflow (Figure~\ref{fig:protocol}).
However, instead of focusing on a single property, we repeated the training for all four target cathode properties, resulting in a total of 16 models.
Furthermore, we tailored these 16 E(3)NN models for each of the nine working ions under consideration, culminating in a total of 144 trained models.
For the AI-experts, we trained 10 classifiers for each working ion (totaling 90 classifiers). 
We used MP materials containing the working ion element, but not overlapping with the training database, as a \textit{less biased} dataset.
The two dataset, rapresenting the two class for the classification are then translated into the material-space using descriptors made of 694 composition-based and structure-based features by means of Matminer.
A detailed explanation of pre-processing and featurization of all ML models, training and testing settings, and the whole protocol layout is provided in Section~\ref{sec:methods}.

Table~\ref{tab:cathodes_regressor} reports the total amount of data available in MP for training and the relative ensemble of E(3)NNs models performance for each cathode class evaluated in the testing set (20\% of the specialized database $M^E$).
The results for each one of the 16 models are reported in the supplementary Tables~\red{S6-S9}.
The performance of each regressor varies across working ions due to differences in dataset size and years of investigation for each cathode type.
Li cathodes are the most prevalent in the MP database, with 2,440 crystal materials. 
For these, the committee of models shows that the average voltage is the property best predicted by the E(3)NN model, with $R^2=0.733$. 
In contrast, the stability energies $\Delta E_{\mathrm{charge}}$ and $\Delta E_{\mathrm{discharge}}$ show lower performance, with $R^2=0.315$ and $R^2=0.263$, respectively.
These results highlight that, despite E(3)NN's proven ability to predict energy, forces~\cite{batzner_e3-equivariant_2022}, and phonon density of states~\cite{chen_direct_2021}, a large energy material database is crucial.
Therefore, the predictions presented here should be interpreted with caution, as further expansion of the cathode material database is necessary to enhance model performance.
The worst cases, which seem extremely hard to interpolate, are K, Cs, Al, Rb, and Y cathodes due to their small training set sizes ($< 150$).
\begin{table}[h!]
	\caption{
    The table reports the total number of candidate cathodes identified by AI-experts within GNoME per working ion $\left( \left| G^E \right| \right)$ and a summary of classification testing performances (AUC of Receiver Operating Characteristic (ROC) and Precision of the classifiers) for the committee of 10 GBDT AI-experts (ensemble). 
    }
	\centering
	\label{tab:cathodes_classifiers}
    \begin{tabular}{
      >{\arraybackslash\let\newline}m{4mm}
      |>{\centering\arraybackslash}m{15mm}>{\centering\arraybackslash}m{15mm}>{\centering\arraybackslash\let\newline}m{21mm}
      }
        \toprule
        & $\mathbf{AUC}$ &  \textbf{Precision} & \multirow{2}*{$\mathbf{\left| G^E \right|}$} 
        \\
         &  $\left(-\right)$ &  $\left(-\right)$ & \\
       \midrule
       Li &  $0.897$  &  $0.764$  &  $413$  \\
       Na &  $0.890$  &  $0.811$  &  $779$  \\
       Mg &  $0.946$  &  $0.864$  &  $888$  \\
       K &   $0.844$  &  $0.706$  &  $1327$  \\
       Ca &  $0.917$  &  $0.791$  &  $1128$  \\
       Cs &  $0.458$  &  $0.400$  &  $4634$  \\
       Al &  $0.887$  &  $0.857$  &  $6280$  \\
       Rb &  $0.886$  &  $0.857$  &  $3559$  \\
       Y &   $0.833$  &  $0.733$  &  $2235$  \\
       \bottomrule
       \end{tabular}
\end{table}
\begin{figure*}[ht!]
  \centering
  \includegraphics[width=1.\linewidth]{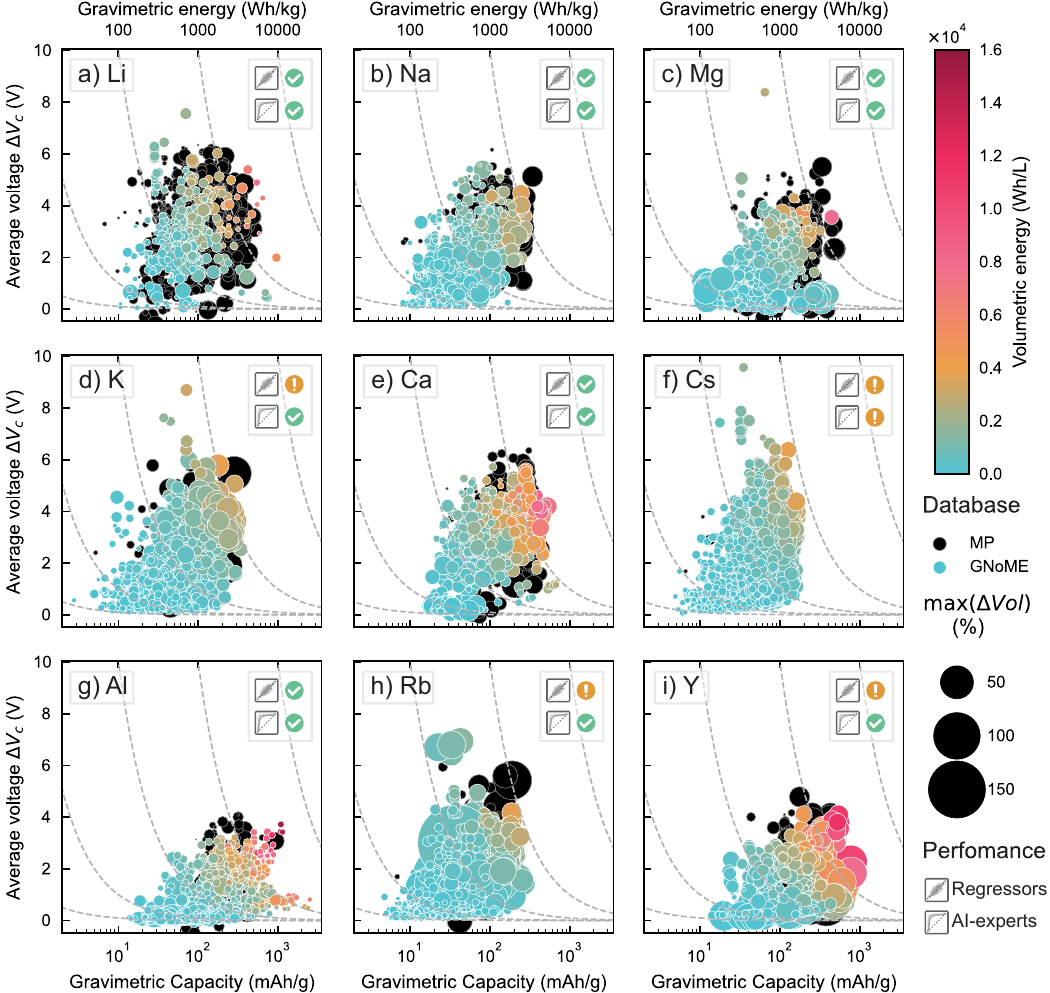}
\caption{
Candidates for battery cathode materials with various charge carriers. Li (a), Na (b), Mg (c), K (d), Ca (e), Cs (f), Al (g), Rb (h), and Y (i).
Each candidate is represented as a point in the scatter plot, showing theoretical gravimetric capacity ($\si{\milli\ampere\hour\per\gram}$, logarithmic scale) versus predicted average voltage difference ($\si{\volt}$) relative to pure element oxidation potential ($\mathrm{X/X^{n+}}$ with $\mathrm{X}$ being the working ion). 
Grey dashed hyperbolas indicate the predicted gravimetric energy ($\si{\watt\hour\per\kilogram}$), noted on the upper axis. 
Dot size represents predicted maximum volume expansion, and the dots are color-coded to represent the predicted volumetric energy ($\si{\watt\hour\per\liter}$).
Due to dataset limitations, model performance varies (see Subsection~\ref{ssec:resutls-batteries}). 
The top right corner legends indicates prediction reliability: a green checkmark (\textcircled{\raisebox{-0.01em}{\scalebox{0.66}{\ding{51}}}})for models with $R^2$ and AUC above 0.5, showing higher accuracy, and a yellow warning (\textcircled{\raisebox{-0.05em}{\scalebox{0.80}{\textbf{!}}}}) for models below this threshold.
}
\label{fig:batteries_result_all}
\end{figure*}

On the other hand, despite the small sample size, the classifiers demonstrate a higher skill in identifying materials belonging to the energy materials space than the more general MP database.
Table~\ref{tab:cathodes_classifiers} reports the AUC and precision metrics on the testing set for each working ion specialized AI-expert ensemble.
Most AI-expert ensembles achieve an $\text{AUC}>0.85$, a clear sign of their robust discriminative capability. 
However, the cathodes for K, Cs, and Y ion batteries present the most challenging sets of materials for the AI-experts to classify.
Detailed metric scores for all 90 classifiers on the test and training sets are reported in Table~\red{S10} in the supplementary information.
\begin{figure*}[ht!]
  \centering
  \includegraphics[width=1.\linewidth]{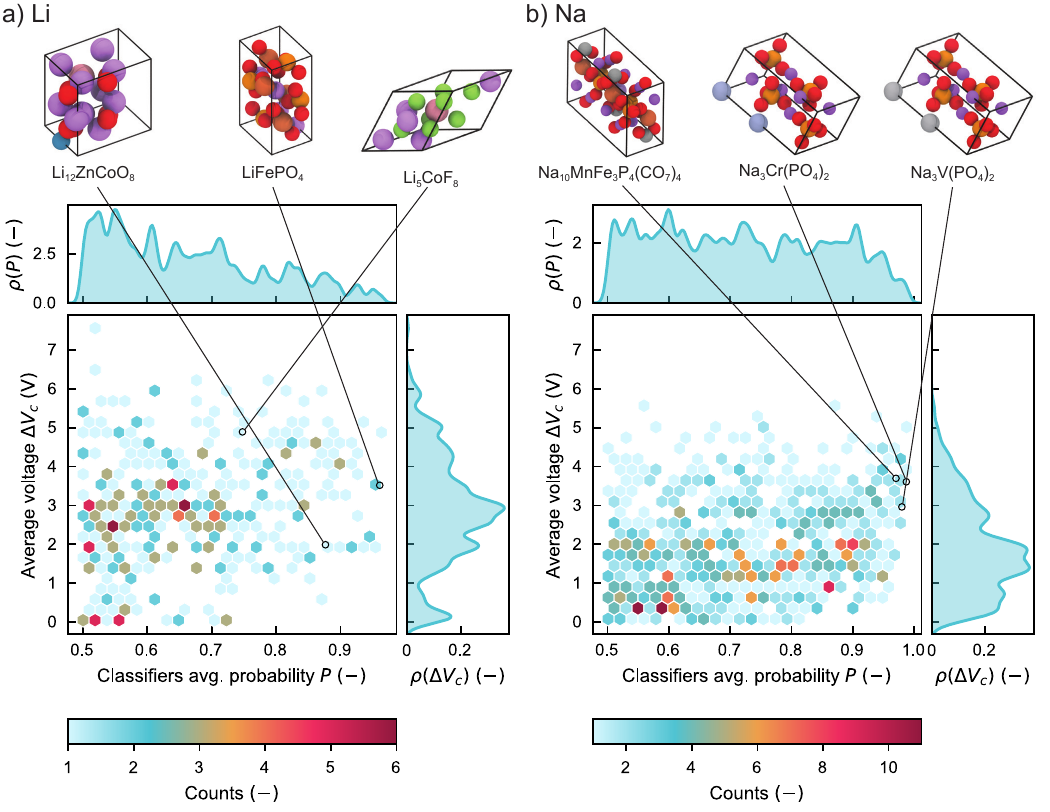}
\caption{ 
Hexagonal plot of cathode candidate materials in the Energy-GNoME database for (a) Li-ion and (b) Na-ion batteries. 
The hexagon colors represent material counts per region, as indicated by the color bar. 
Density distributions, $\rho$, are shown on the plot's top and right, calculated using Gaussian KDE for the average AI-expert probability, $P$, and the average reduction potential, $\Delta V$. 
Three high-ranking screened cathode materials are shown as primitive crystal units above, identified using $R^{B}(y)$ (see Subsubsection~\ref{sssec:m-case-specific-batteries}). 
Atom colors follow the extended CPK~\cite{corey_molecular_1953} scheme by Jmol~\cite{noauthor_jmol}.
}
\label{fig:batteries_result_LiNa}
\end{figure*}

As a result of the screening process, we identify 21,243 GNoME materials with an average probability $P > 0.5$ belonging to the cathode materials space.
Table~\ref{tab:cathodes_classifiers} shows the number of materials, $\left| G^E \right|$, for each specific monovalent and multivalent ion battery.
An interesting result is the inverse proportionality between the number of screened materials and the training set size.
This may indicate that when the protocol is applied to deeply investigate cathode materials, the AI-experts are more skilled in finding biases in the database, narrowing the boundaries of the cathode materials space.
Consequently, the false positive rate when working with small training sets is likely higher compared to cathodes for Li, Na, and Mg ion batteries.
Another possible interpretation is that, in relatively large datasets indicating deep historical investigation, the remaining unexplored materials are fewer than in the case of new \emph{post-lithium} cathodes, which explains the numerous candidates found for these latters.

Regarding the protocol's performance for this case study, we must consider the empirical evidence showing that the accuracy of ML models increases with the size and variety of the training set, independent of the model architecture's complexity~\cite{banko_scaling_2001}.
Therefore, the materials science community can numerically or experimentally validate the possible candidates identified by the AI-experts and enrich the starting specialized materials database.
As depicted in the protocol Figure~\ref{fig:protocol}, these initial results represent just the starting point. 
The next round of ``active learning'' will likely enhance the accuracy and precision of both the regressor and classifier models.

In Figure~\ref{fig:batteries_result_all}, we report all the new potential cathodes for the nine working ions.
In addition to the four properties predicted by the E(3)NN committee, we compute additional properties for the pure cathode material that can aid in decision-making for further investigation.
Namely, we compute the gravimetric and volumetric capacities and energies based on the E(3)NN predictions, the primitive unit cell, and the number of active elements within the cell. 
It is important to note that here, the terms gravimetric and volumetric capacities refer to the mass and volume of the cathode material alone, so as not to be confused with the volumetric and gravimetric properties of the entire battery system (which includes the cathode, anode, electrolyte, etc.).
We compute the gravimetric capacity using the following equation:
\begin{equation}\label{eq:grav_capacity}
  q_g = \dfrac{F}{3.6} \cdot \dfrac{\sum_{i=X}q_i}{\sum_i m_i},
\end{equation}
where $q_g$ is the gravimetric capacity in $\unit{\milli\ampere\hour\per\gram}$, $F$ is the Faraday constant in $\unit{\coulomb\per\mole}$, $m_i$ is the molar mass of each atom inside the material unit cell in $\unit{\gram\per\mole}$, and $q_i$ is the charge carried by the working ion element inside the crystal unit, estimated using the most probable oxidation state of the atom using the bond valence sum method as detailed in Methods Subsubsection~\ref{sssec:m-case-specific-batteries}.
The factor $3.6$ in Eq.~\eqref{eq:grav_capacity} serves as a unit conversion constant, transforming $\unit{\ampere\second\per\gram}$ to  $\unit{\milli\ampere\hour\per\gram}$.
Similarly, we compute the volumetric capacity as follows:
\begin{equation}\label{eq:vol_capacity}
  q_v = \dfrac{F}{3.6\times 10^{-3}} \cdot \dfrac{\sum_{i=X}q_i}{\left|\mathbf{a}\cdot \left(\mathbf{b} \times \mathbf{c} \right)\right|} ,
\end{equation}
where $q_v$ is the volumetric capacity in $\unit{\milli\ampere\hour\per\liter}$, $F$ is the Faraday constant in $\unit{\coulomb\per\mole}$, $\mathbf{a}$, $\mathbf{b}$, and $\mathbf{c}$ are the crystallographic axis vectors with norms measured in $\unit{\meter}$, and $q_i$ is the charge carried by the working ion element inside the crystal unit as in Eq.~\eqref{eq:grav_capacity}.
The factor $3.6\times 10^{-3}$ in Eq.~\eqref{eq:vol_capacity} serves as a unit conversion constant, converting $\unit{\ampere\second\per\meter^3}$ to $\unit{\milli\ampere\hour\per\liter}$.
The relative gravimetric and volumetric energy is then computed using the predicted average voltage $\Delta V_c$:
\begin{equation}\label{eq:grav_vol_energy}
  E_g = \Delta V_c \cdot q_g, \quad E_v = \dfrac{\Delta V_c \cdot q_v}{1000}, 
\end{equation}
where $E_g$ is the gravimetric energy in $\unit{\watt\hour\per\kilogram}$, $q_g$ is the gravimetric capacity computed as in Eq.~\eqref{eq:grav_capacity}, $E_v$ is the volumetric energy in $\unit{\watt\hour\per\liter}$, and $q_v$ is the volumetric capacity computed as in Eq.~\eqref{eq:vol_capacity}.
The division $E_v$ by 1000 converts the units from $\unit{\milli\watt\hour\per\liter}$ to $\unit{\watt\hour\per\liter}$  since the volumetric capacity $q_v$ is measured in $\unit{\milli\ampere\hour\per\liter}$.

Moving our analysis to Li and Na ion batteries, we report in Figure~\ref{fig:batteries_result_LiNa} the distribution of all the Li and Na potential cathodes with respect to the classifier committee's average probability $P$ and the average predicted voltage $\Delta V_c$.
Among the candidates found using the ranking function Eq.~\eqref{eq:batteries_ranking}, we identify $\mathrm{Li_{12}ZnCoO_{8}}$,  $\mathrm{LiFePO_{4}}$ and $\mathrm{Li_{5}CoF_{8}}$. 
$\mathrm{Li_{12}ZnCoO_{8}}$ has a predicted voltage of $\SI{1.99}{\volt}$ and shows an extremely high gravimetric capacity of $\SI{958}{\milli\ampere\hour\per\gram}$, which exceeds the highest capacity commercially available cathode $\mathrm{LiMnO_{2}}$ ($\SI{285}{\milli\ampere\hour\per\gram}$)  and is below the under-investigation $\mathrm{Li_{2}S}$ cathode~\cite{wang_high-performance_2023} with the theoretical gravimetric capacity of $\SI{1675}{\milli\ampere\hour\per\gram}$.
It shows a layered structure common to other cobalt oxide cathodes like $\mathrm{LiCoO_{2}}$.
The model also identifies $\mathrm{Li_{5}CoF_{8}}$ as a possible LIB cathode.
According to the E(3)NN models, the average voltage is $\SI{4.89}{\volt}$, resulting in a theoretical specific energy extremely high at $\SI{2670}{\watt\hour\per\kilogram}$.
However, from the crystal structure, the positions of the lithium ions do not show a fully layered structure, and the high voltage makes it impractical since it exceeds the electrochemical window of common electrolyte compositions ($\sim\SI{4.5}{\volt}$ upper limit for Fluoroethylene-carbonate (FEC) and $\mathrm{LiPF_6}$ electrolyte for high-voltage cathodes~\cite{teufl_implications_2023}). 
This incomplete compliance with the requirements for a good LIB cathode is also predicted by the AI-experts' average probability $P=0.75$  (i.e., there is a 25\% chance that this is not a suitable cathode).
A peculiar result is the identification and presence of $\mathrm{LiFePO_{4}}$ in GNoME, which is already present in the MP database and has been experimentally investigated, explaining the high AI-experts' average probability of $P=0.96$.
Upon examining the data in the crystal files of GNoME and the MP database, the only difference between the two is a Euclidean rotation, which explains the good E(3)NN prediction of the expected voltage $\SI{3.52}{\volt}$ ($\SI{3.79}{\volt}$ for the cathode in the MP database).

For the Na cathode candidates, using the same ranking function, we identify $\mathrm{Na_{10}MnFe_{3}P_{4}(CO_{7})_{4}}$, $\mathrm{Na_3Cr(PO_4)_2}$, and $\mathrm{Na_3V(PO_4)_2}$.
$\mathrm{Na_{10}MnFe_{3}P_{4}(CO_{7})_{4}}$ has a predicted voltage of $\SI{3.70}{\volt}$ and shows good gravimetric capacity of $\SI{250}{\milli\ampere\hour\per\gram}$ and energy density of $\SI{924}{\watt\hour\per\kilogram}$. 
It shows a typical olivine structure with Na aligned in channels. 
Many Na-ion battery cathode materials show similar transition metal and polyanionic frameworks, such as manganese and iron combined with phosphate, like the Na super ionic conductor (NASICON) type $\mathrm{Na_4Fe_3(PO_4)2(P_2O_7)}$ cathode~\cite{subasi_synthesis_2024}.
This explains the high probability that the AI-experts associate with it ($P=0.97$). 
However, the presence of the carbon-oxygen groups $\mathrm{CO_{7}}$ is quite exotic, and to our knowledge, has never been observed in Na cathodes.
The other two top-ranked materials, $\mathrm{Na_3V(PO_4)_2}$ and $\mathrm{Na_3Cr(PO_4)_2}$, are extremely similar, where the V in the first is substituted with Cr, which is next in the periodic table. 
Indeed, they show very close gravimetric capacities of $\SI{259.5}{\milli\ampere\hour\per\gram}$ and $\SI{258.6}{\milli\ampere\hour\per\gram}$, respectively. 
The E(3)NN committee predicted different average voltages: $\SI{2.96}{\volt}$ for $\mathrm{Na_3V(PO_4)_2}$ and $\SI{3.61}{\volt}$ for $\mathrm{Na_3Cr(PO_4)_2}$.
The $\mathrm{Na_3V(PO_4)_2}$ is also similar to an already under-investigation cathode, the NASICON-type polyanion sodium vanadium phosphate (NVP) $\mathrm{Na_3V_2(PO_4)_3}$~\cite{he_research_2023}, which explains the very high probability over the 10 classifiers ($P=0.98$).

\section{Discussion}
\label{sec:conclusions}
The AI protocol presented in this work demonstrates a computationally highly efficient approach for screening vast unexplored material databases such as GNoME to identify promising candidates for energy applications.

We hypothesize that known energy-related materials, such as thermoelectric materials, perovskites, and electrochemical battery cathodes, are affected by biases due to human investigation history. 
Indeed, it is reasonable to assume that the materials published in the literature and later included in specialized databases were not randomly selected and tested (either numerically or experimentally) over time, but rather carefully chosen based on prior knowledge so as to increase the likelihood that such materials would exhibit high performance. 
As a consequence, this accurate selection of materials leads to an uneven and non-uniform representation of the materials space within a given database, ultimately resulting in an anthropogenic bias \cite{Jia_Lynch_Huang_Danielson_Langat_Milder_Ruby_Wang_Friedler_Norquist_etal._2019}.
These biases are invisible in a low-dimensional space like the measurable property space of materials, but we show that they can be identified and exploited by training skilled AI-expert classifiers.
Such classifiers operate in high-dimensional feature spaces for screening.
Indeed, by leveraging complex feature interactions rather than simple acceptance criteria applied to individual features, our method provides a more comprehensive screening process. 
This is particularly evident when compared to previous approaches, such as the one used by  Cerqueira et al.~\cite{cerqueira2024sampling} for superconductors. %

We trained a committee of 10 classifiers, named AI-experts, that learn the biases of the training set and replace human experience in selecting possible energy material candidates. 
Additionally, we trained a committee of 4 decision trees or equivariant neural networks, depending on the database quality, to predict relevant properties for energy materials. 
These properties include figure of merit for thermoelectric materials, band gap for perovskites, and average voltage for cathode materials in batteries.

One of the key advantages of this method is its efficiency in narrowing down the candidate pool size.
This targeted approach allows for a more efficient exploration of the energy materials space, which can potentially save a significant amount of time and resources during experimental and numerical validation. 

A second key advantage is that, given the poor extrapolation capability of ML tools, the property predictions are more robust and reliable within the feature space shared with the training data.

However, a limitation remains in the inability to explicitly measure the false-positive rate for the identified materials, as no direct method for \emph{a priori} quantification has been integrated into the current protocol.
While this limitation does not diminish the overall effectiveness of the protocol, it highlights an area for potential improvement.

We see this work as the first step of a continuous community effort. Further improvements and developments are expected in the future with the very likely discovery of new stable materials and knowledge advancement on the properties of known materials. 
Along this direction, a natural next step for improvement lies in expanding the training dataset.
As more materials are experimentally validated and incorporated into the training set, the accuracy and precision of both the classifier and regressors will improve. 
This planned refinement will enhance the feature space, leading to more accurate and narrowly focused predictions.

Thus, this protocol calls for cooperative ``active learning'' by the community, which will accelerate the discovery of new and high-performing energy materials.

Finally, it is worth noting that - for the sake of simplicity and without a loss of generality of the described methods - in the current work we have predominantly focused on physical and chemical figures of merits. In future development, we expect to include also further material screening taking into account other key aspects such as the expected (eco)toxicity and sustainability. The latter could be possibly achieved by an additional classification step based on available databases~\cite{mancardi_computational_2023}.

\section{Methods}\label{sec:methods}

This Section details our heuristics-driven protocol for evaluating potential new materials for energy applications. 

Our approach is motivated by recent advancements in materials discovery research while addressing critical limitations in applying machine learning in property prediction for materials.

The first advancement is increasing reliance on ML techniques to accelerate materials discovery and optimization processes, which has led to the publication of several specialized databases for various classes of materials for energy applications~\cite{himanen_datadriven_2019,jain2013commentary, hellenbrandt2004inorganic,saal2013materials,draxl2019nomad,curtarolo2012aflowlib,talirz_materials_2020,winther_catalysis-hub.org_2019,zhou_medium-_2019}, such as the four used in this study.
The second advancement stems from the publication of an active learning method by Merchant et al.~\cite{merchant_scaling_2023} that facilitated the discovery of over 380,000 new stable crystal structures, culminating in the GNoME database.

Despite these advancements offering opportunities to identify new energy materials within the GNoME database, we face two significant challenges related to the limitations of current ML models in materials property prediction.

The first limitation involves the reliability of ML models when predicting properties in high-dimensional feature spaces. 
Many models encode the material's properties in these high-dimensional spaces, relying on regressors with thousands or even millions of degrees of freedom. 
Consequently, these models often act as a ``black box'', creating interpretability challenges and significant uncertainty when applied beyond their training domains.
This issue, where ML models struggle to extrapolate, has been documented by Shimakawa et al.~\cite{shimakawa_extrapolative_2024}.
They empirically showed that ML tools perform well when interpolating within the boundaries of the training set but perform poorly when extrapolating beyond them.
The second limitation is the inherent bias present in these databases used for training~\cite{trezza_classification_2024}. 
The source of these biases may be internal due to the material classes' nature or external due to human-driven selection processes. 
Indeed, most materials in these databases were discovered through selection processes guided by human knowledge and intuition, introducing biases into the datasets~\cite{jain_new_2016, trezza_classification_2024}. 
To address and mitigate these issues, under our assumptions, we propose applying regressors only to subsets of data points that share similar biases with the training set so that the model works inside the interpolated region~\cite{trezza_classification_2024}.
\begin{figure}[ht!]
\centering
\includegraphics[width=\onecolumnfig]{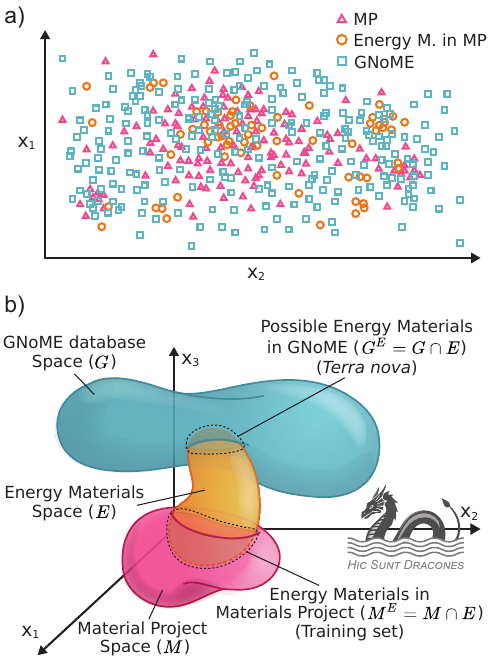}
\caption{
Conceptual illustration of the method.
In low-dimensional space (a), materials are plotted with experimental or chemical properties, creating scattered, often unclassifiable points.
Here, we illustrate the hypothetical case with data from the MP (red triangles), energy-related MP data for battery or perovskite studies (orange circles), and GNoME data (green squares).
In high-dimensional feature space (b), generated from chemical and structural descriptors, distinct n-dimensional regions emerge for MP ($M$) and GNoME ($G$) data.  
We hypothesize the existence of an orange region $E$, where all energy-related materials (e.g., cathodes, perovskites) reside. 
The AI-experts use the intersection $M \cap E$ and the remaining MP data, $M \setminus E$, to define the boundary of $E$. 
This enables the identification of $G \cap E$, the crystals in GNoME with similar properties to $M \cap E$, where regression is more reliable than when applied to the whole $G$ set.
}
\label{fig:method-concept}
\end{figure}

To illustrate this hypothesis, we present the conceptual framework in Figure~\ref{fig:method-concept}, where we assume that the GNoME dataset ($G$), the Materials Project database ($M$), and a specific subset for energy applications ($M^E$) are being used. 
Typically, these datasets differ by orders of magnitude in size, with GNoME containing approximately 380,000 materials, MP containing around 34,000 stable materials, and energy-specific subsets usually comprising thousands of data points.
The energy-specific subset $M^E = M \cap E$ represents the intersection of known materials from the MP dataset and a hypothetical set of all possible materials suited for a given energy application, expressed as $E$.
This hypothetical set $E$ is defined not only by the application-specific requirements but also by historical biases inherent in the materials discovery process, including factors such as synthesis feasibility, experimental limitations, and researchers' prior knowledge.

Our primary hypothesis is that the energy-relevant region in the material space overlaps with the GNoME dataset, indicated as $G^E = G \cap E$. 
If we can accurately estimate the boundary of this subset ($\partial E$), we can effectively screen materials in GNoME that have high potential for use in energy applications. 
This boundary estimation is crucial because it will make the regressor predictions more reliable and allow us to focus our computational resources on the most promising candidates within the vast GNoME dataset.
Here, to implement this approach, we propose training a committee of classifiers -- ``AI-experts'' -- assigned with the task of distinguishing materials in the energy-specific subset ($M^E$) from the remaining part of broader MP dataset ($M\setminus E = M\setminus M^E$). 
These classifiers aim to simulate the decision-making processes of human experts in the field, incorporating the knowledge-derived biases we hypothesize exist in the energy-related subset.
Crucially, the distinction between these three main sets ($G$, $M$, and $E$) is challenging in low-dimensional spaces that can only be represented by few theoretical or experimentally measured properties of the crystal, as illustrated in Figure~\ref{fig:method-concept}a. 
Our second hypothesis is that the separation between these sets becomes more evident in high-dimensional feature spaces, where materials are represented by a more comprehensive set of descriptors, as shown in the ideal case displayed in Figure~\ref{fig:method-concept}b. 
These higher-dimensional spaces are more suitable for ML tasks, as they can potentially reveal patterns and relationships not evident in lower-dimensional representations~\cite{manzhos_machine_2023, ward_general-purpose_2016}.

Building on the overview of the entire protocol provided in Section~\ref{sec:protocol}, the following Sections will detail our methodology, including the technical explanation of the pre-processing and dataset featurization steps, the screening process, and the architecture and training of the regressor.

\subsection{Data pre-processing}\label{ssec:m-pre-processing}
The data pre-processing employed in this work consists of building and cleaning the specialized energy material database.

Various sources were used for the four test cases: Materials Project and published literature database, depending on the specific energy material class under investigation.
For materials available in the MP database, we utilized their API (application programming interface)~\cite{ong_materials_2015} through the python library \texttt{mp-api} to efficiently retrieve relevant information. 
In other cases, we obtain the database from published literature additional material or associated repositories. 

All the specialized ``energy'' material databases were then cleaned to enhance data quality. 
This process encompassed handling missing values, standardizing units, and unifying file formats.
Due to the diverse nature of the training databases used in this study, data cleaning procedures were specifically tailored to each dataset.
Detailed descriptions of these case-specific procedures can be found in Subsection~\ref{ssec:m-case-specific-info}.

\subsection{Materials featurization}\label{ssec:m-materials-featurization}

Depending on the model used for property prediction, we have two possible ways to translate materials into machine-readable data. 
When the data flows through the \textit{structure pipeline}, for regression tasks we perform a Structural Encoding (SE) process proposed by Chen et al.~\cite{chen_direct_2021}, which involves creating a periodic graph representation of the atomic structure of the crystal, with chemical information embedded in the graph using one-hot encoding.
The graph representation consists of $N_n$ nodes $A_i$, each representing an atom in the crystal unit, and $N_e$ edges $e_{ij}$ storing the interatomic distances $r_{ij}$ between neighboring atoms within a cutoff radius $r_{\text{cut}}$ set to $\SI{5}{\angstrom}$ . 
Periodic Boundary Conditions (PBC) are applied to identify neighboring atoms, resulting in a periodic graph.
Chemical information is then encoded at each node using one-hot representation of the atomic element and mass. 
This results in a node feature vector $\mathbf{x}_i=\left\{ x_j \right\} \in \mathbb{R}^{118}$, with the $j$-th element defined as $x_j = m_i \cdot \delta_{ij=Z_i} $, 
where $\delta_{ij}$ is the Kronecker delta, $m_i$ is the atomic mass of atom $A_i$, and $Z_i$ is the atomic number of $A_i$.
For example, if the node $A_i$ represents a Li atom, the corresponding vector would be $\mathbf{x}_i = \left\{ 0, 0, 6.94, 0, \dots, 0 \right\}^{\top}$.
For classification tasks, we featurize each material with a descriptor vector $\mathbf{x}_i\in\mathbb{R}^{694}$, encompassing both composition- and structure- based features.
Such composition-based descriptors include stoichiometric distribution moments, fractional presence of each element within the compound, average number of electrons in each orbital, features related to possible oxidation states, and elemental properties obtained from the Materials-Agnostic Platform for Informatics and Exploration (Magpie) database~\cite{ward_general-purpose_2016}.
Conversely, the structure-based ones come from the Jarvis-ML descriptors~\cite{choudhary_machine_2018}, considering cell and chemical composition.

The database fed to the \textit{composition pipeline} is featurized based on the chemical composition of the materials using the \texttt{matminer} Python library~\cite{ward_matminer_2018}.
In this case, we rely for both regression and classification tasks only on 145 (146 considering the temperature) chemical composition-based features, including stoichiometric features, elemental property statistics, and electronic structure characteristics. 
Further details on both featurization processes can be found in Subsubsection~\ref{sssec:m-case-specific-thermoelectrics}

\subsection{Regressors training}\label{ssec:m-regressors-training}

From the periodic graph representation of the crystal, we utilize a tailored version of the E(3)NN model originally proposed by Chen et al.~\cite{chen_direct_2021} for predicting the phonon density of states (pDOS) across various crystal structures.
This model initially applies a linear transformation to reduce the 118 node features to 64 embedded chemical features.
The data then passes through two layers of graph ``convolutions and gating'' equivariant operations.
The convolution kernel used is a product of learnable radial functions and spherical harmonics of the form:
\begin{equation*}
     K_m^{(l)} (\mathbf{r}_{ij}) = R(r_{ij}) Y_m^{(l)}(\hat{\mathbf{r}}_{ij})
\end{equation*}
where $\mathbf{r}_{ij}$ is the distance vector between $i$-th and $j$-th atoms, $r_{ij}$ and $\hat{\mathbf{r}}_{ij}=\mathbf{r}_{ij}/r_{ij}$ its associated norm and direction vector.
Therefore, a crucial set of hyperparameters includes the maximum order of the spherical harmonics, which was set to $l_{max}=3$, and the radial function defined as a fully connected neural network (FNN):
\begin{equation*}
     R(r_{ij}) = \sum_h W_{kh}\sigma\left(\sum_q W_{hq} B_q\left(r_{ij}\right)\right),
\end{equation*}
where $W$ is the weight matrix of the input and hidden layer, $B_q$ are the radial basis functions, and $\sigma$ is the activation function.
For the training of the pure and mixed models for the perovskite materials, the optimal maximum order of the spherical harmonics was found to be $l_{max}=2$.
We use 10 equally distanced (from $\SI{0}{\angstrom}$ to $r_{cut}=\SI{5}{\angstrom}$) Gaussian radial basis functions, 100 neurons for the hidden layer, and the Sigmoid Linear Unit (SiLU) as the activation function ($\sigma(x)=x \left(1+e^{-x}\right)^{-1}$).
The data then passes through a final single graph convolution layer before summing all the resulting embedded features tensor from all atoms into a single one (sum-pooling). 
Then, the output passes through a Rectified Linear Unit (ReLU) activation layer, and the average of the ReLU outputs (mean-pooling) is the final scalar quantity.
For additional details on the mathematics of the graph neural network, we invite the reader to refer to the original work by Chen et al.~\cite{chen_direct_2021}, the \texttt{e3nn} Python library articles and documentation~\cite{e3nn_paper, e3nn}, and related publications~\cite{thomas2018tensorfieldnetworksrotation, weiler20183dsteerablecnnslearning, kondor2018clebschgordannetsfullyfourier, batzner_e3-equivariant_2022}.
The model is then trained for 100 epochs using the AdamW optimizer~\cite{loshchilov2019decoupledweightdecayregularization}, with an exponentially decaying learning rate and $L_1$-norm (Mean Absolute Error (MAE)) as loss function.
The training behavior primarily depends on the initial learning rate $\eta_{0}$, the weight decay $\lambda$ for the optimizer, and the exponential base of the learning rate decay $\beta$ ($\eta_{i}=\eta_{0}\beta^{i}$ for $i$-th epoch).
Specifically, for the cathode materials, we found optimal $\eta_{0} = \num{1e-3}$, $\lambda =  \num{0.1}$, and $\beta = \num{0.99}$; for the perovskites, we used instead $\eta_{0} = \num{0.01}$, $\lambda = \num{0.01}$, and $\beta = \num{0.96}$.
During training, 100 epochs are sufficient to reach a minimum of the loss function. 
We then select the model with the minimum loss on the test set.

For the composition-based regression models, we adopt the GBDT method~\cite{friedman_greedy_2001}.
Specifically, we set up a pipeline by means of the \texttt{GradientBoostingRegressor} object based on the python ML-library \texttt{scikit-learn} \cite{scikit-learn}.
In particular, before the actual training, we perform reduction of data dimensionality by adopting Recursive Feature Elimination (RFE) \cite{guyon_no_2002}.
The key hyperparameters of the resulting ML models are the number of boosting stages to perform ($N_b$) and the number of selected features to use after the RFE step ($N_f$). 
For each AI-expert training, we perform a k-fold (with $k=4$ number of split datasets) cross-validated grid search on the hyperparameter space:
\begin{equation*}
     \begin{split}
     (N_b, N_f) \in & \left\{50, 100, 250, 500\right\} \\
     & \times \left\{ 
          \frac{\left| M^{E} \right|}{40},  
          \frac{\left| M^{E} \right|}{20},  
          \frac{\left| M^{E} \right|}{10} 
          \right\}.
     \end{split}
\end{equation*}

Regardless of the material representation, we train a committee of 4 regressors. 
This approach provides robust predictions of various material properties and offers the additional benefit of measuring the average deviation among the 4 models' predictions. 
This average deviation is not a direct measure of the uncertainty, in the absence of other \emph{a priori} knowledge about the properties, it serves as an indicator of the reliability of the predictions.

\subsection{AI-experts training}\label{ssec:m-classifier-training}

We hypothesize that the specialized ``energy'' material set $M^E$ is affected by biases due to the specific criteria used to select these materials, which in turn stem from the experts' knowledge. 
By leveraging these biases, we can train ML classifiers -- referred to as AI-experts -- to operate in the high-dimensional space of composition-based descriptors, effectively distinguish materials belonging to $M^E$ from a randomly constructed set $M^{NE} \subset M \setminus M^E$.
To achieve this, we define two classes for our binary classification task:
\begin{itemize}
     \item \textit{Class 1}: Materials within the specialized set $M^E$.
     \item \textit{Class 0}: A randomly selected set of materials $M^{NE}$ from the Materials Project database that do not overlap with $M^E$ (i.e., $M^{NE} \subset M \setminus M^E$), removing possible \textit{polymorphs} crystals, and ensuring that the two classes have the same cardinality $\left| M^E \right| = \left| M^{NE} \right|$ to maintain balance between the classes.
\end{itemize}
Our AI-experts consist of a committee of 10 binary classifiers. 
For this work, we found GBDTs~\cite{friedman_greedy_2001} efficient and sufficiently accurate.
Using the python ML-library \texttt{scikit-learn}~\cite{scikit-learn} pipeline construction, we pre-processed the data before feeding it to the GBDT (\texttt{GradientBoostingClassifier} object in \texttt{scikit-learn}). 
Indeed, before feeding the data into the GBDT classifiers, we standardized the input features to ensure zero average and unit variance for all features. 
Then, the data dimension is reduced by performing RFE~\cite{guyon_no_2002}, to constrain that the dimension of the composition-based descriptor is an order of magnitude lower than the cardinality of the specialized material set $M^E$.
The key hyperparameters of the resulting ML models are the number of boosting stages to perform ($N_b$) and the number of selected features to use after the RFE step ($N_f$). 
For each AI-expert training, we performed a k-fold (with $k=4$ number of split datasets) cross-validated grid search on the hyperparameter space:
\begin{equation*}
     \begin{split}
     (N_b, N_f) \in & \left\{50, 100, 250, 500\right\} \\
     & \times \left\{ 
          \frac{\left| M^{E} \right|}{40},  
          \frac{\left| M^{E} \right|}{20},  
          \frac{\left| M^{E} \right|}{10} 
          \right\}.
     \end{split}
\end{equation*}

From the resulting skilled AI-experts, for each material we obtain a value between 0 and 1, which can be interpreted as a probability $P_i$ that the input material $x$ falls inside the region of the investigated energy material set $M^E$ (class 1). 
Under our hypothesis, this also represents the probability of the input material $x$ being inside the energy material region $E$.
This can also be interpreted as the AI-experts approximating the boundary $\partial E$ with a smooth transition, represented by the classifier's output.

\subsection{Screening}\label{ssec:m-screening}

The specialized AI-experts can then be used to screen the GNoME materials $y \in G$. 
After the composition featurization described earlier and classification processing, we can impose the acceptance criteria:
\begin{equation*}
     \begin{split}
     P(y \in M^E) &= \frac{1}{N}\sum_{i=1}^N P_i(y \in M^E) \\
     &\approx P(y \in E) > 0.5,
     \end{split}
\end{equation*}
where $N=10$ is the number of AI-experts, and $P_i$ is the prediction of the single AI-expert interpreted as the probability of the GNoME material $y$ belonging to ``class 1'', which by construction is equivalent to the probability $P_i(y \in M^E) \approx P_i(y \in E) $.
This ensemble approach allows us to effectively utilize the biases present in $M^E$ to identify potential candidate materials within $G$ that share similar characteristics, which should attenuate the extrapolation problem of the regression models and the benefits of reducing the pool size of materials to investigate as potential new materials.

\subsection{Case-specific pre- and post-processing}\label{ssec:m-case-specific-info}

\subsubsection{Thermoelectric materials}\label{sssec:m-case-specific-thermoelectrics}

For the thermoelectric materials investigation, within the ESTM database, after normalizing the stoichiometry of the formulae, some of the materials appeared multiple times with the same $T$ value, along with different measured $zT$ values. 
In such cases we retained the average $zT$ value only for those instances showing a RSD over such $zT$ values less than 20\%.
This brought the number of materials from the original 870 to 869 (Ag$_{100}$Bi$_{63}$Nb$_7$Sb$_{30}$Se$_{200}$ is the only material reported with two highly different $zT$ values for each of the temperatures it is listed with, namely \SI{323}{\kelvin}, \SI{423}{\kelvin}, \SI{523}{\kelvin}, \SI{623}{\kelvin}, \SI{723}{\kelvin}, \SI{823}{\kelvin}), along with a reduction of the $T$-based replicas from 5,101 to 5,061.

Also, the materials for “class 0” were randomly selected within MP, with the only constraint that these compositions were not already present in the ESTM database.

As already mentioned above, such compounds are featurized with 145 composition-based features encompassing stoichiometric attributes (based on the ratios of elements), elemental property statistics (including the mean, absolute deviation, minimum, and maximum of 22 atomic properties such as atomic number and atomic radii), electronic structure attributes (which represent the average fraction of electrons in the $s$, $p$, $d$, and $f$ valence shells for all elements in the compound), and ionic compound attributes (indicating the possibility of forming an ionic compound, assuming all elements exist in a single oxidation state), adding the temperature $T$ which plays as the 146$^{\mathrm{th}}$ feature.

For plotting and ranking, we used the following function:
\begin{equation}\label{eq:thermoelectrics_ranking}
\begin{split}
R^{T}(x)  & = w_1 \cdot n\left( P(x) \right)  \\
& + w_2 \cdot n\left(  zT(x) \right)\  \\
& + w_3 \cdot n\left( \sigma\left( zT(x) \right)\ \right),  \\
\end{split}
\end{equation}
where:
\begin{description}
     \item[$n(x)$] is the min-max normalization function $n(x) = \frac{x-\min (x)}{\max (x)- \min (x)}$;
     \item[$P(x)$] is the average probability from the AI-experts;
     \item[$ zT(x) $]  is the ``figure of merit'' predicted by the regressors;
     \item[$\sigma\left( zT(x) \right)$]  is the standard deviation of the predicted ``figure of merit'';
     \item $\left[ w_{1}, w_{2}, w_{3} \right]$ are arbitrarily chosen weights, respectively equal to [2, 3, -1].
\end{description}
Note that using negative weights, $w_i$, penalizes that specific candidate property -- in this case study, the deviation of the regressor committee prediction, $\sigma\left( zT(x) \right)$.

\subsubsection{Perovskite}\label{sssec:m-case-specific-perovskite}

For the perovskite materials investigation, the materials for ``class 1'' in MP were selected with the constraint of being suitable for PV applications.
First, the material should not be a metal, as the valence and conduction bands must not overlap.
Furthermore, the material should not possess magnetic properties: these would enhance the probability of self-trapping of charge carriers, thus resulting in reduced carrier mobility and increased recombination rates~\cite{haas_magnetic_1970,ren_tuning_2020}.
Finally, the material should display a band gap in the range $\SI{0.0}{\electronvolt}<E_{\mathrm{g}}\leq\SI{2.5}{\electronvolt}$~\cite{hu_review_2019,horantner_potential_2017,liu_highly_2021}.
The latter constraint was also applied when randomly selecting the non-perovskite materials used to enhance the dataset for the mixed model for regression.
On the contrary, for the training of the AI-experts, the materials for ``class 0'' were randomly selected without any constraints.
The adoption of the same constraints would have had a negative impact on the classification capabilities of the models, as having only non-metallic and non-magnetic materials would have resulted in a partial loss of the knowledge leveraged by the AI-experts.
At the other end of the spectrum, including only metallic and magnetic materials in ``class 0'' would have introduced an undesired bias that materials not displaying these properties have higher probabilities of being considered possible candidates.

For plotting and ranking, we used the following function:
\begin{equation}\label{eq:perovskite_ranking}
\begin{split}
R^{P}(x)  & = w_1 \cdot n\left( P(x) \right)  \\
& + w_2 \cdot n\left( \sigma\left( P(x) \right)\right)  \\
& + w_3 \cdot n\left( \sigma\left( E_g(x) \right)\ \right) \\
& + w_4 \cdot n\left( \left|E_g(x) - 1.34\right| \right), \\
\end{split}
\end{equation}
where:
\begin{description}
     \item[$n(x)$] is the min-max normalization function $n(x) = \frac{x-\min (x)}{\max (x)- \min (x)}$;
     \item[$P(x)$] is the average probability from the AI-experts;
     \item[$\sigma\left( P(x) \right)$] is the standard deviation of the AI-experts prediction;
     \item[$\sigma\left( E_g(x) \right)$] is the standard deviation of the predicted band gap;
     \item[$\left| E_g(x) - 1.34 \right|$] is the distance of the predicted band gap from the ideal $\SI{1.34}{\electronvolt}$ value~\cite{konstantakou_critical_2017,zhou_highly_2020};
     \item $\left[ w_{1}, w_{2}, w_{3}, w_{4}\right]$ are arbitrarily chosen weights, respectively equal to [4, -2, -2, -1].
\end{description}
Note that using negative weights, $w_i$, penalizes that specific candidate property -- in this case study, the deviation of the regressor committee prediction, $\sigma\left( zT(x) \right)$, and AI-experts $\sigma\left( P(x) \right)$ and the  distance for ideal $\SI{1.34}{\electronvolt}$ band gap.

\subsubsection{Cathodes}\label{sssec:m-case-specific-batteries}

For the cathode materials database, a crucial preprocessing step was outlier detection to eliminate possible data anomalies. 
Given the small training sets available for some working ions, noise produced by irregularities could strongly degrade ML model performance.
We employed the Interquartile Range (IQR) method~\cite{upton_understanding_1996}. 
For each property of interest for our material class, we calculated the first quartile ($Q_1$) and third quartile ($Q_3$), and determined the IQR as $IQR=Q_3 - Q_1$.
Thus the data points that fell outside the range $\left[Q_1-k\cdot IQR, Q_3+k\cdot IQR \right]$ were removed since they were considered a possible outlier.
The scale value $k$ for the IQR is usually set at 1.5. 
Since an actual high-performance material can be mistaken as an outlier, we set $k$ conservatively to 3.0 to retain a broader range of data.

For the cathode materials investigation, the materials for ``class 0'' were randomly selected with the constraint that the working-ion (Li, Na, Mg, K, Ca, Cs, Al, Rb, Y) is present in the composition. 
This constraint was implemented to prevent the classifier from developing a bias that materials containing the specific working-ion are automatically considered as possible candidates, which could potentially overestimate their number.
This condition was also applied after the AI-experts' classification process.

To determine the oxidation state for each working ion within the unit crystal cell of the candidate cathodes, we employ the Python library \texttt{pymatgen}~\cite{ong_python_2013}, specifically its \texttt{oxi\_state\_guesses} function.
The algorithm first computes the bond valence sum $V_i$ of all symmetrically distinct $i$-th sites using the method and parameters defined by O'Keefe and Brese~\cite{okeefe_atom_1991} with the formula:
\begin{equation*}
     V_i = \sum_{j}^{j\in N} e^{\dfrac{R_{ij}-d_{ij}}{b}},
\end{equation*}
where $b$ is a "universal" constant equal to $\SI{0.37}{\angstrom}$, $d_{ij}$ is the bond length between the $i$-th atom and the $j$-th atom in the neighborhood set $N$, and $R_{ij}$ is the bond valence parameter defined as:
\begin{equation*}
     R_{ij} = \left( r_i + r_j -\dfrac{r_i r_j\left(\sqrt{c_i} - \sqrt{c_j}\right)^2}{c_i r_i  + c_j r_j }\right) \left( 1-\delta_{ij} \right).
\end{equation*}
Here, $\delta_{ij}$ is the Kronecker delta, $r_i$ is the "size" parameter, and $c_i$ is a second parameter related to electronegativity, both tabulated for each element.
The posterior probabilities of all oxidation states $X^{n+}$ of the atom element is computed using Bayesian inference:
\begin{equation*}
     P\left(X^{n+} | V_i\right) = P\left(V_i | X^{n+} \right) \cdot P\left( X^{n+} \right),
\end{equation*}
where the likelihood probability $P\left(V_i | X^{n+} \right)$ is modeled as a Gaussian distribution with mean and standard deviation determined from an analysis of the Inorganic Crystal Structure Database (ICSD), and the prior probability $P\left( X^{n+}, \right)$ is tabulated and derived from a frequency analysis of the ICSD.
Finally, the algorithm computes and selects the most probable oxidation state combination that results in a charge-neutral cell.

For plotting and ranking, we used the following function:
\begin{equation}\label{eq:batteries_ranking}
\begin{split}
R^{B}(x)  & = w_1 \cdot n\left( P(x) \right) \\
          & + w_2 \cdot n\left( \sigma\left( P(x) \right)\right)  \\
          & + w_3 \cdot n\left( \sigma\left( \Delta V_c(x) \right)\ \right) \\
          & + w_4 \cdot n\left( q(x) \right) \\
          & + w_5 \cdot n\left( e(x) \right)  \\
          & + w_6 \cdot n\left(  \max \left(\Delta Vol(x)\right)  \right)  \\
          & + w_7 \cdot n\left( \Delta E_{charge}(x) \right)  \\
          & + w_8 \cdot n\left( \Delta E_{discharge}(x) \right).   
\end{split}
\end{equation}
Where:
\begin{description}
     \item[$n(x)$] is the min-max normalization function $n(x) = \frac{x-\min (x)}{\max (x)- \min (x)}$;
     \item[$P(x)$] is the average probability from the AI-experts;
     \item[$\sigma\left( P(x) \right)$] is the standard deviation of the AI-experts prediction;
     \item[$\sigma\left( \Delta V_c(x) \right)$]  is the standard deviation of the ``average voltage'' of the regressor prediction;
     \item[$q(x)$] is the predicted ``gravimetric capacity'';
     \item[$e(x)$] is the predicted ``gravimetric energy'';
     \item[$\max\left(\Delta Vol(x)\right)$] is the predicted maximum voltage change during the phase transition;
     \item[$\Delta E_{charge}(x)$] is the predicted ``energy above the hull'' of the charged state (i.e. the stability of charge);
     \item[$\Delta E_{discharge}(x)$] is the predicted ``energy above the hull'' of the discharged state (i.e. the stability of discharge);
     \item $\left[ w_{1}, \dots, w_{8} \right]$ are arbitrarily chosen weights, respectively equal to [4, -2, -2, 2, 2, -0.5, -0.5, -0.5].
\end{description}
Note that using negative weights, $w_i$, penalizes that specific candidate property -- in this case study, the deviation of the regressor committee prediction for the average voltage, $\sigma\left( V_c(x) \right)$, and AI-experts $\sigma\left( P(x) \right)$, the predicted expansion, $\max\left(\Delta Vol(x)\right)$, and the instability in charged, $\Delta E_{charge}(x)$, and discharged states, $\Delta E_{discharge}(x)$.
 
\backmatter

\section*{Acknowledgements}\label{sec:acknowledgements}
We acknowledge ISCRA (\texttt{IsB29\_NEXT-LIB}) for awarding this project access to the LEONARDO supercomputer, owned by the EuroHPC Joint Undertaking, hosted by CINECA (Italy). %
\section*{CRediT authorship contribution statement}\label{sec:author-contributions-statement}
\textbf{PDA}: Conceptualization, Data curation, Formal Analysis, Investigation, Methodology, Project administration, Resources, Software, Visualization, Writing -- original draft; 
\textbf{GT}: Conceptualization, Data curation, Formal Analysis, Investigation, Methodology, Software, Writing -- original draft; 
\textbf{GB}: Data curation, Formal Analysis, Investigation, Software, Writing -- original draft; 
\textbf{PA}: Funding acquisition, Supervision, Writing -- review \& editing;
\textbf{EC}: Conceptualization, Funding acquisition, Methodology, Project administration, Resources, Supervision, Writing -- review \& editing.
All authors read and approved the final manuscript.
\section*{Additional information}\label{sec:additional-information}

Together with the material presented in this article, there is a supporting information document that provide additional plots and data on the training and testing of the 156 regressors and 120 classifiers used in this study. 
This supporting information document can be accessed at \url{https://TBD}.
\\
\textbf{Code availability}
\\
The entire workflow is managed through Jupyter notebooks, with additional installation requirements and routines provided and maintained in the GitHub repository (\url{https://TBD}) and Zenodo repository (\url{https://TBD}).
\\
\textbf{Data availability}
\\
All GNoME database screening results (Energy-GNoME), including metadata and crystal structure files, are available in our GitHub (\url{https://TBD}) and Zenodo (\url{https://TBD}) repositories. Furthermore, the Energy-GNoME database can be interactively explored using our web application at \url{https://TBD}.
\\
\textbf{Competing interests}
\\
The authors declare no competing financial interest. 
\bibliography{biblio}

\begin{thebibliography}{10}
\expandafter\ifx\csname url\endcsname\relax
  \def\url#1{\burl{#1}}\fi
\expandafter\ifx\csname urlprefix\endcsname\relax\def\urlprefix{URL }\fi
\providecommand{\bibinfo}[2]{#2}
\providecommand{\eprint}[2][]{\url{#2}}
\providecommand{\doi}[1]{\url{https://doi.org/#1}}
\bibcommenthead

\bibitem{Sun2019}
\bibinfo{author}{Sun, H.}, \bibinfo{author}{Edziah, B.~K.}, \bibinfo{author}{Sun, C.} \& \bibinfo{author}{Kporsu, A.~K.}
\newblock \bibinfo{title}{Institutional quality, green innovation and energy efficiency}.
\newblock \emph{\bibinfo{journal}{Energy Policy}} \textbf{\bibinfo{volume}{135}} (\bibinfo{year}{2019}).

\bibitem{kojima2009organometal}
\bibinfo{author}{Kojima, A.}, \bibinfo{author}{Teshima, K.}, \bibinfo{author}{Shirai, Y.} \& \bibinfo{author}{Miyasaka, T.}
\newblock \bibinfo{title}{Organometal halide perovskites as visible-light sensitizers for photovoltaic cells}.
\newblock \emph{\bibinfo{journal}{Journal of the american chemical society}} \textbf{\bibinfo{volume}{131}}, \bibinfo{pages}{6050--6051} (\bibinfo{year}{2009}).

\bibitem{green2014emergence}
\bibinfo{author}{Green, M.~A.}, \bibinfo{author}{Ho-Baillie, A.} \& \bibinfo{author}{Snaith, H.~J.}
\newblock \bibinfo{title}{The emergence of perovskite solar cells}.
\newblock \emph{\bibinfo{journal}{Nature photonics}} \textbf{\bibinfo{volume}{8}}, \bibinfo{pages}{506--514} (\bibinfo{year}{2014}).

\bibitem{snyder2008complex}
\bibinfo{author}{Snyder, G.~J.} \& \bibinfo{author}{Toberer, E.~S.}
\newblock \bibinfo{title}{Complex thermoelectric materials}.
\newblock \emph{\bibinfo{journal}{Nature materials}} \textbf{\bibinfo{volume}{7}}, \bibinfo{pages}{105--114} (\bibinfo{year}{2008}).

\bibitem{Bell2008}
\bibinfo{author}{Bell, L.~E.}
\newblock \bibinfo{title}{Cooling, heating, generating power, and recovering waste heat with thermoelectric systems} (\bibinfo{year}{2008}).

\bibitem{whittingham_electrical_1976}
\bibinfo{author}{Whittingham, M.~S.}
\newblock \bibinfo{title}{Electrical {Energy} {Storage} and {Intercalation} {Chemistry}}.
\newblock \emph{\bibinfo{journal}{Science}} \textbf{\bibinfo{volume}{192}}, \bibinfo{pages}{1126--1127} (\bibinfo{year}{1976}).
\newblock \urlprefix\url{https://www.science.org/doi/10.1126/science.192.4244.1126}.

\bibitem{mizushima_lixcoo2_1980}
\bibinfo{author}{Mizushima, K.}, \bibinfo{author}{Jones, P.}, \bibinfo{author}{Wiseman, P.} \& \bibinfo{author}{Goodenough, J.}
\newblock \bibinfo{title}{{LixCoO2} (0{\textless}x{\textless}-1): {A} new cathode material for batteries of high energy density}.
\newblock \emph{\bibinfo{journal}{Materials Research Bulletin}} \textbf{\bibinfo{volume}{15}}, \bibinfo{pages}{783--789} (\bibinfo{year}{1980}).
\newblock \urlprefix\url{https://linkinghub.elsevier.com/retrieve/pii/0025540880900124}.

\bibitem{lazzari_cyclable_1980}
\bibinfo{author}{Lazzari, M.} \& \bibinfo{author}{Scrosati, B.}
\newblock \bibinfo{title}{A {Cyclable} {Lithium} {Organic} {Electrolyte} {Cell} {Based} on {Two} {Intercalation} {Electrodes}}.
\newblock \emph{\bibinfo{journal}{Journal of The Electrochemical Society}} \textbf{\bibinfo{volume}{127}}, \bibinfo{pages}{773--774} (\bibinfo{year}{1980}).
\newblock \urlprefix\url{https://iopscience.iop.org/article/10.1149/1.2129753}.

\bibitem{kittner2017energy}
\bibinfo{author}{Kittner, N.}, \bibinfo{author}{Lill, F.} \& \bibinfo{author}{Kammen, D.~M.}
\newblock \bibinfo{title}{Energy storage deployment and innovation for the clean energy transition}.
\newblock \emph{\bibinfo{journal}{Nature Energy}} \textbf{\bibinfo{volume}{2}}, \bibinfo{pages}{1--6} (\bibinfo{year}{2017}).

\bibitem{halkos2020reviewing}
\bibinfo{author}{Halkos, G.~E.} \& \bibinfo{author}{Gkampoura, E.-C.}
\newblock \bibinfo{title}{Reviewing usage, potentials, and limitations of renewable energy sources}.
\newblock \emph{\bibinfo{journal}{Energies}} \textbf{\bibinfo{volume}{13}}, \bibinfo{pages}{2906} (\bibinfo{year}{2020}).

\bibitem{nadeem2018comparative}
\bibinfo{author}{Nadeem, F.}, \bibinfo{author}{Hussain, S.~S.}, \bibinfo{author}{Tiwari, P.~K.}, \bibinfo{author}{Goswami, A.~K.} \& \bibinfo{author}{Ustun, T.~S.}
\newblock \bibinfo{title}{Comparative review of energy storage systems, their roles, and impacts on future power systems}.
\newblock \emph{\bibinfo{journal}{IEEE access}} \textbf{\bibinfo{volume}{7}}, \bibinfo{pages}{4555--4585} (\bibinfo{year}{2018}).

\bibitem{hohenberg1964inhomogeneous}
\bibinfo{author}{Hohenberg, P.} \& \bibinfo{author}{Kohn, W.}
\newblock \bibinfo{title}{Inhomogeneous electron gas}.
\newblock \emph{\bibinfo{journal}{Physical review}} \textbf{\bibinfo{volume}{136}}, \bibinfo{pages}{B864} (\bibinfo{year}{1964}).

\bibitem{kohn1965self}
\bibinfo{author}{Kohn, W.} \& \bibinfo{author}{Sham, L.~J.}
\newblock \bibinfo{title}{Self-consistent equations including exchange and correlation effects}.
\newblock \emph{\bibinfo{journal}{Physical review}} \textbf{\bibinfo{volume}{140}}, \bibinfo{pages}{A1133} (\bibinfo{year}{1965}).

\bibitem{Nandy2022}
\bibinfo{author}{Nandy, A.} \emph{et~al.}
\newblock \bibinfo{title}{Mofsimplify, machine learning models with extracted stability data of three thousand metal–organic frameworks}.
\newblock \emph{\bibinfo{journal}{Scientific Data}} \textbf{\bibinfo{volume}{9}} (\bibinfo{year}{2022}).

\bibitem{back2024accelerated}
\bibinfo{author}{Back, S.} \emph{et~al.}
\newblock \bibinfo{title}{Accelerated chemical science with ai}.
\newblock \emph{\bibinfo{journal}{Digital Discovery}} \textbf{\bibinfo{volume}{3}}, \bibinfo{pages}{23--33} (\bibinfo{year}{2024}).

\bibitem{schrier2023pursuit}
\bibinfo{author}{Schrier, J.}, \bibinfo{author}{Norquist, A.~J.}, \bibinfo{author}{Buonassisi, T.} \& \bibinfo{author}{Brgoch, J.}
\newblock \bibinfo{title}{In pursuit of the exceptional: Research directions for machine learning in chemical and materials science}.
\newblock \emph{\bibinfo{journal}{Journal of the American Chemical Society}} \textbf{\bibinfo{volume}{145}}, \bibinfo{pages}{21699--21716} (\bibinfo{year}{2023}).

\bibitem{zdeborova_new_2017}
\bibinfo{author}{Zdeborová, L.}
\newblock \bibinfo{title}{New tool in the box}.
\newblock \emph{\bibinfo{journal}{Nature Physics}} \textbf{\bibinfo{volume}{13}}, \bibinfo{pages}{420--421} (\bibinfo{year}{2017}).
\newblock \urlprefix\url{https://www.nature.com/articles/nphys4053}.

\bibitem{himanen_datadriven_2019}
\bibinfo{author}{Himanen, L.}, \bibinfo{author}{Geurts, A.}, \bibinfo{author}{Foster, A.~S.} \& \bibinfo{author}{Rinke, P.}
\newblock \bibinfo{title}{Data‐{Driven} {Materials} {Science}: {Status}, {Challenges}, and {Perspectives}}.
\newblock \emph{\bibinfo{journal}{Advanced Science}} \textbf{\bibinfo{volume}{6}}, \bibinfo{pages}{1900808} (\bibinfo{year}{2019}).
\newblock \urlprefix\url{https://onlinelibrary.wiley.com/doi/10.1002/advs.201900808}.

\bibitem{lee_cyber-physical_2015}
\bibinfo{author}{Lee, J.}, \bibinfo{author}{Bagheri, B.} \& \bibinfo{author}{Kao, H.-A.}
\newblock \bibinfo{title}{A {Cyber}-{Physical} {Systems} architecture for {Industry} 4.0-based manufacturing systems}.
\newblock \emph{\bibinfo{journal}{Manufacturing Letters}} \textbf{\bibinfo{volume}{3}}, \bibinfo{pages}{18--23} (\bibinfo{year}{2015}).
\newblock \urlprefix\url{https://linkinghub.elsevier.com/retrieve/pii/S221384631400025X}.

\bibitem{casini_machine_2024}
\bibinfo{author}{Casini, M.} \emph{et~al.}
\newblock \bibinfo{title}{Machine {Learning} and image analysis towards improved energy management in {Industry} 4.0: a practical case study on quality control}.
\newblock \emph{\bibinfo{journal}{Energy Efficiency}} \textbf{\bibinfo{volume}{17}}, \bibinfo{pages}{48} (\bibinfo{year}{2024}).
\newblock \urlprefix\url{https://link.springer.com/10.1007/s12053-024-10228-7}.

\bibitem{li_methods_2023}
\bibinfo{author}{Li, J.}, \bibinfo{author}{Herdem, M.~S.}, \bibinfo{author}{Nathwani, J.} \& \bibinfo{author}{Wen, J.~Z.}
\newblock \bibinfo{title}{Methods and applications for {Artificial} {Intelligence}, {Big} {Data}, {Internet} of {Things}, and {Blockchain} in smart energy management}.
\newblock \emph{\bibinfo{journal}{Energy and AI}} \textbf{\bibinfo{volume}{11}}, \bibinfo{pages}{100208} (\bibinfo{year}{2023}).
\newblock \urlprefix\url{https://www.sciencedirect.com/science/article/pii/S2666546822000544}.

\bibitem{casini_current_2024}
\bibinfo{author}{Casini, M.}, \bibinfo{author}{De~Angelis, P.}, \bibinfo{author}{Chiavazzo, E.} \& \bibinfo{author}{Bergamasco, L.}
\newblock \bibinfo{title}{Current trends on the use of deep learning methods for image analysis in energy applications}.
\newblock \emph{\bibinfo{journal}{Energy and AI}} \textbf{\bibinfo{volume}{15}}, \bibinfo{pages}{100330} (\bibinfo{year}{2024}).
\newblock \urlprefix\url{https://linkinghub.elsevier.com/retrieve/pii/S2666546823001027}.

\bibitem{koroteev_artificial_2021}
\bibinfo{author}{Koroteev, D.} \& \bibinfo{author}{Tekic, Z.}
\newblock \bibinfo{title}{Artificial intelligence in oil and gas upstream: {Trends}, challenges, and scenarios for the future}.
\newblock \emph{\bibinfo{journal}{Energy and AI}} \textbf{\bibinfo{volume}{3}}, \bibinfo{pages}{100041} (\bibinfo{year}{2021}).
\newblock \urlprefix\url{https://www.sciencedirect.com/science/article/pii/S2666546820300410}.

\bibitem{liu_machine_2021}
\bibinfo{author}{Liu, Y.}, \bibinfo{author}{Esan, O.~C.}, \bibinfo{author}{Pan, Z.} \& \bibinfo{author}{An, L.}
\newblock \bibinfo{title}{Machine learning for advanced energy materials}.
\newblock \emph{\bibinfo{journal}{Energy and AI}} \textbf{\bibinfo{volume}{3}}, \bibinfo{pages}{100049} (\bibinfo{year}{2021}).
\newblock \urlprefix\url{https://www.sciencedirect.com/science/article/pii/S2666546821000033}.

\bibitem{jain2013commentary}
\bibinfo{author}{Jain, A.} \emph{et~al.}
\newblock \bibinfo{title}{Commentary: {The} {Materials} {Project}: {A} materials genome approach to accelerating materials innovation}.
\newblock \emph{\bibinfo{journal}{APL Materials}} \textbf{\bibinfo{volume}{1}}, \bibinfo{pages}{011002} (\bibinfo{year}{2013}).
\newblock \urlprefix\url{https://pubs.aip.org/apm/article/1/1/011002/119685/Commentary-The-Materials-Project-A-materials}.

\bibitem{hellenbrandt2004inorganic}
\bibinfo{author}{Hellenbrandt, M.}
\newblock \bibinfo{title}{The {Inorganic} {Crystal} {Structure} {Database} ({ICSD})—{Present} and {Future}}.
\newblock \emph{\bibinfo{journal}{Crystallography Reviews}} \textbf{\bibinfo{volume}{10}}, \bibinfo{pages}{17--22} (\bibinfo{year}{2004}).
\newblock \urlprefix\url{http://www.tandfonline.com/doi/abs/10.1080/08893110410001664882}.

\bibitem{saal2013materials}
\bibinfo{author}{Saal, J.~E.}, \bibinfo{author}{Kirklin, S.}, \bibinfo{author}{Aykol, M.}, \bibinfo{author}{Meredig, B.} \& \bibinfo{author}{Wolverton, C.}
\newblock \bibinfo{title}{Materials {Design} and {Discovery} with {High}-{Throughput} {Density} {Functional} {Theory}: {The} {Open} {Quantum} {Materials} {Database} ({OQMD})}.
\newblock \emph{\bibinfo{journal}{JOM}} \textbf{\bibinfo{volume}{65}}, \bibinfo{pages}{1501--1509} (\bibinfo{year}{2013}).
\newblock \urlprefix\url{http://link.springer.com/10.1007/s11837-013-0755-4}.

\bibitem{draxl2019nomad}
\bibinfo{author}{Draxl, C.} \& \bibinfo{author}{Scheffler, M.}
\newblock \bibinfo{title}{The {NOMAD} laboratory: from data sharing to artificial intelligence}.
\newblock \emph{\bibinfo{journal}{Journal of Physics: Materials}} \textbf{\bibinfo{volume}{2}}, \bibinfo{pages}{036001} (\bibinfo{year}{2019}).
\newblock \urlprefix\url{https://iopscience.iop.org/article/10.1088/2515-7639/ab13bb}.

\bibitem{curtarolo2012aflowlib}
\bibinfo{author}{Curtarolo, S.} \emph{et~al.}
\newblock \bibinfo{title}{{AFLOWLIB}.{ORG}: {A} distributed materials properties repository from high-throughput ab initio calculations}.
\newblock \emph{\bibinfo{journal}{Computational Materials Science}} \textbf{\bibinfo{volume}{58}}, \bibinfo{pages}{227--235} (\bibinfo{year}{2012}).
\newblock \urlprefix\url{https://linkinghub.elsevier.com/retrieve/pii/S0927025612000687}.

\bibitem{butler_machine_2018}
\bibinfo{author}{Butler, K.~T.}, \bibinfo{author}{Davies, D.~W.}, \bibinfo{author}{Cartwright, H.}, \bibinfo{author}{Isayev, O.} \& \bibinfo{author}{Walsh, A.}
\newblock \bibinfo{title}{Machine learning for molecular and materials science}.
\newblock \emph{\bibinfo{journal}{Nature}} \textbf{\bibinfo{volume}{559}}, \bibinfo{pages}{547--555} (\bibinfo{year}{2018}).
\newblock \urlprefix\url{https://www.nature.com/articles/s41586-018-0337-2}.

\bibitem{tabor_accelerating_2018}
\bibinfo{author}{Tabor, D.~P.} \emph{et~al.}
\newblock \bibinfo{title}{Accelerating the discovery of materials for clean energy in the era of smart automation}.
\newblock \emph{\bibinfo{journal}{Nature Reviews Materials}} \textbf{\bibinfo{volume}{3}}, \bibinfo{pages}{5--20} (\bibinfo{year}{2018}).
\newblock \urlprefix\url{https://www.nature.com/articles/s41578-018-0005-z}.

\bibitem{Fanourgakis2020}
\bibinfo{author}{Fanourgakis, G.~S.}, \bibinfo{author}{Gkagkas, K.}, \bibinfo{author}{Tylianakis, E.} \& \bibinfo{author}{Froudakis, G.~E.}
\newblock \bibinfo{title}{A universal machine learning algorithm for large-scale screening of materials}.
\newblock \emph{\bibinfo{journal}{Journal of the American Chemical Society}} \textbf{\bibinfo{volume}{142}}, \bibinfo{pages}{3814--3822} (\bibinfo{year}{2020}).

\bibitem{Trezza_Bergamasco_Fasano_Chiavazzo_2022}
\bibinfo{author}{Trezza, G.}, \bibinfo{author}{Bergamasco, L.}, \bibinfo{author}{Fasano, M.} \& \bibinfo{author}{Chiavazzo, E.}
\newblock \bibinfo{title}{Minimal crystallographic descriptors of sorption properties in hypothetical mofs and role in sequential learning optimization}.
\newblock \emph{\bibinfo{journal}{npj Computational Materials}} \textbf{\bibinfo{volume}{8}}, \bibinfo{pages}{123} (\bibinfo{year}{2022}).
\newblock \urlprefix\url{https://www.nature.com/articles/s41524-022-00806-7}.

\bibitem{cerqueira2024sampling}
\bibinfo{author}{Cerqueira, T.~F.}, \bibinfo{author}{Sanna, A.} \& \bibinfo{author}{Marques, M.~A.}
\newblock \bibinfo{title}{Sampling the materials space for conventional superconducting compounds}.
\newblock \emph{\bibinfo{journal}{Advanced Materials}} \textbf{\bibinfo{volume}{36}} (\bibinfo{year}{2024}).

\bibitem{moses2021machine}
\bibinfo{author}{Moses, I.~A.} \emph{et~al.}
\newblock \bibinfo{title}{Machine learning screening of metal-ion battery electrode materials}.
\newblock \emph{\bibinfo{journal}{ACS Applied Materials \& Interfaces}} \textbf{\bibinfo{volume}{13}}, \bibinfo{pages}{53355--53362} (\bibinfo{year}{2021}).

\bibitem{rutt_expanding_2022}
\bibinfo{author}{Rutt, A.} \emph{et~al.}
\newblock \bibinfo{title}{Expanding the {Material} {Search} {Space} for {Multivalent} {Cathodes}}.
\newblock \emph{\bibinfo{journal}{ACS Applied Materials \& Interfaces}} \textbf{\bibinfo{volume}{14}}, \bibinfo{pages}{44367--44376} (\bibinfo{year}{2022}).
\newblock \urlprefix\url{https://pubs.acs.org/doi/10.1021/acsami.2c11733}.

\bibitem{rong_efficient_2016}
\bibinfo{author}{Rong, Z.}, \bibinfo{author}{Kitchaev, D.}, \bibinfo{author}{Canepa, P.}, \bibinfo{author}{Huang, W.} \& \bibinfo{author}{Ceder, G.}
\newblock \bibinfo{title}{An efficient algorithm for finding the minimum energy path for cation migration in ionic materials}.
\newblock \emph{\bibinfo{journal}{The Journal of Chemical Physics}} \textbf{\bibinfo{volume}{145}}, \bibinfo{pages}{074112} (\bibinfo{year}{2016}).
\newblock \urlprefix\url{https://pubs.aip.org/jcp/article/145/7/074112/810162/An-efficient-algorithm-for-finding-the-minimum}.

\bibitem{Wang2023}
\bibinfo{author}{Wang, H.}, \bibinfo{author}{Ouyang, R.}, \bibinfo{author}{Chen, W.} \& \bibinfo{author}{Pasquarello, A.}
\newblock \bibinfo{title}{High-quality data enabling universality of band gap descriptor and discovery of photovoltaic perovskites}.
\newblock \emph{\bibinfo{journal}{Journal of the American Chemical Society}}  (\bibinfo{year}{2023}).

\bibitem{kim2018machine}
\bibinfo{author}{Kim, K.} \emph{et~al.}
\newblock \bibinfo{title}{Machine-learning-accelerated high-throughput materials screening: Discovery of novel quaternary heusler compounds}.
\newblock \emph{\bibinfo{journal}{Physical Review Materials}} \textbf{\bibinfo{volume}{2}}, \bibinfo{pages}{123801} (\bibinfo{year}{2018}).

\bibitem{kang2020machine}
\bibinfo{author}{Kang, P.}, \bibinfo{author}{Liu, Z.}, \bibinfo{author}{Abou-Rachid, H.} \& \bibinfo{author}{Guo, H.}
\newblock \bibinfo{title}{Machine-learning assisted screening of energetic materials}.
\newblock \emph{\bibinfo{journal}{The Journal of Physical Chemistry A}} \textbf{\bibinfo{volume}{124}}, \bibinfo{pages}{5341--5351} (\bibinfo{year}{2020}).

\bibitem{wang2009pubchem}
\bibinfo{author}{Wang, Y.} \emph{et~al.}
\newblock \bibinfo{title}{Pubchem: a public information system for analyzing bioactivities of small molecules}.
\newblock \emph{\bibinfo{journal}{Nucleic acids research}} \textbf{\bibinfo{volume}{37}}, \bibinfo{pages}{W623--W633} (\bibinfo{year}{2009}).

\bibitem{rao2022machine}
\bibinfo{author}{Rao, Z.} \emph{et~al.}
\newblock \bibinfo{title}{Machine learning--enabled high-entropy alloy discovery}.
\newblock \emph{\bibinfo{journal}{Science}} \textbf{\bibinfo{volume}{378}}, \bibinfo{pages}{78--85} (\bibinfo{year}{2022}).

\bibitem{merchant_scaling_2023}
\bibinfo{author}{Merchant, A.} \emph{et~al.}
\newblock \bibinfo{title}{Scaling deep learning for materials discovery}.
\newblock \emph{\bibinfo{journal}{Nature}} \textbf{\bibinfo{volume}{624}}, \bibinfo{pages}{80--85} (\bibinfo{year}{2023}).
\newblock \urlprefix\url{https://www.nature.com/articles/s41586-023-06735-9}.

\bibitem{trezza_classification_2024}
\bibinfo{author}{Trezza, G.} \& \bibinfo{author}{Chiavazzo, E.}
\newblock \bibinfo{title}{Classification-based detection and quantification of cross-domain data bias in materials discovery} (\bibinfo{year}{2024}).
\newblock \urlprefix\url{https://arxiv.org/abs/2311.09891}.
\newblock \eprint{2311.09891}.

\bibitem{sootsman2009new}
\bibinfo{author}{Sootsman, J.~R.}, \bibinfo{author}{Chung, D.~Y.} \& \bibinfo{author}{Kanatzidis, M.~G.}
\newblock \bibinfo{title}{New and old concepts in thermoelectric materials}.
\newblock \emph{\bibinfo{journal}{Angewandte Chemie International Edition}} \textbf{\bibinfo{volume}{48}}, \bibinfo{pages}{8616--8639} (\bibinfo{year}{2009}).

\bibitem{corey_molecular_1953}
\bibinfo{author}{Corey, R.~B.} \& \bibinfo{author}{Pauling, L.}
\newblock \bibinfo{title}{Molecular {Models} of {Amino} {Acids}, {Peptides}, and {Proteins}}.
\newblock \emph{\bibinfo{journal}{Review of Scientific Instruments}} \textbf{\bibinfo{volume}{24}}, \bibinfo{pages}{621--627} (\bibinfo{year}{1953}).
\newblock \urlprefix\url{https://pubs.aip.org/rsi/article/24/8/621/441115/Molecular-Models-of-Amino-Acids-Peptides-and}.

\bibitem{noauthor_jmol}
\bibinfo{title}{Jmol: an open-source {Java} viewer for chemical structures in {3D}} (\bibinfo{year}{2024}).
\newblock \urlprefix\url{https://jmol.sourceforge.net/}.

\bibitem{na2022public}
\bibinfo{author}{Na, G.~S.} \& \bibinfo{author}{Chang, H.}
\newblock \bibinfo{title}{A public database of thermoelectric materials and system-identified material representation for data-driven discovery}.
\newblock \emph{\bibinfo{journal}{npj Computational Materials}} \textbf{\bibinfo{volume}{8}}, \bibinfo{pages}{214} (\bibinfo{year}{2022}).

\bibitem{ward2018matminer}
\bibinfo{author}{Ward, L.} \emph{et~al.}
\newblock \bibinfo{title}{Matminer: An open source toolkit for materials data mining}.
\newblock \emph{\bibinfo{journal}{Computational Materials Science}} \textbf{\bibinfo{volume}{152}}, \bibinfo{pages}{60--69} (\bibinfo{year}{2018}).

\bibitem{ward2016general}
\bibinfo{author}{Ward, L.}, \bibinfo{author}{Agrawal, A.}, \bibinfo{author}{Choudhary, A.} \& \bibinfo{author}{Wolverton, C.}
\newblock \bibinfo{title}{A general-purpose machine learning framework for predicting properties of inorganic materials}.
\newblock \emph{\bibinfo{journal}{npj Computational Materials}} \textbf{\bibinfo{volume}{2}}, \bibinfo{pages}{1--7} (\bibinfo{year}{2016}).

\bibitem{roy_review_2020}
\bibinfo{author}{Roy, P.}, \bibinfo{author}{Kumar~Sinha, N.}, \bibinfo{author}{Tiwari, S.} \& \bibinfo{author}{Khare, A.}
\newblock \bibinfo{title}{A review on perovskite solar cells: {Evolution} of architecture, fabrication techniques, commercialization issues and status}.
\newblock \emph{\bibinfo{journal}{Solar Energy}} \textbf{\bibinfo{volume}{198}}, \bibinfo{pages}{665--688} (\bibinfo{year}{2020}).
\newblock \urlprefix\url{https://linkinghub.elsevier.com/retrieve/pii/S0038092X20300888}.

\bibitem{nair_recent_2020}
\bibinfo{author}{Nair, S.}, \bibinfo{author}{Patel, S.~B.} \& \bibinfo{author}{Gohel, J.~V.}
\newblock \bibinfo{title}{Recent trends in efficiency-stability improvement in perovskite solar cells}.
\newblock \emph{\bibinfo{journal}{Materials Today Energy}} \textbf{\bibinfo{volume}{17}}, \bibinfo{pages}{100449} (\bibinfo{year}{2020}).
\newblock \urlprefix\url{https://linkinghub.elsevier.com/retrieve/pii/S246860692030068X}.

\bibitem{osterrieder_autonomous_2023}
\bibinfo{author}{Osterrieder, T.} \emph{et~al.}
\newblock \bibinfo{title}{Autonomous optimization of an organic solar cell in a 4-dimensional parameter space}.
\newblock \emph{\bibinfo{journal}{Energy \& Environmental Science}} \textbf{\bibinfo{volume}{16}}, \bibinfo{pages}{3984--3993} (\bibinfo{year}{2023}).
\newblock \urlprefix\url{https://xlink.rsc.org/?DOI=D3EE02027D}.

\bibitem{hu_review_2019}
\bibinfo{author}{Hu, Z.} \emph{et~al.}
\newblock \bibinfo{title}{A {Review} on {Energy} {Band}‐{Gap} {Engineering} for {Perovskite} {Photovoltaics}}.
\newblock \emph{\bibinfo{journal}{Solar RRL}} \textbf{\bibinfo{volume}{3}}, \bibinfo{pages}{1900304} (\bibinfo{year}{2019}).
\newblock \urlprefix\url{https://onlinelibrary.wiley.com/doi/10.1002/solr.201900304}.

\bibitem{horantner_potential_2017}
\bibinfo{author}{Hörantner, M.~T.} \emph{et~al.}
\newblock \bibinfo{title}{The {Potential} of {Multijunction} {Perovskite} {Solar} {Cells}}.
\newblock \emph{\bibinfo{journal}{ACS Energy Letters}} \textbf{\bibinfo{volume}{2}}, \bibinfo{pages}{2506--2513} (\bibinfo{year}{2017}).
\newblock \urlprefix\url{https://pubs.acs.org/doi/10.1021/acsenergylett.7b00647}.

\bibitem{liu_highly_2021}
\bibinfo{author}{Liu, X.} \emph{et~al.}
\newblock \bibinfo{title}{Highly efficient wide-band-gap perovskite solar cells fabricated by sequential deposition method}.
\newblock \emph{\bibinfo{journal}{Nano Energy}} \textbf{\bibinfo{volume}{86}}, \bibinfo{pages}{106114} (\bibinfo{year}{2021}).
\newblock \urlprefix\url{https://linkinghub.elsevier.com/retrieve/pii/S2211285521003700}.

\bibitem{larcher_towards_2015}
\bibinfo{author}{Larcher, D.} \& \bibinfo{author}{Tarascon, J.-M.}
\newblock \bibinfo{title}{Towards greener and more sustainable batteries for electrical energy storage}.
\newblock \emph{\bibinfo{journal}{Nature Chemistry}} \textbf{\bibinfo{volume}{7}}, \bibinfo{pages}{19--29} (\bibinfo{year}{2015}).
\newblock \urlprefix\url{https://www.nature.com/articles/nchem.2085}.

\bibitem{canepa_odyssey_2017}
\bibinfo{author}{Canepa, P.} \emph{et~al.}
\newblock \bibinfo{title}{Odyssey of {Multivalent} {Cathode} {Materials}: {Open} {Questions} and {Future} {Challenges}}.
\newblock \emph{\bibinfo{journal}{Chemical Reviews}} \textbf{\bibinfo{volume}{117}}, \bibinfo{pages}{4287--4341} (\bibinfo{year}{2017}).
\newblock \urlprefix\url{https://pubs.acs.org/doi/10.1021/acs.chemrev.6b00614}.

\bibitem{manthiram_reflection_2020}
\bibinfo{author}{Manthiram, A.}
\newblock \bibinfo{title}{A reflection on lithium-ion battery cathode chemistry}.
\newblock \emph{\bibinfo{journal}{Nature Communications}} \textbf{\bibinfo{volume}{11}}, \bibinfo{pages}{1550} (\bibinfo{year}{2020}).
\newblock \urlprefix\url{https://www.nature.com/articles/s41467-020-15355-0}.

\bibitem{batzner_e3-equivariant_2022}
\bibinfo{author}{Batzner, S.} \emph{et~al.}
\newblock \bibinfo{title}{E(3)-equivariant graph neural networks for data-efficient and accurate interatomic potentials}.
\newblock \emph{\bibinfo{journal}{Nature Communications}} \textbf{\bibinfo{volume}{13}}, \bibinfo{pages}{2453} (\bibinfo{year}{2022}).
\newblock \urlprefix\url{https://www.nature.com/articles/s41467-022-29939-5}.

\bibitem{chen_direct_2021}
\bibinfo{author}{Chen, Z.} \emph{et~al.}
\newblock \bibinfo{title}{Direct {Prediction} of {Phonon} {Density} of {States} {With} {Euclidean} {Neural} {Networks}}.
\newblock \emph{\bibinfo{journal}{Advanced Science}} \textbf{\bibinfo{volume}{8}}, \bibinfo{pages}{2004214} (\bibinfo{year}{2021}).
\newblock \urlprefix\url{https://onlinelibrary.wiley.com/doi/10.1002/advs.202004214}.

\bibitem{banko_scaling_2001}
\bibinfo{author}{Banko, M.} \& \bibinfo{author}{Brill, E.}
\newblock \bibinfo{title}{Scaling to very very large corpora for natural language disambiguation} (\bibinfo{year}{2001}).
\newblock \urlprefix\url{http://portal.acm.org/citation.cfm?doid=1073012.1073017}.

\bibitem{wang_high-performance_2023}
\bibinfo{author}{Wang, P.} \emph{et~al.}
\newblock \bibinfo{title}{High-{Performance} {Lithium}–{Sulfur} {Batteries} via {Molecular} {Complexation}}.
\newblock \emph{\bibinfo{journal}{Journal of the American Chemical Society}} \textbf{\bibinfo{volume}{145}}, \bibinfo{pages}{18865--18876} (\bibinfo{year}{2023}).
\newblock \urlprefix\url{https://pubs.acs.org/doi/10.1021/jacs.3c05209}.

\bibitem{teufl_implications_2023}
\bibinfo{author}{Teufl, T.} \emph{et~al.}
\newblock \bibinfo{title}{Implications of the {Thermal} {Stability} of {FEC}-{Based} {Electrolytes} for {Li}-{Ion} {Batteries}}.
\newblock \emph{\bibinfo{journal}{Journal of The Electrochemical Society}}  (\bibinfo{year}{2023}).
\newblock \urlprefix\url{https://iopscience.iop.org/article/10.1149/1945-7111/acbc52}.

\bibitem{subasi_synthesis_2024}
\bibinfo{author}{Subaşı, Y.} \emph{et~al.}
\newblock \bibinfo{title}{Synthesis and characterization of a crystalline {Na} $_{\textrm{4}}$ {Fe} $_{\textrm{3}}$ ({PO} $_{\textrm{4}}$ ) $_{\textrm{2}}$ ({P} $_{\textrm{2}}$ {O} $_{\textrm{7}}$ ) cathode material for sodium-ion batteries}.
\newblock \emph{\bibinfo{journal}{Journal of Materials Chemistry A}} \textbf{\bibinfo{volume}{12}}, \bibinfo{pages}{23506--23517} (\bibinfo{year}{2024}).
\newblock \urlprefix\url{https://xlink.rsc.org/?DOI=D4TA03554B}.

\bibitem{he_research_2023}
\bibinfo{author}{He, F.} \emph{et~al.}
\newblock \bibinfo{title}{Research {Progress} on {Electrochemical} {Properties} of {Na} $_{\textrm{3}}$ {V} $_{\textrm{2}}$ ({PO} $_{\textrm{4}}$ ) $_{\textrm{3}}$ as {Cathode} {Material} for {Sodium}-{Ion} {Batteries}}.
\newblock \emph{\bibinfo{journal}{Industrial \& Engineering Chemistry Research}} \textbf{\bibinfo{volume}{62}}, \bibinfo{pages}{3444--3464} (\bibinfo{year}{2023}).
\newblock \urlprefix\url{https://pubs.acs.org/doi/10.1021/acs.iecr.2c04054}.

\bibitem{Jia_Lynch_Huang_Danielson_Langat_Milder_Ruby_Wang_Friedler_Norquist_etal._2019}
\bibinfo{author}{Jia, X.} \emph{et~al.}
\newblock \bibinfo{title}{Anthropogenic biases in chemical reaction data hinder exploratory inorganic synthesis}.
\newblock \emph{\bibinfo{journal}{Nature}} \textbf{\bibinfo{volume}{573}}, \bibinfo{pages}{251–255} (\bibinfo{year}{2019}).
\newblock \urlprefix\url{https://www.nature.com/articles/s41586-019-1540-5}.

\bibitem{mancardi_computational_2023}
\bibinfo{author}{Mancardi, G.} \emph{et~al.}
\newblock \bibinfo{title}{A computational view on nanomaterial intrinsic and extrinsic features for nanosafety and sustainability}.
\newblock \emph{\bibinfo{journal}{Materials Today}} \textbf{\bibinfo{volume}{67}}, \bibinfo{pages}{344--370} (\bibinfo{year}{2023}).
\newblock \urlprefix\url{https://linkinghub.elsevier.com/retrieve/pii/S1369702123001803}.

\bibitem{talirz_materials_2020}
\bibinfo{author}{Talirz, L.} \emph{et~al.}
\newblock \bibinfo{title}{Materials {Cloud}, a platform for open computational science}.
\newblock \emph{\bibinfo{journal}{Scientific Data}} \textbf{\bibinfo{volume}{7}}, \bibinfo{pages}{299} (\bibinfo{year}{2020}).
\newblock \urlprefix\url{https://www.nature.com/articles/s41597-020-00637-5}.

\bibitem{winther_catalysis-hub.org_2019}
\bibinfo{author}{Winther, K.~T.} \emph{et~al.}
\newblock \bibinfo{title}{Catalysis-{Hub}.org, an open electronic structure database for surface reactions}.
\newblock \emph{\bibinfo{journal}{Scientific Data}} \textbf{\bibinfo{volume}{6}}, \bibinfo{pages}{75} (\bibinfo{year}{2019}).
\newblock \urlprefix\url{https://www.nature.com/articles/s41597-019-0081-y}.

\bibitem{zhou_medium-_2019}
\bibinfo{author}{Zhou, C.} \& \bibinfo{author}{Wu, S.}
\newblock \bibinfo{title}{Medium- and high-temperature latent heat thermal energy storage: {Material} database, system review, and corrosivity assessment}.
\newblock \emph{\bibinfo{journal}{International Journal of Energy Research}} \textbf{\bibinfo{volume}{43}}, \bibinfo{pages}{621--661} (\bibinfo{year}{2019}).
\newblock \urlprefix\url{https://onlinelibrary.wiley.com/doi/10.1002/er.4216}.

\bibitem{shimakawa_extrapolative_2024}
\bibinfo{author}{Shimakawa, H.}, \bibinfo{author}{Kumada, A.} \& \bibinfo{author}{Sato, M.}
\newblock \bibinfo{title}{Extrapolative prediction of small-data molecular property using quantum mechanics-assisted machine learning}.
\newblock \emph{\bibinfo{journal}{npj Computational Materials}} \textbf{\bibinfo{volume}{10}}, \bibinfo{pages}{11} (\bibinfo{year}{2024}).
\newblock \urlprefix\url{https://www.nature.com/articles/s41524-023-01194-2}.

\bibitem{jain_new_2016}
\bibinfo{author}{Jain, A.}, \bibinfo{author}{Hautier, G.}, \bibinfo{author}{Ong, S.~P.} \& \bibinfo{author}{Persson, K.}
\newblock \bibinfo{title}{New opportunities for materials informatics: {Resources} and data mining techniques for uncovering hidden relationships}.
\newblock \emph{\bibinfo{journal}{Journal of Materials Research}} \textbf{\bibinfo{volume}{31}}, \bibinfo{pages}{977--994} (\bibinfo{year}{2016}).
\newblock \urlprefix\url{https://doi.org/10.1557/jmr.2016.80}.

\bibitem{manzhos_machine_2023}
\bibinfo{author}{Manzhos, S.}, \bibinfo{author}{Tsuda, S.} \& \bibinfo{author}{Ihara, M.}
\newblock \bibinfo{title}{Machine learning in computational chemistry: interplay between (non)linearity, basis sets, and dimensionality}.
\newblock \emph{\bibinfo{journal}{Physical Chemistry Chemical Physics}} \textbf{\bibinfo{volume}{25}}, \bibinfo{pages}{1546--1555} (\bibinfo{year}{2023}).
\newblock \urlprefix\url{https://xlink.rsc.org/?DOI=D2CP04155C}.

\bibitem{ward_general-purpose_2016}
\bibinfo{author}{Ward, L.}, \bibinfo{author}{Agrawal, A.}, \bibinfo{author}{Choudhary, A.} \& \bibinfo{author}{Wolverton, C.}
\newblock \bibinfo{title}{A general-purpose machine learning framework for predicting properties of inorganic materials}.
\newblock \emph{\bibinfo{journal}{npj Computational Materials}} \textbf{\bibinfo{volume}{2}}, \bibinfo{pages}{16028} (\bibinfo{year}{2016}).
\newblock \urlprefix\url{https://www.nature.com/articles/npjcompumats201628}.

\bibitem{ong_materials_2015}
\bibinfo{author}{Ong, S.~P.} \emph{et~al.}
\newblock \bibinfo{title}{The {Materials} {Application} {Programming} {Interface} ({API}): {A} simple, flexible and efficient {API} for materials data based on {REpresentational} {State} {Transfer} ({REST}) principles}.
\newblock \emph{\bibinfo{journal}{Computational Materials Science}} \textbf{\bibinfo{volume}{97}}, \bibinfo{pages}{209--215} (\bibinfo{year}{2015}).
\newblock \urlprefix\url{https://linkinghub.elsevier.com/retrieve/pii/S0927025614007113}.

\bibitem{choudhary_machine_2018}
\bibinfo{author}{Choudhary, K.}, \bibinfo{author}{DeCost, B.} \& \bibinfo{author}{Tavazza, F.}
\newblock \bibinfo{title}{Machine learning with force-field-inspired descriptors for materials: {Fast} screening and mapping energy landscape}.
\newblock \emph{\bibinfo{journal}{Physical Review Materials}} \textbf{\bibinfo{volume}{2}}, \bibinfo{pages}{083801} (\bibinfo{year}{2018}).
\newblock \urlprefix\url{https://link.aps.org/doi/10.1103/PhysRevMaterials.2.083801}.

\bibitem{ward_matminer_2018}
\bibinfo{author}{Ward, L.} \emph{et~al.}
\newblock \bibinfo{title}{Matminer: {An} open source toolkit for materials data mining}.
\newblock \emph{\bibinfo{journal}{Computational Materials Science}} \textbf{\bibinfo{volume}{152}}, \bibinfo{pages}{60--69} (\bibinfo{year}{2018}).
\newblock \urlprefix\url{https://www.sciencedirect.com/science/article/pii/S0927025618303252}.

\bibitem{e3nn_paper}
\bibinfo{author}{Geiger, M.} \& \bibinfo{author}{Smidt, T.}
\newblock \bibinfo{title}{e3nn: Euclidean neural networks} (\bibinfo{year}{2022}).
\newblock \urlprefix\url{https://arxiv.org/abs/2207.09453}.

\bibitem{e3nn}
\bibinfo{author}{Geiger, M.} \emph{et~al.}
\newblock \bibinfo{title}{Euclidean neural networks: e3nn}, \bibinfo{version}{0.5.0} (\bibinfo{year}{2022}).
\newblock \urlprefix\url{https://doi.org/10.5281/zenodo.6459381}.

\bibitem{thomas2018tensorfieldnetworksrotation}
\bibinfo{author}{Thomas, N.} \emph{et~al.}
\newblock \bibinfo{title}{Tensor field networks: Rotation- and translation-equivariant neural networks for 3d point clouds} (\bibinfo{year}{2018}).
\newblock \urlprefix\url{https://arxiv.org/abs/1802.08219}.
\newblock \eprint{1802.08219}.

\bibitem{weiler20183dsteerablecnnslearning}
\bibinfo{author}{Weiler, M.}, \bibinfo{author}{Geiger, M.}, \bibinfo{author}{Welling, M.}, \bibinfo{author}{Boomsma, W.} \& \bibinfo{author}{Cohen, T.}
\newblock \bibinfo{title}{3d steerable cnns: Learning rotationally equivariant features in volumetric data} (\bibinfo{year}{2018}).
\newblock \urlprefix\url{https://arxiv.org/abs/1807.02547}.
\newblock \eprint{1807.02547}.

\bibitem{kondor2018clebschgordannetsfullyfourier}
\bibinfo{author}{Kondor, R.}, \bibinfo{author}{Lin, Z.} \& \bibinfo{author}{Trivedi, S.}
\newblock \bibinfo{title}{Clebsch-gordan nets: a fully fourier space spherical convolutional neural network} (\bibinfo{year}{2018}).
\newblock \urlprefix\url{https://arxiv.org/abs/1806.09231}.
\newblock \eprint{1806.09231}.

\bibitem{loshchilov2019decoupledweightdecayregularization}
\bibinfo{author}{Loshchilov, I.} \& \bibinfo{author}{Hutter, F.}
\newblock \bibinfo{title}{Decoupled weight decay regularization} (\bibinfo{year}{2019}).
\newblock \urlprefix\url{https://arxiv.org/abs/1711.05101}.
\newblock \eprint{1711.05101}.

\bibitem{friedman_greedy_2001}
\bibinfo{author}{Friedman, J.~H.}
\newblock \bibinfo{title}{Greedy function approximation: {A} gradient boosting machine.}
\newblock \emph{\bibinfo{journal}{The Annals of Statistics}} \textbf{\bibinfo{volume}{29}} (\bibinfo{year}{2001}).

\bibitem{scikit-learn}
\bibinfo{author}{Pedregosa, F.} \emph{et~al.}
\newblock \bibinfo{title}{Scikit-learn: Machine learning in {P}ython}.
\newblock \emph{\bibinfo{journal}{Journal of Machine Learning Research}} \textbf{\bibinfo{volume}{12}}, \bibinfo{pages}{2825--2830} (\bibinfo{year}{2011}).

\bibitem{guyon_no_2002}
\bibinfo{author}{Guyon, I.}, \bibinfo{author}{Weston, J.}, \bibinfo{author}{Barnhill, S.} \& \bibinfo{author}{Vapnik, V.}
\newblock \bibinfo{title}{Gene selection for cancer classification using support vector machines}.
\newblock \emph{\bibinfo{journal}{Machine Learning}} \textbf{\bibinfo{volume}{46}}, \bibinfo{pages}{389--422} (\bibinfo{year}{2002}).
\newblock \urlprefix\url{http://link.springer.com/10.1023/A:1012487302797}.

\bibitem{haas_magnetic_1970}
\bibinfo{author}{Haas, C.}
\newblock \bibinfo{title}{Magnetic semiconductors}.
\newblock \emph{\bibinfo{journal}{C R C Critical Reviews in Solid State Sciences}} \textbf{\bibinfo{volume}{1}}, \bibinfo{pages}{47--98} (\bibinfo{year}{1970}).
\newblock \urlprefix\url{http://www.tandfonline.com/doi/abs/10.1080/10408437008243418}.

\bibitem{ren_tuning_2020}
\bibinfo{author}{Ren, L.} \emph{et~al.}
\newblock \bibinfo{title}{Tuning {Magnetism} and {Photocurrent} in {Mn}-{Doped} {Organic}–{Inorganic} {Perovskites}}.
\newblock \emph{\bibinfo{journal}{The Journal of Physical Chemistry Letters}} \textbf{\bibinfo{volume}{11}}, \bibinfo{pages}{2577--2584} (\bibinfo{year}{2020}).
\newblock \urlprefix\url{https://pubs.acs.org/doi/10.1021/acs.jpclett.0c00034}.

\bibitem{konstantakou_critical_2017}
\bibinfo{author}{Konstantakou, M.} \& \bibinfo{author}{Stergiopoulos, T.}
\newblock \bibinfo{title}{A critical review on tin halide perovskite solar cells}.
\newblock \emph{\bibinfo{journal}{Journal of Materials Chemistry A}} \textbf{\bibinfo{volume}{5}}, \bibinfo{pages}{11518--11549} (\bibinfo{year}{2017}).
\newblock \urlprefix\url{https://xlink.rsc.org/?DOI=C7TA00929A}.

\bibitem{zhou_highly_2020}
\bibinfo{author}{Zhou, X.} \emph{et~al.}
\newblock \bibinfo{title}{Highly {Efficient} and {Stable} {GABr}‐{Modified} {Ideal}‐{Bandgap} (1.35 {eV}) {Sn}/{Pb} {Perovskite} {Solar} {Cells} {Achieve} 20.63\% {Efficiency} with a {Record} {Small} \textit{{V}} $_{\textrm{oc}}$ {Deficit} of 0.33 {V}}.
\newblock \emph{\bibinfo{journal}{Advanced Materials}} \textbf{\bibinfo{volume}{32}}, \bibinfo{pages}{1908107} (\bibinfo{year}{2020}).
\newblock \urlprefix\url{https://onlinelibrary.wiley.com/doi/10.1002/adma.201908107}.

\bibitem{upton_understanding_1996}
\bibinfo{author}{Upton, G. J.~G.} \& \bibinfo{author}{Cook, I.}
\newblock \emph{\bibinfo{title}{Understanding statistics}} \bibinfo{edition}{1. publ} edn (\bibinfo{publisher}{Oxford University Press}, \bibinfo{address}{Oxford}, \bibinfo{year}{1996}).

\bibitem{ong_python_2013}
\bibinfo{author}{Ong, S.~P.} \emph{et~al.}
\newblock \bibinfo{title}{Python {Materials} {Genomics} (pymatgen): {A} robust, open-source python library for materials analysis}.
\newblock \emph{\bibinfo{journal}{Computational Materials Science}} \textbf{\bibinfo{volume}{68}}, \bibinfo{pages}{314--319} (\bibinfo{year}{2013}).
\newblock \urlprefix\url{https://linkinghub.elsevier.com/retrieve/pii/S0927025612006295}.

\bibitem{okeefe_atom_1991}
\bibinfo{author}{O'Keefe, M.} \& \bibinfo{author}{Brese, N.~E.}
\newblock \bibinfo{title}{Atom sizes and bond lengths in molecules and crystals}.
\newblock \emph{\bibinfo{journal}{Journal of the American Chemical Society}} \textbf{\bibinfo{volume}{113}}, \bibinfo{pages}{3226--3229} (\bibinfo{year}{1991}).
\newblock \urlprefix\url{https://pubs.acs.org/doi/abs/10.1021/ja00009a002}.

\end{thebibliography}

\end{document}



\title{
    Energy-GNoME: A Living Database of Selected Materials for Energy Applications
    \\
    (Supplementary information)
}

\author[1]{Paolo De Angelis}
\author[1]{Giovanni Trezza}
\author[1]{Giulio Barletta}
\author[1,2]{Pietro Asinari}
\author[1]{Eliodoro Chiavazzo \footnote{Corresponding autor: \href{mailto:eliodoro.chiavazzo@polito.it}{eliodoro.chiavazzo@polito.it}}}

\affil[1]{Department of Energy, Politecnico di Torino, Corso Duca degli Abruzzi, 24, Torino, 10129, Italy}
\affil[2]{Istituto Nazionale di Ricerca Metrologica, Strada delle Cacce, 91, Torino, 10135, Italy}

\maketitle
\tableofcontents
\listoffigures
\listoftables
\clearpage
%

\section{Thermoelectrics models performance}\label{sec:thermoelectrics}

\subsection{Regressors ($zT$)}\label{ssec:thermoelectrics_regressors}

\begin{figure}[H]
    \centering
    \includegraphics[width=1\linewidth]{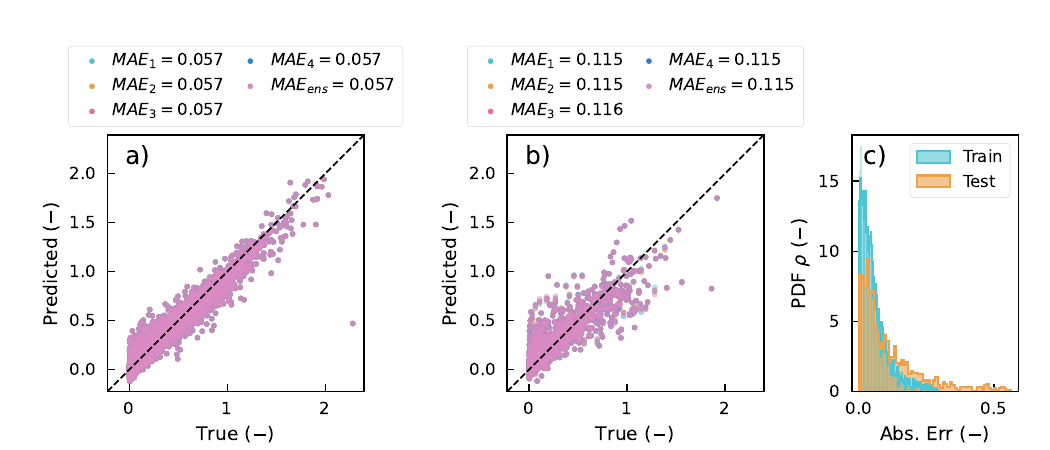}
    \caption[Results for thermoelectrics regressors.]{
        Results for thermoelectrics regressors.
        a) Parity plot showing the true against predicted figure of merit $zT$ for the 4 trained regressors (highlighting the corresponding mean absolute error $MAE_i$), together with the ensemble case, in which each material is represented by the average prediction across the four trained models for training set and b) for testing set.
        c) Absolute error distribution over training and testing sets.
    }
    \label{fig:thermoelectrics_parity}
\end{figure}
\begin{table}[H]
	\caption[Performances of the 4 thermoelectrics regressors.]{
        Performances of the 4 thermoelectrics regressors, along with the resulting ensemble model, in terms of $R^2$, MAE, RMSE for both training and testing.
        }
	\centering
	\label{tab:thermoelectrics_parity}
    \begin{tabular}{p{30mm}>{\raggedright\let\newline}cccc|c}
        \toprule
        & Model 1 & Model 2 & Model 3 & Model 4 & Ens. Model \\
       \midrule
       $R^2$ train $\mathrm{(-)}$ & 0.941 & 0.941 & 0.941 & 0.941 & 0.941 \\
       $\mathrm{MAE}$ train $\mathrm{(-)}$ & 0.057 & 0.057 & 0.057 & 0.057 & 0.057 \\
       $\mathrm{RMSE}$ train $\mathrm{(-)}$ & 0.086 & 0.086 & 0.086 & 0.086 & 0.086 \\
       $R^2$ test $\mathrm{(-)}$ & 0.730 & 0.731 & 0.728 & 0.727 & 0.730 \\
       $\mathrm{MAE}$ test $\mathrm{(-)}$ & 0.115 & 0.115 & 0.116 & 0.115 & 0.115 \\
       $\mathrm{RMSE}$ test $\mathrm{(-)}$ & 0.170 & 0.170 & 0.171 & 0.171 & 0.170 \\
       \bottomrule
       \end{tabular}
\end{table}

\subsection{AI-experts}\label{ssec:thermoelectrics_classifiers}
\begin{figure}[H]
    \centering
    \includegraphics[width=1\linewidth]{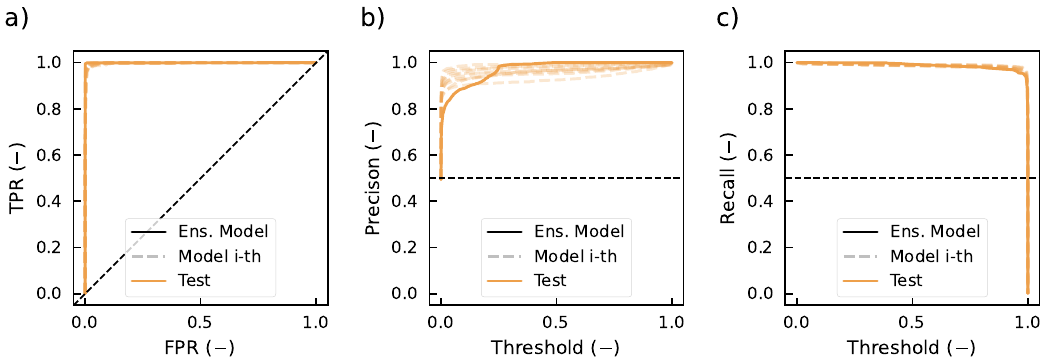}
    \caption[Results for thermoelectrics AI-experts models.]{
    Results for thermoelectrics AI-experts models.
    a) ROC, b) precision and c) recall curves over the training/testing sets of all the 10 classifiers, together with the ensemble model curve.
    }
    \label{fig:thermoelectrics_roc}
\end{figure}
\begin{table}[H]
	\caption[Performances for thermoelectrics of the 10 AI-experts models.]{
        Performances for thermoelectrics of the 10 AI-experts models, for each of the 10 classifier sand  the resulting ensemble model, in terms of AUC of ROC curve, precision and recall for  testing.
        }
	\centering
	\label{tab:thermoelectrics_classifiers}
    \begin{tabular}{p{18mm}>{\raggedright\let\newline}cccccccccc|c}
        \toprule
        & M. 1 & M. 2 & M. 3 & M. 4 & M. 5 & M. 6 & M. 7 & M. 8 & M. 9 & M. 10 & Ens. Model \\
       \midrule
        $\mathrm{AUC}$ test $\mathrm{(-)}$ & 0.999 & 0.999 & 1.000 & 0.998 & 0.998 & 0.999 & 0.999 & 0.998 & 0.998 & 0.999 & 1.000 \\
        $\mathrm{Precision}$ test $\mathrm{(-)}$ & 0.967 & 0.960 & 0.993 & 0.967 & 0.966 & 0.986 & 0.984 & 0.953 & 0.925 & 0.976 & 1.000 \\
        $\mathrm{Recall}$ test $\mathrm{(-)}$ & 0.991 & 0.986 & 0.989 & 0.992 & 0.987 & 0.992 & 0.986 & 0.988 & 0.987 & 0.989 & 0.992 \\       
        \bottomrule
       \end{tabular}
\end{table}

\section{Perovskites models performance}\label{sec:perovskites}

\subsection{Regressors ``Pure-model'' ($E_g$)}\label{ssec:perovskites-puremodel}
\begin{figure}[H]
    \centering
    \includegraphics[width=1\linewidth]{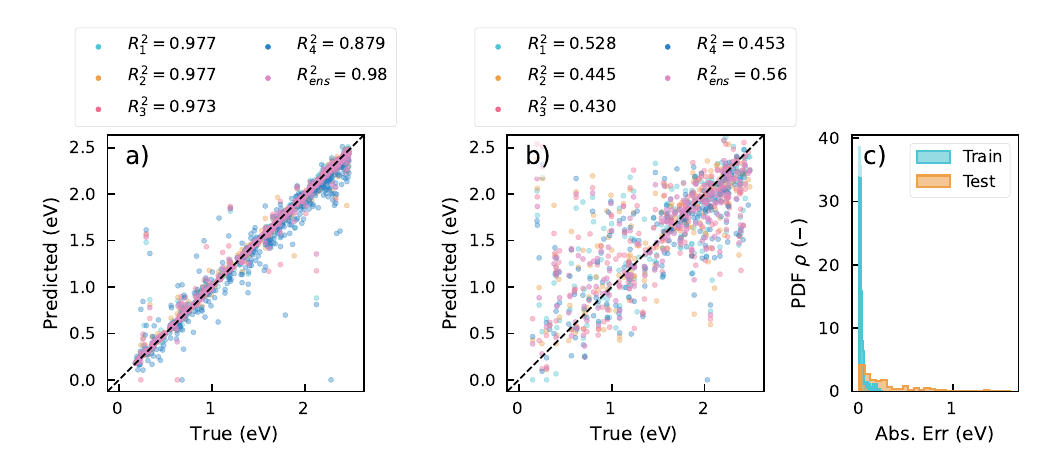}
    \caption[Results for perovskites ``Pure-model'' regressors.]{
        Results for perovskites ``Pure-model'' regressors.
        a) Parity plot showing the true against predicted band gap $E_g$ for the 4 trained regressors (highlighting the corresponding mean absolute error $MAE_i$), together with the ensemble case, in which each material is represented by the average prediction across the four trained models for training set and b) for testing set.
        c) Absolute error distribution over training and testing sets.
    }
    \label{fig:perovskite_puremodel_parity}
\end{figure}
\begin{table}[H]
	\caption[Performances of the 4 perovskites ``Pure-model'' regressors.]{
        Performances of the 4 perovskites ``Pure-model'' regressors, along with the resulting ensemble model, in terms of $R^2$, MAE, RMSE for both training and testing.
        }
	\centering
	\label{tab:perovskite_puremodel_parity}
    \begin{tabular}{p{30mm}>{\raggedright\let\newline}cccc|c}
        \toprule
        & Model 1 & Model 2 & Model 3 & Model 4 & Ens. Model \\
        \midrule
        $R^2$ train $\mathrm{(-)}$ & 0.977 & 0.977 & 0.973 & 0.879 & 0.981 \\
        $\mathrm{MAE}$ train $\mathrm{(eV)}$ & 0.034 & 0.046 & 0.032 & 0.135 & 0.032 \\
        $\mathrm{RMSE}$ train $\mathrm{(eV)}$ & 0.101 & 0.100 & 0.108 & 0.231 & 0.091 \\
        $R^2$ test $\mathrm{(-)}$ & 0.528 & 0.445 & 0.430 & 0.453 & 0.557 \\
        $\mathrm{MAE}$ test $\mathrm{(eV)}$ & 0.291 & 0.307 & 0.311 & 0.326 & 0.273 \\
        $\mathrm{RMSE}$ test $\mathrm{(eV)}$ & 0.436 & 0.473 & 0.479 & 0.469 & 0.422 \\
        \bottomrule
       \end{tabular}
\end{table}

\subsection{Regressors ``Mixed-model'' ($E_g$)}\label{ssec:perovskites-mixedmodel}
\begin{figure}[H]
    \centering
    \includegraphics[width=1\linewidth]{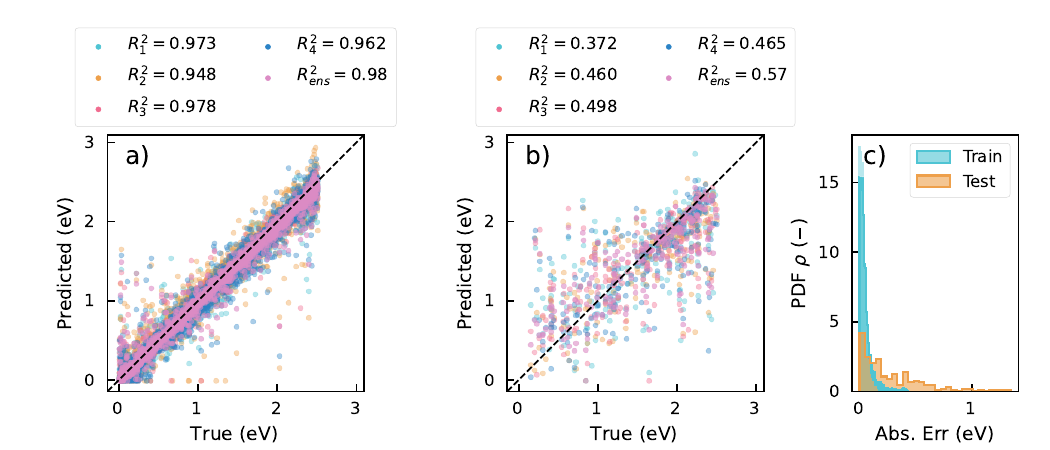}
    \caption[Results for perovskites ``Mixed-model'' regressors.]{
        Results for perovskites ``Mixed-model'' regressors.
        a) Parity plot showing the true against predicted band gap $E_g$ for the 4 trained regressors (highlighting the corresponding mean absolute error $MAE_i$), together with the ensemble case, in which each material is represented by the average prediction across the four trained models for training set and b) for testing set.
        c) Absolute error distribution over training and testing sets.
    }
    \label{fig:perovskite_mixedmodel_parity}
\end{figure}
\begin{table}[H]
	\caption[Performances of the 4 perovskites ``Mixed-model'' regressors.]{
        Performances of the 4 perovskites ``Mixed-model'' regressors, along with the resulting ensemble model, in terms of $R^2$, MAE, RMSE for both training and testing.
        }
	\centering
	\label{tab:perovskite_mixedmodel_parity}
    \begin{tabular}{p{30mm}>{\raggedright\let\newline}cccc|c}
        \toprule
        & Model 1 & Model 2 & Model 3 & Model 4 & Ens. Model \\
        \midrule
        $R^2$ train $\mathrm{(-)}$ & 0.973 & 0.948 & 0.978 & 0.962 & 0.982 \\
        $\mathrm{MAE}$ train $\mathrm{(eV)}$ & 0.077 & 0.105 & 0.061 & 0.091 & 0.054 \\
        $\mathrm{RMSE}$ train $\mathrm{(eV)}$ & 0.124 & 0.172 & 0.112 & 0.148 & 0.102 \\
        $R^2$ test $\mathrm{(-)}$ & 0.372 & 0.460 & 0.498 & 0.465 & 0.572 \\
        $\mathrm{MAE}$ test $\mathrm{(eV)}$ & 0.348 & 0.332 & 0.305 & 0.307 & 0.289 \\
        $\mathrm{RMSE}$ test $\mathrm{(eV)}$ & 0.503 & 0.466 & 0.450 & 0.464 & 0.415 \\
        \bottomrule
       \end{tabular}
\end{table}

\subsection{AI-experts}\label{ssec:perovskites-classifiers}
\begin{figure}[H]
    \centering
    \includegraphics[width=1\linewidth]{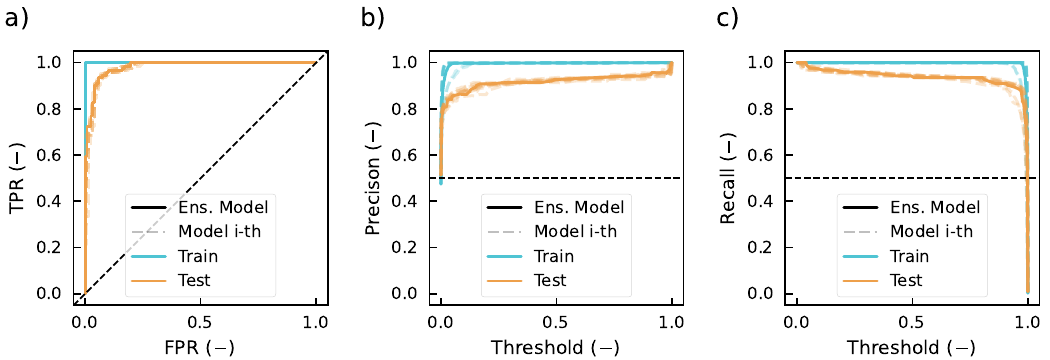}
    \caption[Results for perovskites AI-experts models.]{
    Results for perovskites AI-experts models.
    a) ROC, b) precision and c) recall curves over the training/testing sets of all the 10 classifiers, together with the ensemble model curve.
    }
    \label{fig:perovskite_roc}
\end{figure}
\begin{table}[H]
	\caption[Performances for perovskites of the 10 AI-experts models.]{
        Performances for perovskites of the 10 AI-experts models, for each of the 10 classifier sand  the resulting ensemble model, in terms of AUC of ROC curve, precision and recall for  testing.
        }
	\centering
	\label{tab:perovskite_classifiers}
    \begin{tabular}{p{18mm}>{\raggedright\let\newline}cccccccccc|c}
        \toprule
        & M. 1 & M. 2 & M. 3 & M. 4 & M. 5 & M. 6 & M. 7 & M. 8 & M. 9 & M. 10 & Ens. Model \\
        \midrule
        $\mathrm{AUC}$ train $\mathrm{(-)}$ & 1.000 & 1.000 & 1.000 & 1.000 & 1.000 & 1.000 & 1.000 & 1.000 & 1.000 & 1.000 & 1.000 \\
        $\mathrm{Precision}$ train $\mathrm{(-)}$ & 1.000 & 1.000 & 1.000 & 1.000 & 1.000 & 1.000 & 1.000 & 1.000 & 1.000 & 1.000 & 1.000 \\
        $\mathrm{Recall}$ train $\mathrm{(-)}$ & 1.000 & 1.000 & 1.000 & 1.000 & 1.000 & 1.000 & 1.000 & 1.000 & 1.000 & 1.000 & 1.000 \\
        $\mathrm{AUC}$ test $\mathrm{(-)}$ & 0.983 & 0.977 & 0.977 & 0.979 & 0.976 & 0.977 & 0.980 & 0.978 & 0.980 & 0.979 & 0.982 \\
        $\mathrm{Precision}$ test $\mathrm{(-)}$ & 0.936 & 0.927 & 0.928 & 0.921 & 0.928 & 0.913 & 0.935 & 0.928 & 0.927 & 0.920 & 0.927 \\
        $\mathrm{Recall}$ test $\mathrm{(-)}$ & 0.951 & 0.935 & 0.943 & 0.943 & 0.943 & 0.943 & 0.943 & 0.943 & 0.935 & 0.935 & 0.935 \\
        \bottomrule
       \end{tabular}
\end{table}

\section{Batteries models performance}\label{sec:batteries}

\subsection{Regressors ($\Delta V_c$)}\label{ssec:batteries_average_voltage_regressors}

\begin{figure}[H]
    \centering
    \includegraphics[width=1\linewidth]{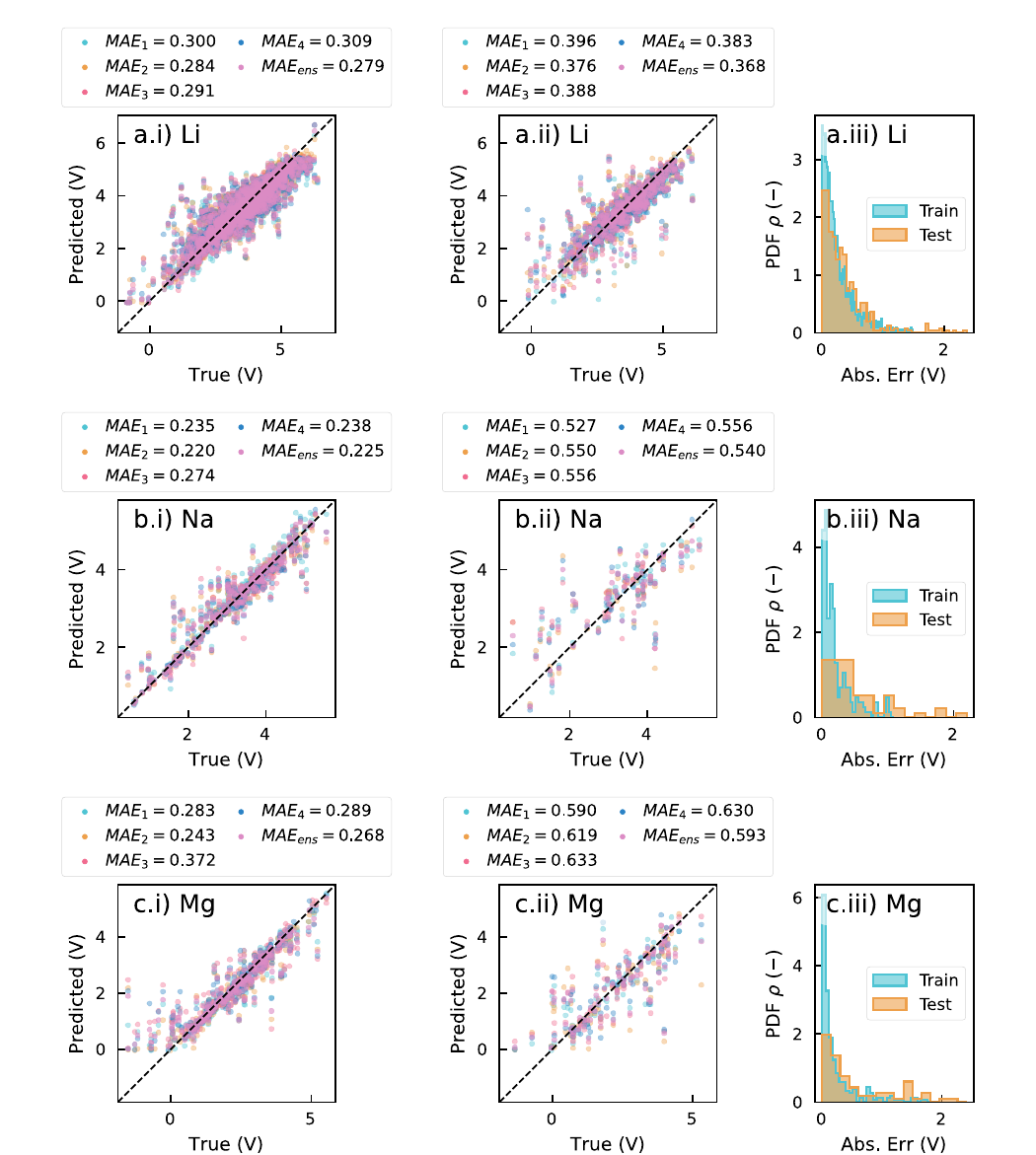}
    \caption[Results for Li, Na, Mg, K, Ca, Cs, Al, Rb, and Y cathode materials $\Delta V_c$  regressors.]{
        Results for Li (a), Na (b), Mg (c), K (d), Ca (e), Cs (f), Al (g), Rb (h), and Y (i) cathode materials $\Delta V_c$  regressors.
        i) Parity plot showing the true against predicted avarege voltage $\Delta V_c$ for the 4 trained regressors (highlighting the corresponding mean absolute error $MAE_i$), together with the ensemble case, in which each material is represented by the average prediction across the four trained models for training set and ii) for testing set.
        iii) Absolute error distribution over training and testing sets.
    }
    \label{fig:batteries_average_voltage_parity_1}
\end{figure}
\begin{figure}[H]
    \addtocounter{figure}{-1}
    \centering
    \includegraphics[width=1\linewidth]{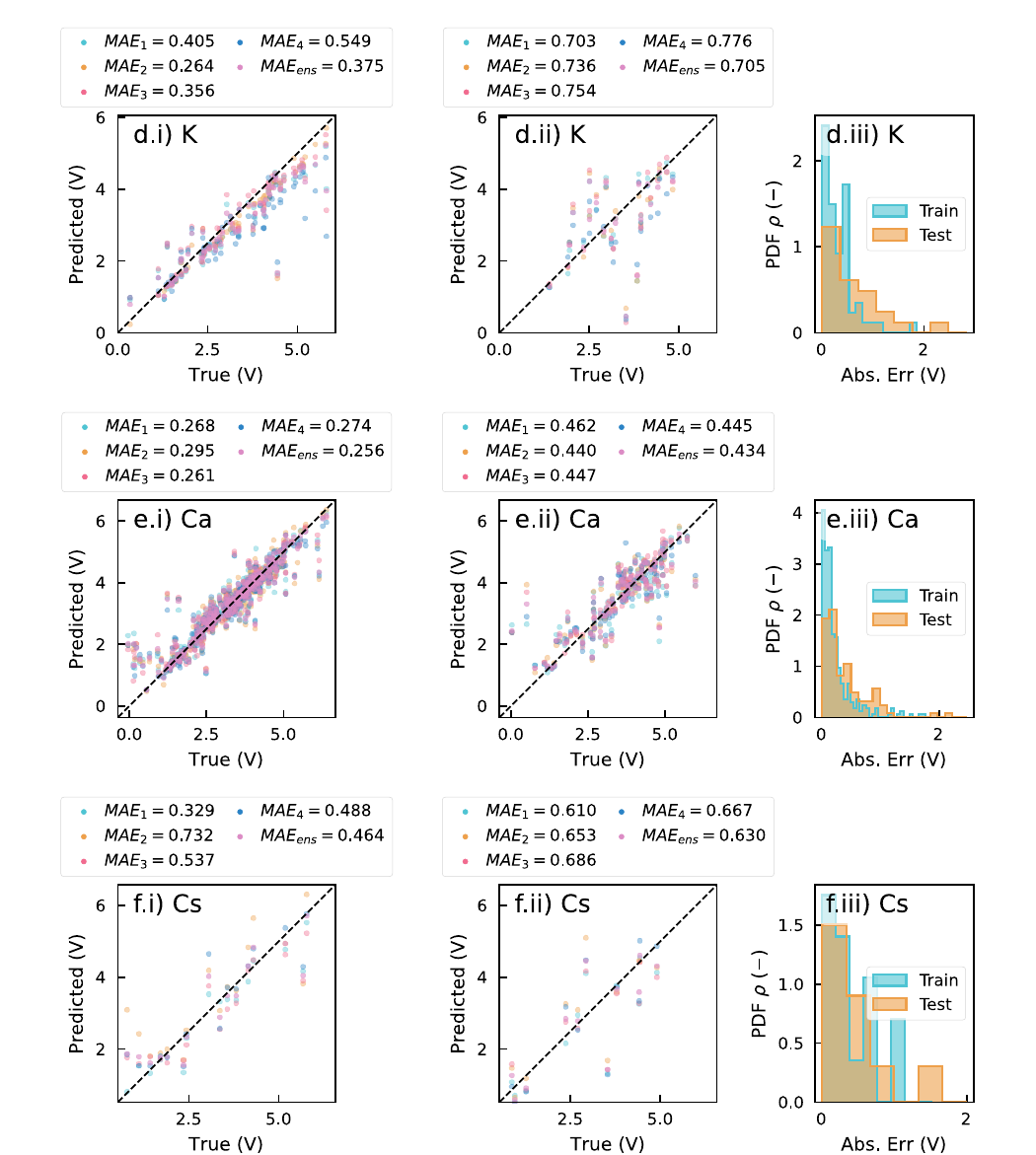}
    \caption[]{
        Results for Li (a), Na (b), Mg (c), K (d), Ca (e), Cs (f), Al (g), Rb (h), and Y (i) cathode materials $\Delta V_c$  regressors.
        i) Parity plot showing the true against predicted avarege voltage $\Delta V_c$ for the 4 trained regressors (highlighting the corresponding mean absolute error $MAE_i$), together with the ensemble case, in which each material is represented by the average prediction across the four trained models for training set and ii) for testing set.
        iii) Absolute error distribution over training and testing sets.
        (Continued)
    }
    \label{fig:batteries_average_voltage_parity_2}
\end{figure}
\begin{figure}[H]
    \addtocounter{figure}{-1}
    \centering
    \includegraphics[width=1\linewidth]{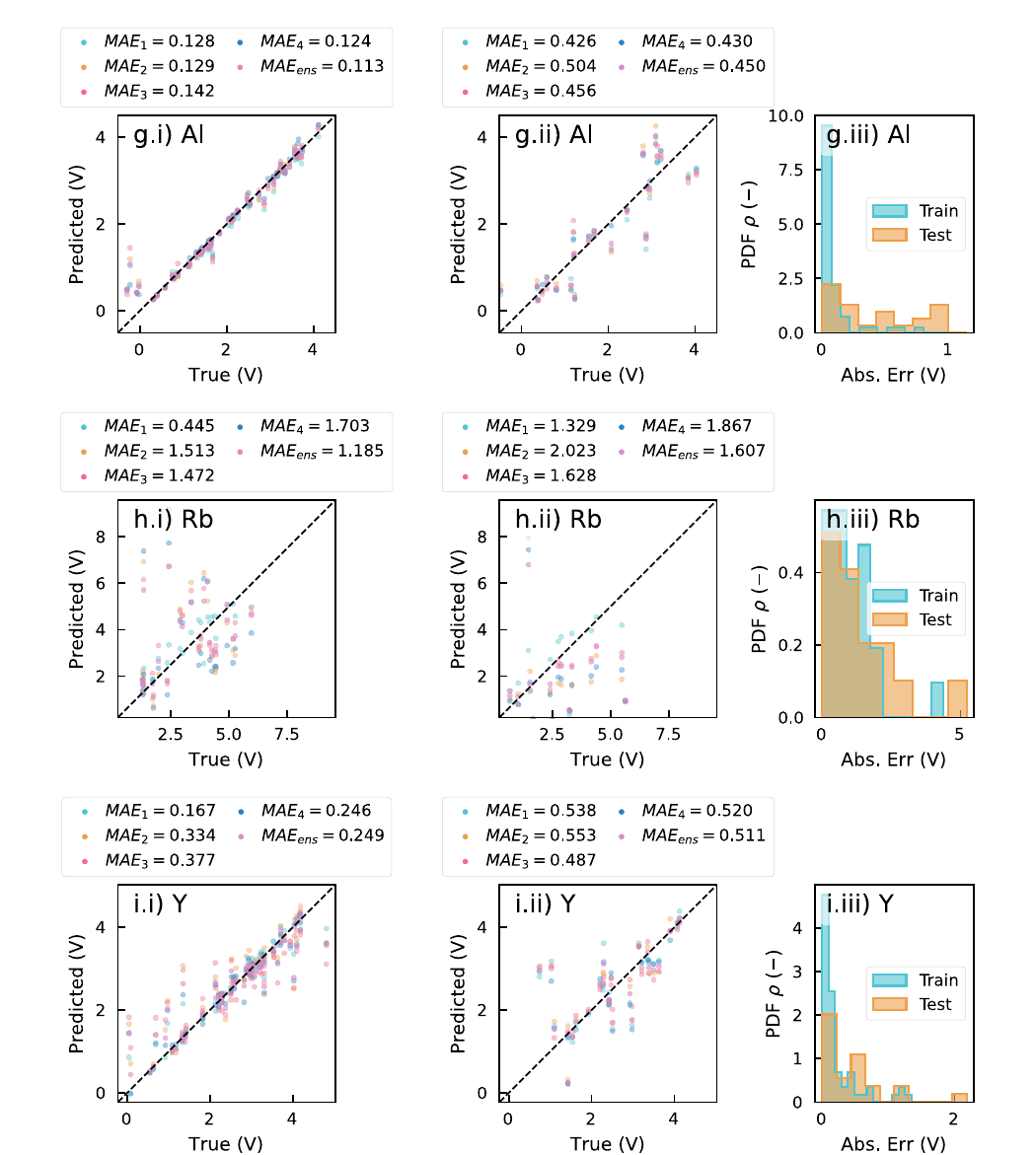}
    \caption[]{
        Results for Li (a), Na (b), Mg (c), K (d), Ca (e), Cs (f), Al (g), Rb (h), and Y (i) cathode materials $\Delta V_c$  regressors.
        i) Parity plot showing the true against predicted avarege voltage $\Delta V_c$ for the 4 trained regressors (highlighting the corresponding mean absolute error $MAE_i$), together with the ensemble case, in which each material is represented by the average prediction across the four trained models for training set and ii) for testing set.
        iii) Absolute error distribution over training and testing sets.
        (Continued)
    }
    \label{fig:batteries_average_voltage_parity_3}
\end{figure}
\begin{longtable}{c|p{30mm}>{\raggedright\let\newline}cccc|c}
    \caption[Performances of the 4 cathode materials per the 9 working ions $\Delta V_c$ regressors.]{
        Performances of the 4 cathode materials per the 9 working ions $\Delta V_c$ regressors (for a total of 36 distinct models), along with the resulting ensemble model, in terms of $R^2$, MAE, RMSE for both training and testing.
        }\label{tab:batteries_average_voltage_parity} \\
    \toprule
    & & Model 1 & Model 2 & Model 3 & Model 4 & Ens. Model \\
    \midrule
    \endfirsthead
    \caption[]{
        Performances of the 4 cathode materials per the 9 working ions $\Delta V_c$ regressors (for a total of 36 distinct models), along with the resulting ensemble model, in terms of $R^2$, MAE, RMSE for both training and testing. (Continued)
    } \\
    \toprule
    & & Model 1 & Model 2 & Model 3 & Model 4 & Ens. Model \\
    \midrule
    \endhead
    
    \endfoot
    \multirow{6}*{Li} &
    $R^2$ train $\mathrm{(-)}$ & 0.807 & 0.821 & 0.821 & 0.802 & 0.828 \\*
    & $\mathrm{MAE}$ train $\mathrm{(V)}$ & 0.300 & 0.284 & 0.291 & 0.309 & 0.279 \\*
    & $\mathrm{RMSE}$ train $\mathrm{(V)}$ & 0.451 & 0.435 & 0.435 & 0.458 & 0.427 \\*
    & $R^2$ test $\mathrm{(-)}$ & 0.696 & 0.729 & 0.711 & 0.708 & 0.733 \\*
    & $\mathrm{MAE}$ test $\mathrm{(V)}$ & 0.396 & 0.376 & 0.388 & 0.383 & 0.368 \\*
    & $\mathrm{RMSE}$ test $\mathrm{(V)}$ & 0.600 & 0.567 & 0.585 & 0.588 & 0.563 \\
    \nopagebreak
    \midrule
    \multirow{6}*{Na} &$R^2$ train $\mathrm{(-)}$ & 0.900 & 0.897 & 0.864 & 0.886 & 0.898 \\*
    & $\mathrm{MAE}$ train $\mathrm{(V)}$ & 0.235 & 0.220 & 0.274 & 0.238 & 0.225 \\*
    & $\mathrm{RMSE}$ train $\mathrm{(V)}$ & 0.343 & 0.349 & 0.400 & 0.367 & 0.346 \\*
    & $R^2$ test $\mathrm{(-)}$ & 0.645 & 0.558 & 0.567 & 0.577 & 0.606 \\*
    & $\mathrm{MAE}$ test $\mathrm{(V)}$ & 0.527 & 0.550 & 0.556 & 0.556 & 0.540 \\*
    & $\mathrm{RMSE}$ test $\mathrm{(V)}$ & 0.707 & 0.789 & 0.780 & 0.771 & 0.744 \\
    \nopagebreak
    \midrule
    \multirow{6}*{Mg} & $R^2$ train $\mathrm{(-)}$ & 0.871 & 0.878 & 0.773 & 0.831 & 0.855 \\*
    & $\mathrm{MAE}$ train $\mathrm{(V)}$ & 0.283 & 0.243 & 0.372 & 0.289 & 0.268 \\*
    & $\mathrm{RMSE}$ train $\mathrm{(V)}$ & 0.474 & 0.462 & 0.628 & 0.543 & 0.502 \\*
    & $R^2$ test $\mathrm{(-)}$ & 0.625 & 0.550 & 0.591 & 0.578 & 0.619 \\*
    & $\mathrm{MAE}$ test $\mathrm{(V)}$ & 0.590 & 0.619 & 0.633 & 0.630 & 0.593 \\*
    & $\mathrm{RMSE}$ test $\mathrm{(V)}$ & 0.870 & 0.952 & 0.907 & 0.922 & 0.876 \\
    \nopagebreak
    \midrule
    \multirow{6}*{K} & $R^2$ train $\mathrm{(-)}$ & 0.815 & 0.882 & 0.809 & 0.662 & 0.825 \\*
    & $\mathrm{MAE}$ train $\mathrm{(V)}$ & 0.405 & 0.264 & 0.356 & 0.549 & 0.375 \\*
    & $\mathrm{RMSE}$ train $\mathrm{(V)}$ & 0.574 & 0.458 & 0.583 & 0.776 & 0.558 \\*
    & $R^2$ test $\mathrm{(-)}$ & -0.186 & -0.232 & -0.337 & -0.156 & -0.173 \\*
    & $\mathrm{MAE}$ test $\mathrm{(V)}$ & 0.703 & 0.736 & 0.754 & 0.776 & 0.705 \\*
    & $\mathrm{RMSE}$ test $\mathrm{(V)}$ & 1.029 & 1.049 & 1.093 & 1.016 & 1.023 \\
    \nopagebreak
    \midrule
    \multirow{6}*{Ca} & $R^2$ train $\mathrm{(-)}$ & 0.857 & 0.840 & 0.866 & 0.862 & 0.866 \\*
    & $\mathrm{MAE}$ train $\mathrm{(V)}$ & 0.268 & 0.295 & 0.261 & 0.274 & 0.256 \\*
    & $\mathrm{RMSE}$ train $\mathrm{(V)}$ & 0.445 & 0.472 & 0.431 & 0.438 & 0.431 \\*
    & $R^2$ test $\mathrm{(-)}$ & 0.666 & 0.664 & 0.672 & 0.701 & 0.695 \\*
    & $\mathrm{MAE}$ test $\mathrm{(V)}$ & 0.462 & 0.440 & 0.447 & 0.445 & 0.434 \\*
    & $\mathrm{RMSE}$ test $\mathrm{(V)}$ & 0.684 & 0.686 & 0.678 & 0.647 & 0.654 \\
    \nopagebreak
    \midrule
    \multirow{6}*{Cs} & $R^2$ train $\mathrm{(-)}$ & 0.894 & 0.587 & 0.811 & 0.800 & 0.823 \\*
    & $\mathrm{MAE}$ train $\mathrm{(V)}$ & 0.329 & 0.732 & 0.537 & 0.488 & 0.464 \\*
    & $\mathrm{RMSE}$ train $\mathrm{(V)}$ & 0.500 & 0.986 & 0.666 & 0.686 & 0.646 \\*
    & $R^2$ test $\mathrm{(-)}$ & 0.594 & 0.510 & 0.568 & 0.517 & 0.594 \\*
    & $\mathrm{MAE}$ test $\mathrm{(V)}$ & 0.610 & 0.653 & 0.686 & 0.667 & 0.630 \\*
    & $\mathrm{RMSE}$ test $\mathrm{(V)}$ & 0.865 & 0.950 & 0.892 & 0.943 & 0.865 \\
    \nopagebreak
    \midrule
    \multirow{6}*{Al} & $R^2$ train $\mathrm{(-)}$ & 0.951 & 0.944 & 0.936 & 0.966 & 0.954 \\*
    & $\mathrm{MAE}$ train $\mathrm{(V)}$ & 0.128 & 0.129 & 0.142 & 0.124 & 0.113 \\*
    & $\mathrm{RMSE}$ train $\mathrm{(V)}$ & 0.256 & 0.274 & 0.293 & 0.213 & 0.249 \\*
    & $R^2$ test $\mathrm{(-)}$ & 0.797 & 0.737 & 0.773 & 0.795 & 0.783 \\*
    & $\mathrm{MAE}$ test $\mathrm{(V)}$ & 0.426 & 0.504 & 0.456 & 0.430 & 0.450 \\*
    & $\mathrm{RMSE}$ test $\mathrm{(V)}$ & 0.553 & 0.629 & 0.584 & 0.555 & 0.571 \\
    \nopagebreak
    \midrule
    \multirow{6}*{Rb} & $R^2$ train $\mathrm{(-)}$ & 0.833 & -0.961 & -1.346 & -1.356 & -0.290 \\*
    & $\mathrm{MAE}$ train $\mathrm{(V)}$ & 0.445 & 1.513 & 1.472 & 1.703 & 1.185 \\*
    & $\mathrm{RMSE}$ train $\mathrm{(V)}$ & 0.579 & 1.986 & 2.172 & 2.177 & 1.611 \\*
    & $R^2$ test $\mathrm{(-)}$ & -0.275 & -1.880 & -1.241 & -1.398 & -0.944 \\*
    & $\mathrm{MAE}$ test $\mathrm{(V)}$ & 1.329 & 2.023 & 1.628 & 1.867 & 1.607 \\*
    & $\mathrm{RMSE}$ test $\mathrm{(V)}$ & 1.778 & 2.673 & 2.358 & 2.439 & 2.196 \\
    \nopagebreak
    \midrule
    \multirow{6}*{Y} & $R^2$ train $\mathrm{(-)}$ & 0.941 & 0.786 & 0.728 & 0.867 & 0.859 \\*
    & $\mathrm{MAE}$ train $\mathrm{(V)}$ & 0.167 & 0.334 & 0.377 & 0.246 & 0.249 \\*
    & $\mathrm{RMSE}$ train $\mathrm{(V)}$ & 0.272 & 0.518 & 0.585 & 0.409 & 0.421 \\*
    & $R^2$ test $\mathrm{(-)}$ & 0.284 & 0.324 & 0.390 & 0.363 & 0.365 \\*
    & $\mathrm{MAE}$ test $\mathrm{(V)}$ & 0.538 & 0.553 & 0.487 & 0.520 & 0.511 \\*
    & $\mathrm{RMSE}$ test $\mathrm{(V)}$ & 0.804 & 0.781 & 0.743 & 0.759 & 0.757 \\
    \nopagebreak
    \bottomrule
\end{longtable}

\subsection{Regressors ($\max(\Delta Vol)$)}\label{ssec:batteries_max_delta_volume_regressors}

\begin{figure}[H]
    \centering
    \includegraphics[width=1\linewidth]{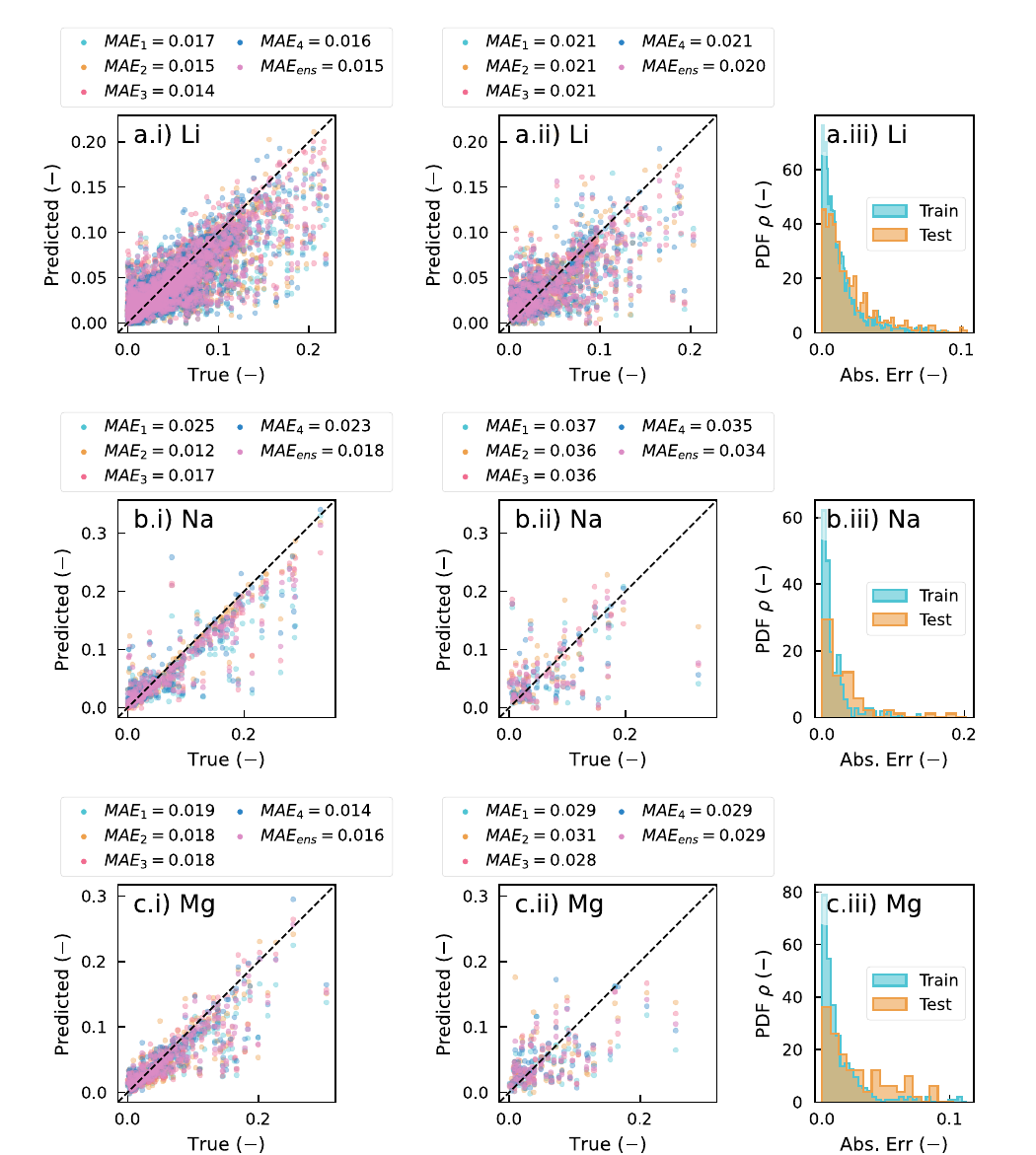}
    \caption[Results for Li, Na, Mg, K, Ca, Cs, Al, Rb, and Y cathode materials $\max(\Delta Vol)$  regressors.]{
        Results for Li (a), Na (b), Mg (c), K (d), Ca (e), Cs (f), Al (g), Rb (h), and Y (i) cathode materials $\max(\Delta Vol)$  regressors.
        i) Parity plot showing the true against predicted maximum voltage expansion $\max(\Delta Vol)$for the 4 trained regressors (highlighting the corresponding mean absolute error $MAE_i$), together with the ensemble case, in which each material is represented by the average prediction across the four trained models for training set and ii) for testing set.
        iii) Absolute error distribution over training and testing sets.
    }
    \label{fig:batteries_max_delta_volume_parity_1}
\end{figure}
\begin{figure}[H]
    \addtocounter{figure}{-1}
    \centering
    \includegraphics[width=1\linewidth]{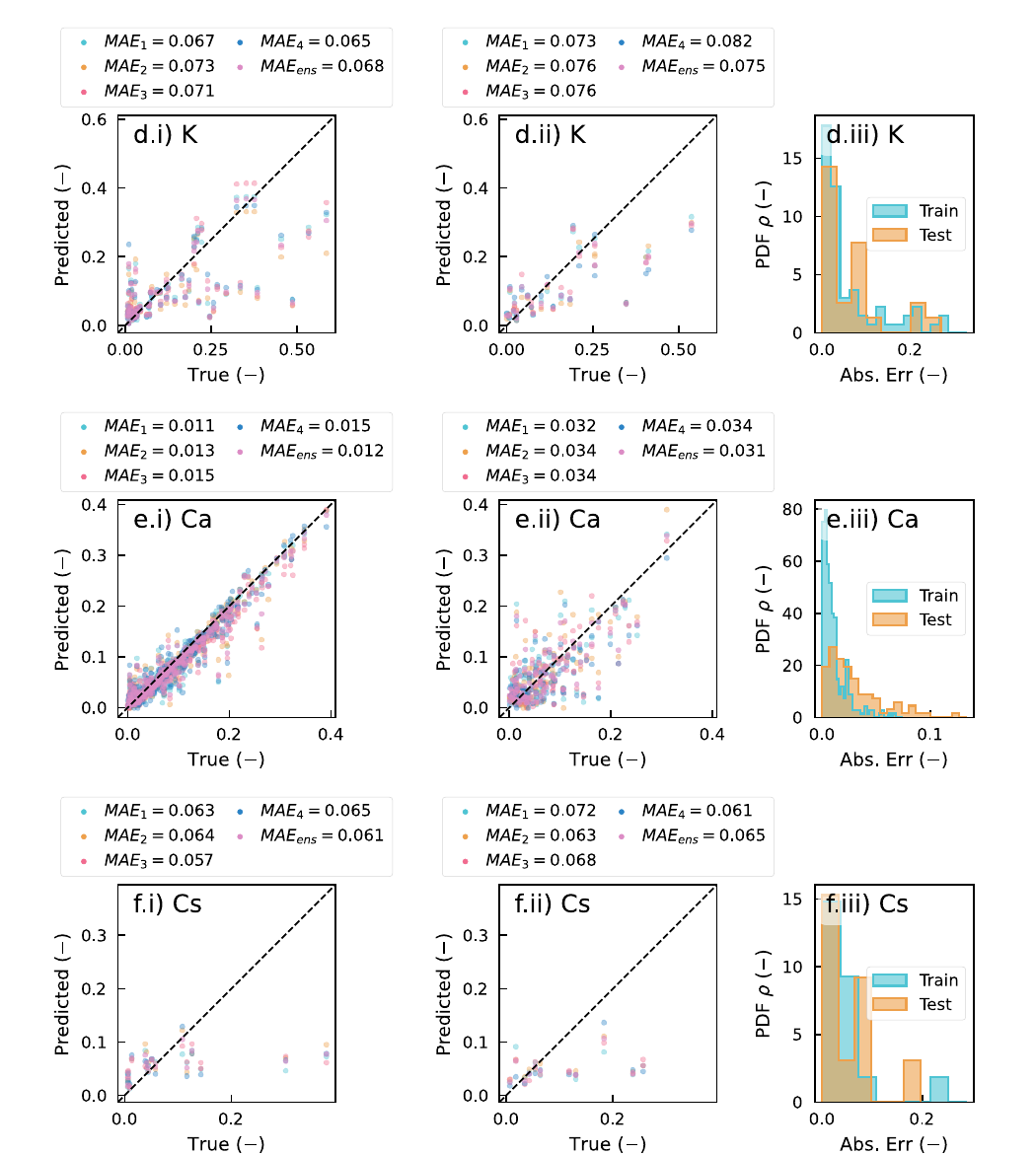}
    \caption[]{
        Results for Li (a), Na (b), Mg (c), K (d), Ca (e), Cs (f), Al (g), Rb (h), and Y (i) cathode materials $\max(\Delta Vol)$  regressors.
        i) Parity plot showing the true against predicted maximum voltage expansion $\max(\Delta Vol)$ for the 4 trained regressors (highlighting the corresponding mean absolute error $MAE_i$), together with the ensemble case, in which each material is represented by the average prediction across the four trained models for training set and ii) for testing set.
        iii) Absolute error distribution over training and testing sets.
        (Continued)
    }
    \label{fig:batteries_max_delta_volume_parity_2}
\end{figure}
\begin{figure}[H]
    \addtocounter{figure}{-1}
    \centering
    \includegraphics[width=1\linewidth]{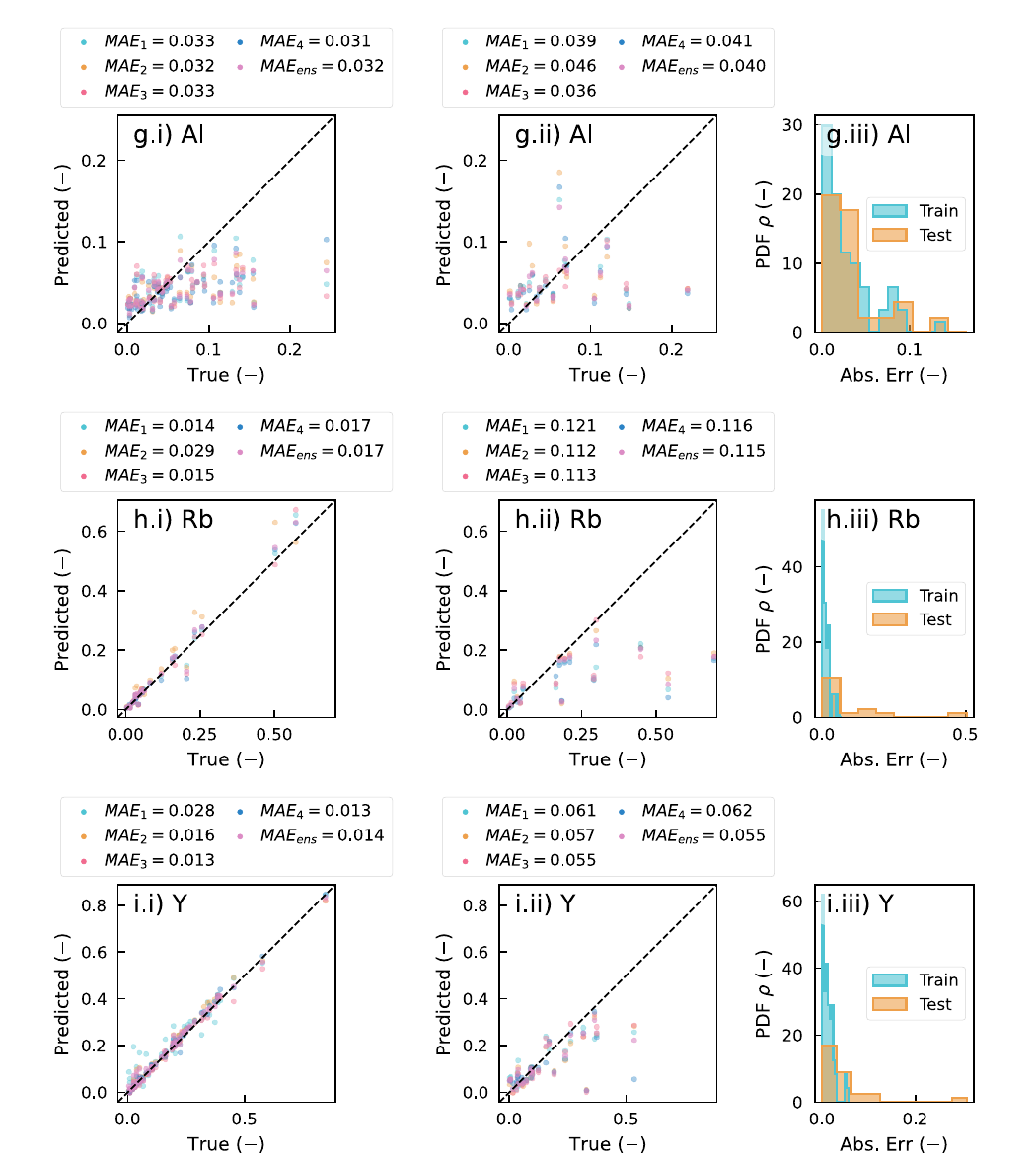}
    \caption[]{
        Results for Li (a), Na (b), Mg (c), K (d), Ca (e), Cs (f), Al (g), Rb (h), and Y (i) cathode materials $\max(\Delta Vol)$  regressors.
        i) Parity plot showing the true against predicted maximum voltage expansion $\max(\Delta Vol)$ for the 4 trained regressors (highlighting the corresponding mean absolute error $MAE_i$), together with the ensemble case, in which each material is represented by the average prediction across the four trained models for training set and ii) for testing set.
        iii) Absolute error distribution over training and testing sets.
        (Continued)
    }
    \label{fig:batteries_max_delta_volume_parity_3}
\end{figure}
\begin{longtable}{c|p{30mm}>{\raggedright\let\newline}cccc|c}
    \caption[Performances of the 4 cathode materials per the 9 working ions$\max(\Delta Vol)$ regressors.]{
        Performances of the 4 cathode materials per the 9 working ions $\max(\Delta Vol)$ regressors (for a total of 36 distinct models), along with the resulting ensemble model, in terms of $R^2$, MAE, RMSE for both training and testing.
        }\label{tab:batteries_average_voltage_parity}\label{tab:batteries_max_delta_volume_parity} \\
    \toprule
    & & Model 1 & Model 2 & Model 3 & Model 4 & Ens. Model \\
    \midrule
    \endfirsthead
    \caption[]{
        Performances of the 4 cathode materials per the 9 working ions $\max(\Delta Vol)$ regressors (for a total of 36 distinct models), along with the resulting ensemble model, in terms of $R^2$, MAE, RMSE for both training and testing. (Continued)
    } \\
    \toprule
    & & Model 1 & Model 2 & Model 3 & Model 4 & Ens. Model \\
    \midrule
    \endhead
    
    \endfoot
    \multirow{6}*{Li} &
    $R^2$ train $\mathrm{(-)}$ & 0.575 & 0.639 & 0.694 & 0.624 & 0.673 \\*
& $\mathrm{MAE}$ train $\mathrm{(-)}$ & 0.017 & 0.015 & 0.014 & 0.016 & 0.015 \\*
& $\mathrm{RMSE}$ train $\mathrm{(-)}$ & 0.027 & 0.024 & 0.023 & 0.025 & 0.023 \\*
& $R^2$ test $\mathrm{(-)}$ & 0.260 & 0.380 & 0.341 & 0.442 & 0.442 \\*
& $\mathrm{MAE}$ test $\mathrm{(-)}$ & 0.021 & 0.021 & 0.021 & 0.021 & 0.020 \\*
& $\mathrm{RMSE}$ test $\mathrm{(-)}$ & 0.035 & 0.032 & 0.033 & 0.030 & 0.030 \\
    \nopagebreak
    \midrule
    \multirow{6}*{Na} &
    $R^2$ train $\mathrm{(-)}$ & 0.683 & 0.876 & 0.819 & 0.713 & 0.806 \\*
& $\mathrm{MAE}$ train $\mathrm{(-)}$ & 0.025 & 0.012 & 0.017 & 0.023 & 0.018 \\*
& $\mathrm{RMSE}$ train $\mathrm{(-)}$ & 0.041 & 0.026 & 0.031 & 0.039 & 0.032 \\*
& $R^2$ test $\mathrm{(-)}$ & 0.055 & 0.235 & 0.162 & 0.146 & 0.205 \\*
& $\mathrm{MAE}$ test $\mathrm{(-)}$ & 0.037 & 0.036 & 0.036 & 0.035 & 0.034 \\*
& $\mathrm{RMSE}$ test $\mathrm{(-)}$ & 0.060 & 0.054 & 0.057 & 0.057 & 0.055 \\
    \nopagebreak
    \midrule
    \multirow{6}*{Mg} & 
    $R^2$ train $\mathrm{(-)}$ & 0.601 & 0.672 & 0.645 & 0.762 & 0.704 \\*
& $\mathrm{MAE}$ train $\mathrm{(-)}$ & 0.019 & 0.018 & 0.018 & 0.014 & 0.016 \\*
& $\mathrm{RMSE}$ train $\mathrm{(-)}$ & 0.032 & 0.029 & 0.030 & 0.025 & 0.028 \\*
& $R^2$ test $\mathrm{(-)}$ & 0.222 & 0.200 & 0.372 & 0.306 & 0.335 \\*
& $\mathrm{MAE}$ test $\mathrm{(-)}$ & 0.029 & 0.031 & 0.028 & 0.029 & 0.029 \\*
& $\mathrm{RMSE}$ test $\mathrm{(-)}$ & 0.043 & 0.044 & 0.039 & 0.041 & 0.040 \\
    \nopagebreak
    \midrule
    \multirow{6}*{K} & 
    $R^2$ train $\mathrm{(-)}$ & 0.426 & 0.328 & 0.404 & 0.447 & 0.414 \\*
& $\mathrm{MAE}$ train $\mathrm{(-)}$ & 0.067 & 0.073 & 0.071 & 0.065 & 0.068 \\*
& $\mathrm{RMSE}$ train $\mathrm{(-)}$ & 0.108 & 0.117 & 0.110 & 0.106 & 0.109 \\*
& $R^2$ test $\mathrm{(-)}$ & 0.484 & 0.473 & 0.423 & 0.340 & 0.440 \\*
& $\mathrm{MAE}$ test $\mathrm{(-)}$ & 0.073 & 0.076 & 0.076 & 0.082 & 0.075 \\*
& $\mathrm{RMSE}$ test $\mathrm{(-)}$ & 0.104 & 0.105 & 0.110 & 0.118 & 0.109 \\
    \nopagebreak
    \midrule
    \multirow{6}*{Ca} & 
    $R^2$ train $\mathrm{(-)}$ & 0.942 & 0.921 & 0.912 & 0.915 & 0.941 \\*
& $\mathrm{MAE}$ train $\mathrm{(-)}$ & 0.011 & 0.013 & 0.015 & 0.015 & 0.012 \\*
& $\mathrm{RMSE}$ train $\mathrm{(-)}$ & 0.018 & 0.022 & 0.023 & 0.022 & 0.019 \\*
& $R^2$ test $\mathrm{(-)}$ & 0.491 & 0.477 & 0.479 & 0.472 & 0.550 \\*
& $\mathrm{MAE}$ test $\mathrm{(-)}$ & 0.032 & 0.034 & 0.034 & 0.034 & 0.031 \\*
& $\mathrm{RMSE}$ test $\mathrm{(-)}$ & 0.046 & 0.047 & 0.047 & 0.047 & 0.043 \\
    \nopagebreak
    \midrule
    \multirow{6}*{Cs} & 
    $R^2$ train $\mathrm{(-)}$ & -0.011 & 0.096 & 0.055 & 0.027 & 0.050 \\*
& $\mathrm{MAE}$ train $\mathrm{(-)}$ & 0.063 & 0.064 & 0.057 & 0.065 & 0.061 \\*
& $\mathrm{RMSE}$ train $\mathrm{(-)}$ & 0.105 & 0.099 & 0.102 & 0.103 & 0.102 \\*
& $R^2$ test $\mathrm{(-)}$ & -0.388 & -0.213 & -0.231 & -0.211 & -0.246 \\*
& $\mathrm{MAE}$ test $\mathrm{(-)}$ & 0.072 & 0.063 & 0.068 & 0.061 & 0.065 \\*
& $\mathrm{RMSE}$ test $\mathrm{(-)}$ & 0.099 & 0.092 & 0.093 & 0.092 & 0.094 \\
    \nopagebreak
    \midrule
    \multirow{6}*{Al} & 
    $R^2$ train $\mathrm{(-)}$ & 0.070 & 0.140 & -0.006 & 0.190 & 0.119 \\*
& $\mathrm{MAE}$ train $\mathrm{(-)}$ & 0.033 & 0.032 & 0.033 & 0.031 & 0.032 \\*
& $\mathrm{RMSE}$ train $\mathrm{(-)}$ & 0.048 & 0.046 & 0.050 & 0.045 & 0.047 \\*
& $R^2$ test $\mathrm{(-)}$ & -0.171 & -0.433 & -0.049 & -0.303 & -0.186 \\*
& $\mathrm{MAE}$ test $\mathrm{(-)}$ & 0.039 & 0.046 & 0.036 & 0.041 & 0.040 \\*
& $\mathrm{RMSE}$ test $\mathrm{(-)}$ & 0.058 & 0.064 & 0.055 & 0.061 & 0.059 \\
    \nopagebreak
    \midrule
    \multirow{6}*{Rb} & 
    $R^2$ train $\mathrm{(-)}$ & 0.976 & 0.913 & 0.962 & 0.966 & 0.970 \\*
& $\mathrm{MAE}$ train $\mathrm{(-)}$ & 0.014 & 0.029 & 0.015 & 0.017 & 0.017 \\*
& $\mathrm{RMSE}$ train $\mathrm{(-)}$ & 0.022 & 0.042 & 0.028 & 0.026 & 0.025 \\*
& $R^2$ test $\mathrm{(-)}$ & -0.010 & 0.090 & 0.079 & -0.040 & 0.040 \\*
& $\mathrm{MAE}$ test $\mathrm{(-)}$ & 0.121 & 0.112 & 0.113 & 0.116 & 0.115 \\*
& $\mathrm{RMSE}$ test $\mathrm{(-)}$ & 0.197 & 0.187 & 0.188 & 0.200 & 0.192 \\
    \nopagebreak
    \midrule
    \multirow{6}*{Y} & 
    $R^2$ train $\mathrm{(-)}$ & 0.924 & 0.980 & 0.987 & 0.989 & 0.985 \\*
& $\mathrm{MAE}$ train $\mathrm{(-)}$ & 0.028 & 0.016 & 0.013 & 0.013 & 0.014 \\*
& $\mathrm{RMSE}$ train $\mathrm{(-)}$ & 0.043 & 0.022 & 0.018 & 0.017 & 0.019 \\*
& $R^2$ test $\mathrm{(-)}$ & 0.474 & 0.553 & 0.555 & 0.222 & 0.488 \\*
& $\mathrm{MAE}$ test $\mathrm{(-)}$ & 0.061 & 0.057 & 0.055 & 0.062 & 0.055 \\*
& $\mathrm{RMSE}$ test $\mathrm{(-)}$ & 0.097 & 0.090 & 0.090 & 0.118 & 0.096 \\
    \nopagebreak
    \bottomrule
\end{longtable}

\subsection{Regressors ($\Delta E_{charge}$)}\label{ssec:batteries_stability_charge_regressors}

\begin{figure}[H]
    \centering
    \includegraphics[width=1\linewidth]{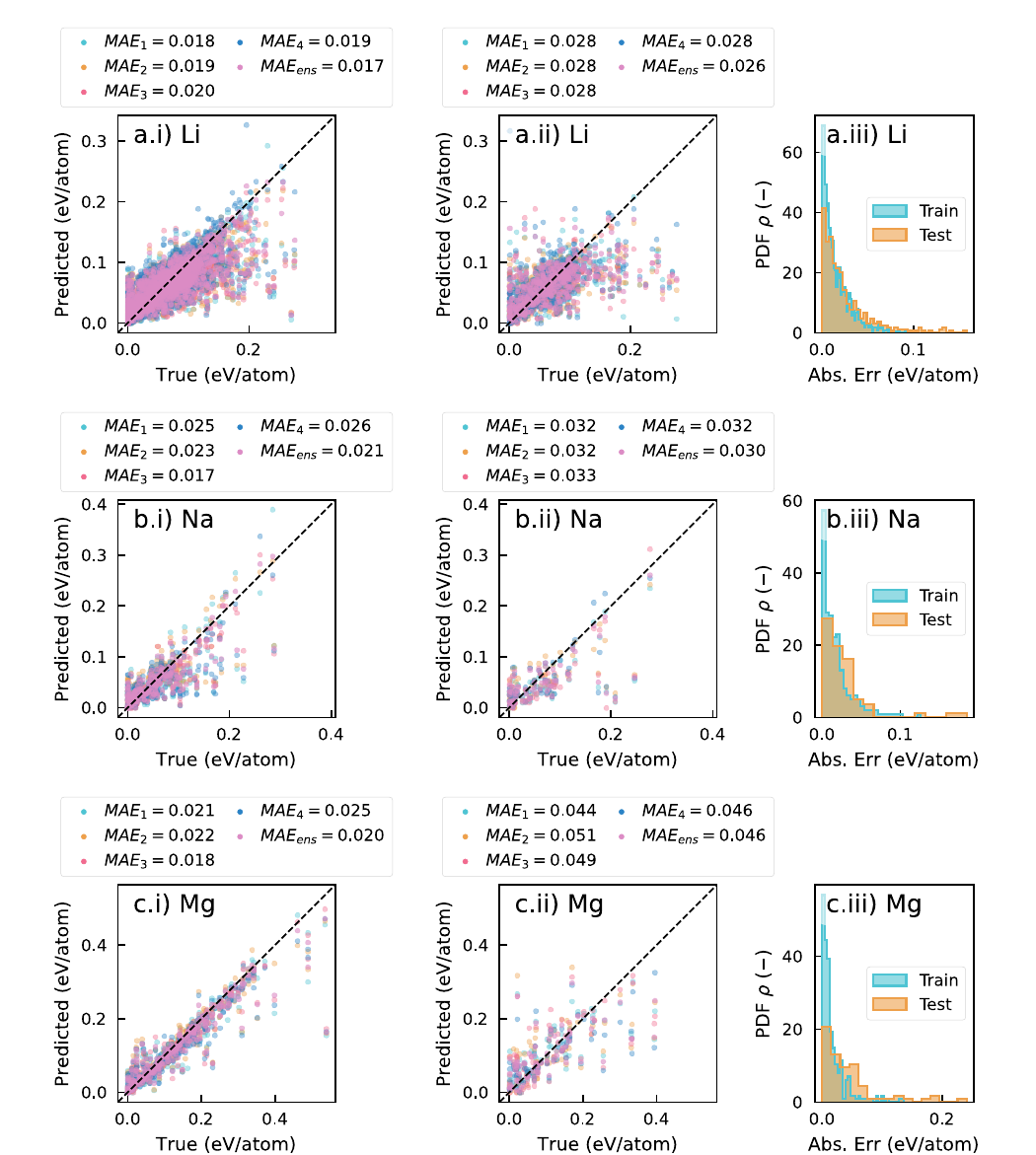}
    \caption[Results for Li, Na, Mg, K, Ca, Cs, Al, Rb, and Y cathode materials $\Delta E_{charge}$  regressors.]{
        Results for Li (a), Na (b), Mg (c), K (d), Ca (e), Cs (f), Al (g), Rb (h), and Y (i) cathode materials $\Delta E_{charge}$  regressors.
        i) Parity plot showing the true against predicted cathode stability in charged state $\Delta E_{charge}$ for the 4 trained regressors (highlighting the corresponding mean absolute error $MAE_i$), together with the ensemble case, in which each material is represented by the average prediction across the four trained models for training set and ii) for testing set.
        iii) Absolute error distribution over training and testing sets.
    }
    \label{fig:batteries_stability_charge_parity_1}
\end{figure}
\begin{figure}[H]
    \addtocounter{figure}{-1}
    \centering
    \includegraphics[width=1\linewidth]{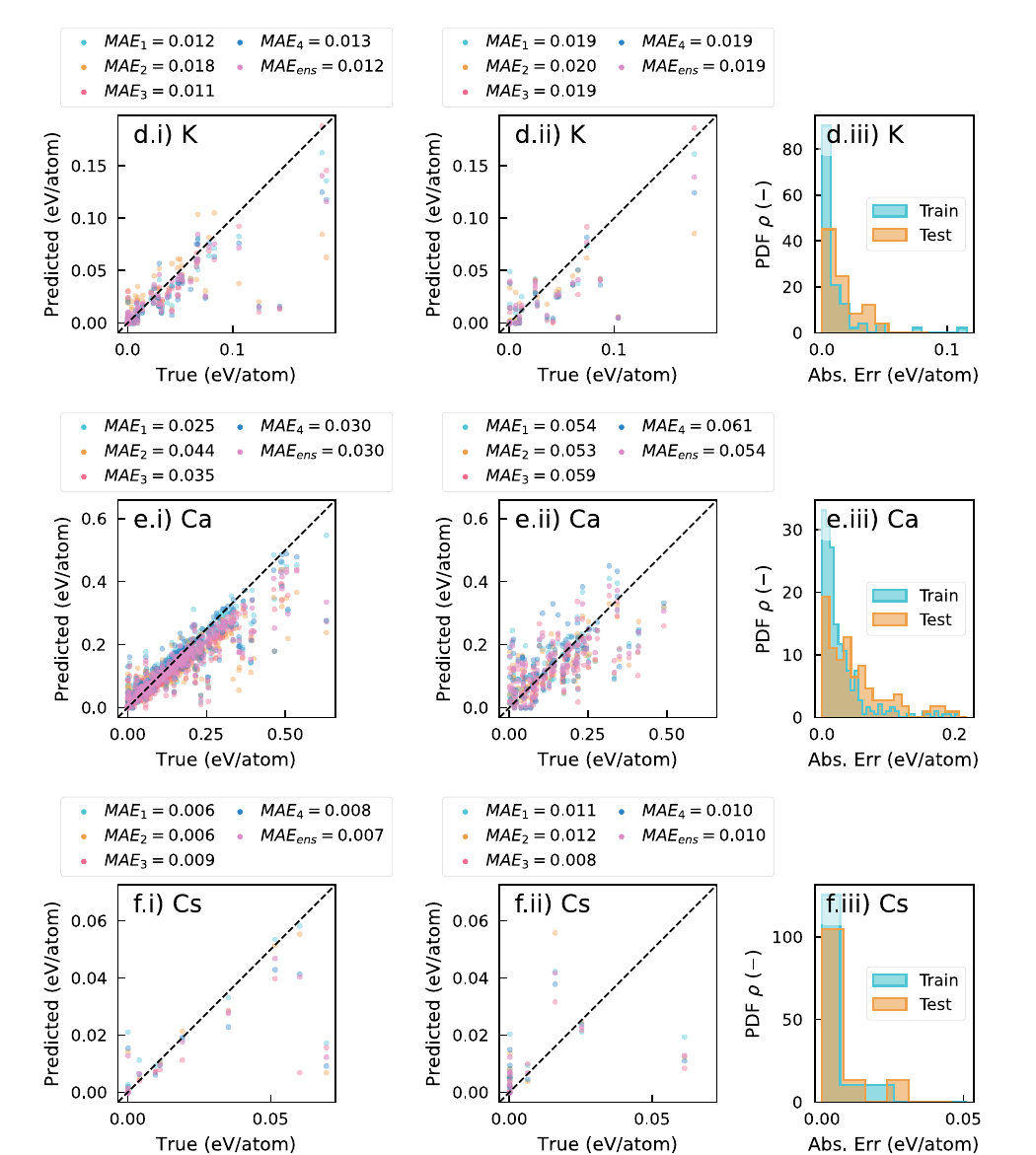}
    \caption[]{
        Results for Li (a), Na (b), Mg (c), K (d), Ca (e), Cs (f), Al (g), Rb (h), and Y (i) cathode materials $\Delta E_{charge}$  regressors.
        i) Parity plot showing the true against predicted cathode stability in charged state $\Delta E_{charge}$ for the 4 trained regressors (highlighting the corresponding mean absolute error $MAE_i$), together with the ensemble case, in which each material is represented by the average prediction across the four trained models for training set and ii) for testing set.
        iii) Absolute error distribution over training and testing sets.
        (Continued)
    }
    \label{fig:batteries_stability_charge_parity_2}
\end{figure}
\begin{figure}[H]
    \addtocounter{figure}{-1}
    \centering
    \includegraphics[width=1\linewidth]{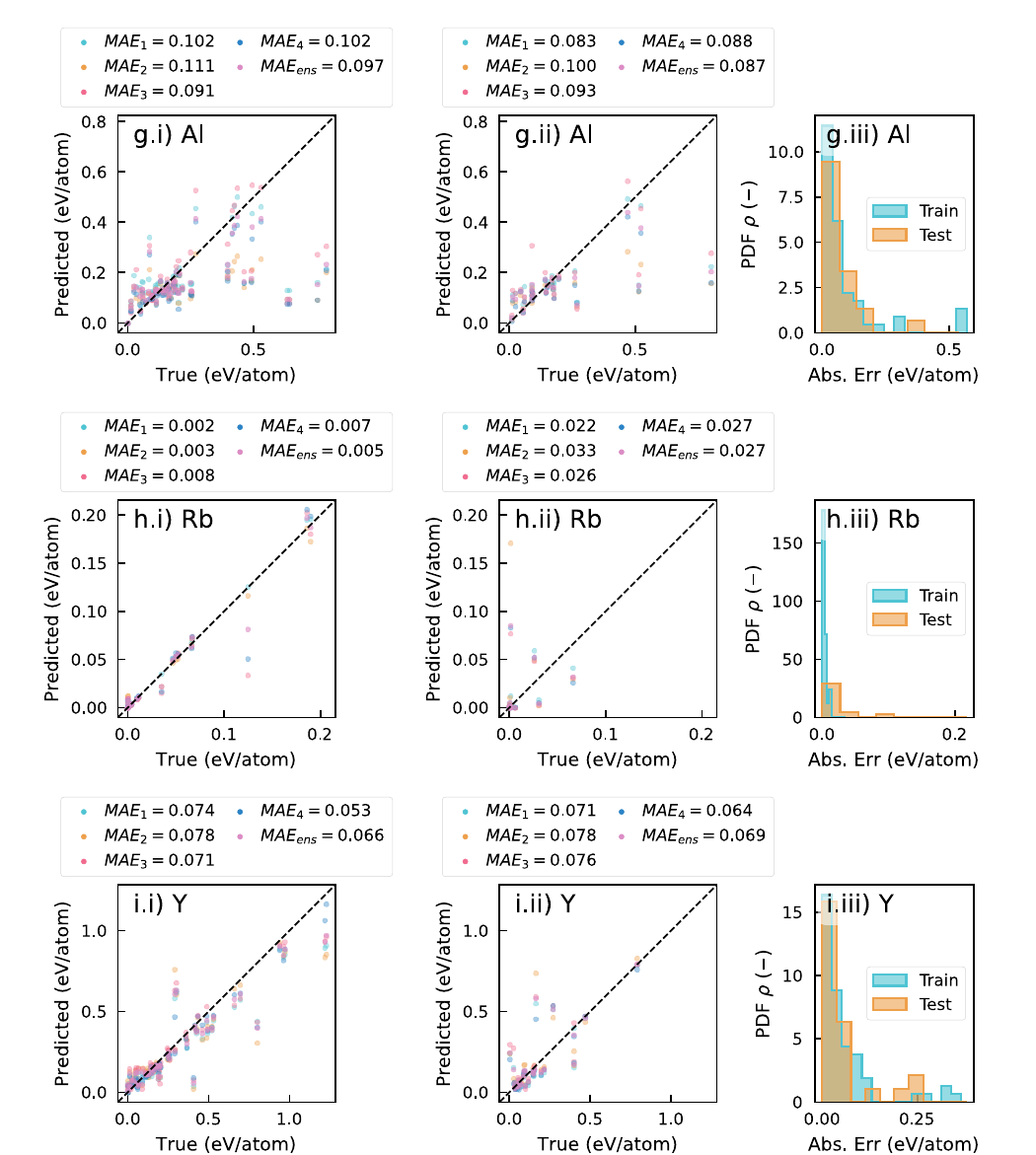}
    \caption[]{
        Results for Li (a), Na (b), Mg (c), K (d), Ca (e), Cs (f), Al (g), Rb (h), and Y (i) cathode materials $\Delta E_{charge}$  regressors.
        i) Parity plot showing the true against predicted cathode stability in charged state $\Delta E_{charge}$ for the 4 trained regressors (highlighting the corresponding mean absolute error $MAE_i$), together with the ensemble case, in which each material is represented by the average prediction across the four trained models for training set and ii) for testing set.
        iii) Absolute error distribution over training and testing sets.
        (Continued)
    }
    \label{fig:batteries_stability_charge_parity_3}
\end{figure}
\begin{longtable}{c|p{30mm}>{\raggedright\let\newline}cccc|c}
    \caption[Performances of the 4 cathode materials per the 9 working ions $\Delta E_{charge}$ regressors.]{
        Performances of the 4 cathode materials per the 9 working ions $\Delta E_{charge}$ regressors (for a total of 36 distinct models), along with the resulting ensemble model, in terms of $R^2$, MAE, RMSE for both training and testing.
        }\label{tab:batteries_stability_charge_parity} \\
    \toprule
    & & Model 1 & Model 2 & Model 3 & Model 4 & Ens. Model \\
    \midrule
    \endfirsthead
    \caption[]{
        Performances of the 4 cathode materials per the 9 working ions $\Delta E_{charge}$ regressors (for a total of 36 distinct models), along with the resulting ensemble model, in terms of $R^2$, MAE, RMSE for both training and testing. (Continued)
    } \\
    \toprule
    & & Model 1 & Model 2 & Model 3 & Model 4 & Ens. Model \\
    \midrule
    \endhead
    
    \endfoot
    \multirow{6}*{Li} &
    $R^2$ train $\mathrm{(-)}$ & 0.598 & 0.573 & 0.536 & 0.591 & 0.621 \\*
& $\mathrm{MAE}$ train $\mathrm{(eV)}$ & 0.018 & 0.019 & 0.020 & 0.019 & 0.017 \\*
& $\mathrm{RMSE}$ train $\mathrm{(eV)}$ & 0.029 & 0.029 & 0.031 & 0.029 & 0.028 \\*
& $R^2$ test $\mathrm{(-)}$ & 0.242 & 0.279 & 0.245 & 0.222 & 0.315 \\*
& $\mathrm{MAE}$ test $\mathrm{(eV)}$ & 0.028 & 0.028 & 0.028 & 0.028 & 0.026 \\*
& $\mathrm{RMSE}$ test $\mathrm{(eV)}$ & 0.043 & 0.042 & 0.043 & 0.044 & 0.041 \\
    \nopagebreak
    \midrule
    \multirow{6}*{Na} &
    $R^2$ train $\mathrm{(-)}$ & 0.558 & 0.604 & 0.737 & 0.508 & 0.654 \\*
& $\mathrm{MAE}$ train $\mathrm{(eV)}$ & 0.025 & 0.023 & 0.017 & 0.026 & 0.021 \\*
& $\mathrm{RMSE}$ train $\mathrm{(eV)}$ & 0.038 & 0.036 & 0.029 & 0.040 & 0.033 \\*
& $R^2$ test $\mathrm{(-)}$ & 0.384 & 0.436 & 0.371 & 0.395 & 0.432 \\*
& $\mathrm{MAE}$ test $\mathrm{(eV)}$ & 0.032 & 0.032 & 0.033 & 0.032 & 0.030 \\*
& $\mathrm{RMSE}$ test $\mathrm{(eV)}$ & 0.051 & 0.049 & 0.052 & 0.051 & 0.049 \\
    \nopagebreak
    \midrule
    \multirow{6}*{Mg} & 
    $R^2$ train $\mathrm{(-)}$ & 0.852 & 0.846 & 0.875 & 0.834 & 0.868 \\*
& $\mathrm{MAE}$ train $\mathrm{(eV)}$ & 0.021 & 0.022 & 0.018 & 0.025 & 0.020 \\*
& $\mathrm{RMSE}$ train $\mathrm{(eV)}$ & 0.040 & 0.041 & 0.037 & 0.042 & 0.038 \\*
& $R^2$ test $\mathrm{(-)}$ & 0.501 & 0.387 & 0.432 & 0.477 & 0.491 \\*
& $\mathrm{MAE}$ test $\mathrm{(eV)}$ & 0.044 & 0.051 & 0.049 & 0.046 & 0.046 \\*
& $\mathrm{RMSE}$ test $\mathrm{(eV)}$ & 0.068 & 0.076 & 0.073 & 0.070 & 0.069 \\
    \nopagebreak
    \midrule
    \multirow{6}*{K} & 
    $R^2$ train $\mathrm{(-)}$ & 0.658 & 0.390 & 0.660 & 0.602 & 0.622 \\*
& $\mathrm{MAE}$ train $\mathrm{(eV)}$ & 0.012 & 0.018 & 0.011 & 0.013 & 0.012 \\*
& $\mathrm{RMSE}$ train $\mathrm{(eV)}$ & 0.024 & 0.033 & 0.024 & 0.026 & 0.026 \\*
& $R^2$ test $\mathrm{(-)}$ & 0.535 & 0.413 & 0.531 & 0.513 & 0.545 \\*
& $\mathrm{MAE}$ test $\mathrm{(eV)}$ & 0.019 & 0.020 & 0.019 & 0.019 & 0.019 \\*
& $\mathrm{RMSE}$ test $\mathrm{(eV)}$ & 0.028 & 0.032 & 0.029 & 0.029 & 0.028 \\
    \nopagebreak
    \midrule
    \multirow{6}*{Ca} & 
    $R^2$ train $\mathrm{(-)}$ & 0.846 & 0.618 & 0.770 & 0.784 & 0.794 \\*
& $\mathrm{MAE}$ train $\mathrm{(eV)}$ & 0.025 & 0.044 & 0.035 & 0.030 & 0.030 \\*
& $\mathrm{RMSE}$ train $\mathrm{(eV)}$ & 0.044 & 0.070 & 0.054 & 0.052 & 0.051 \\*
& $R^2$ test $\mathrm{(-)}$ & 0.469 & 0.477 & 0.367 & 0.331 & 0.477 \\*
& $\mathrm{MAE}$ test $\mathrm{(eV)}$ & 0.054 & 0.053 & 0.059 & 0.061 & 0.054 \\*
& $\mathrm{RMSE}$ test $\mathrm{(eV)}$ & 0.076 & 0.076 & 0.083 & 0.086 & 0.076 \\
    \nopagebreak
    \midrule
    \multirow{6}*{Cs} & 
    $R^2$ train $\mathrm{(-)}$ & 0.629 & 0.520 & 0.319 & 0.491 & 0.555 \\*
& $\mathrm{MAE}$ train $\mathrm{(eV)}$ & 0.006 & 0.006 & 0.009 & 0.008 & 0.007 \\*
& $\mathrm{RMSE}$ train $\mathrm{(eV)}$ & 0.014 & 0.016 & 0.019 & 0.017 & 0.015 \\*
& $R^2$ test $\mathrm{(-)}$ & 0.139 & -0.208 & 0.138 & 0.063 & 0.082 \\*
& $\mathrm{MAE}$ test $\mathrm{(eV)}$ & 0.011 & 0.012 & 0.008 & 0.010 & 0.010 \\*
& $\mathrm{RMSE}$ test $\mathrm{(eV)}$ & 0.017 & 0.020 & 0.017 & 0.017 & 0.017 \\
    \nopagebreak
    \midrule
    \multirow{6}*{Al} & 
    $R^2$ train $\mathrm{(-)}$ & 0.150 & -0.019 & 0.245 & 0.057 & 0.157 \\*
& $\mathrm{MAE}$ train $\mathrm{(eV)}$ & 0.102 & 0.111 & 0.091 & 0.102 & 0.097 \\*
& $\mathrm{RMSE}$ train $\mathrm{(eV)}$ & 0.172 & 0.189 & 0.162 & 0.182 & 0.172 \\*
& $R^2$ test $\mathrm{(-)}$ & 0.381 & 0.148 & 0.395 & 0.256 & 0.335 \\*
& $\mathrm{MAE}$ test $\mathrm{(eV)}$ & 0.083 & 0.100 & 0.093 & 0.088 & 0.087 \\*
& $\mathrm{RMSE}$ test $\mathrm{(eV)}$ & 0.153 & 0.179 & 0.151 & 0.167 & 0.158 \\
    \nopagebreak
    \midrule
    \multirow{6}*{Rb} & 
    $R^2$ train $\mathrm{(-)}$ & 0.996 & 0.988 & 0.872 & 0.911 & 0.968 \\*
& $\mathrm{MAE}$ train $\mathrm{(eV)}$ & 0.002 & 0.003 & 0.008 & 0.007 & 0.005 \\*
& $\mathrm{RMSE}$ train $\mathrm{(eV)}$ & 0.004 & 0.006 & 0.019 & 0.016 & 0.010 \\*
& $R^2$ test $\mathrm{(-)}$ & 0.027 & -0.386 & -0.088 & -0.076 & -0.079 \\*
& $\mathrm{MAE}$ test $\mathrm{(eV)}$ & 0.022 & 0.033 & 0.026 & 0.027 & 0.027 \\*
& $\mathrm{RMSE}$ test $\mathrm{(eV)}$ & 0.062 & 0.074 & 0.065 & 0.065 & 0.065 \\
    \nopagebreak
    \midrule
    \multirow{6}*{Y} & 
    $R^2$ train $\mathrm{(-)}$ & 0.837 & 0.780 & 0.843 & 0.901 & 0.852 \\*
& $\mathrm{MAE}$ train $\mathrm{(eV)}$ & 0.074 & 0.078 & 0.071 & 0.053 & 0.066 \\*
& $\mathrm{RMSE}$ train $\mathrm{(eV)}$ & 0.118 & 0.137 & 0.115 & 0.092 & 0.112 \\*
& $R^2$ test $\mathrm{(-)}$ & 0.510 & 0.298 & 0.379 & 0.635 & 0.493 \\*
& $\mathrm{MAE}$ test $\mathrm{(eV)}$ & 0.071 & 0.078 & 0.076 & 0.064 & 0.069 \\*
& $\mathrm{RMSE}$ test $\mathrm{(eV)}$ & 0.120 & 0.143 & 0.135 & 0.103 & 0.122 \\
    \nopagebreak
    \bottomrule
\end{longtable}

\subsection{Regressors ($\Delta E_{discharge}$)}\label{ssec:batteries_stability_discharge_regressors}

\begin{figure}[H]
    \centering
    \includegraphics[width=1\linewidth]{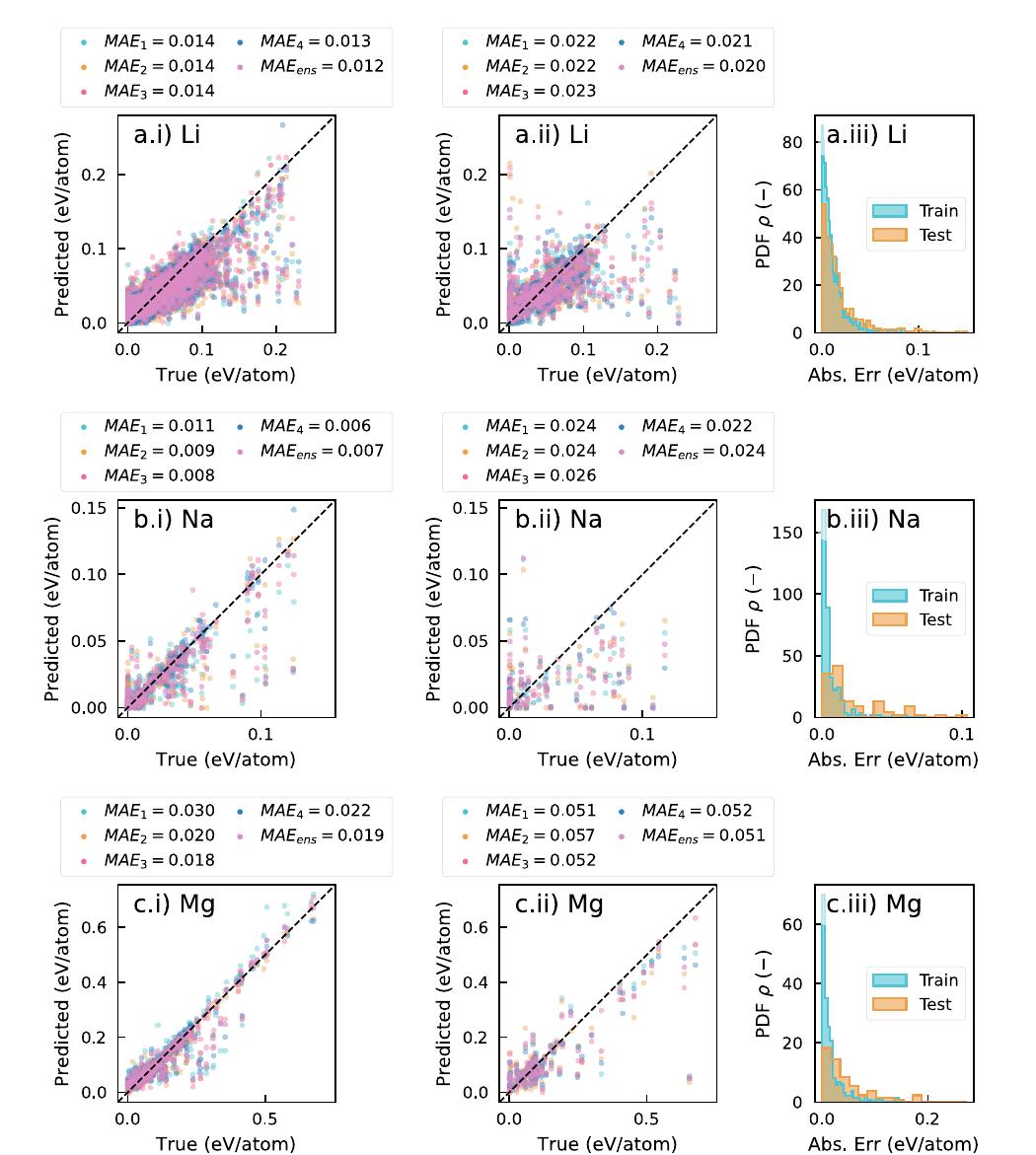}
    \caption[Results for Li, Na, Mg, K, Ca, Cs, Al, Rb, and Y cathode materials $\Delta E_{discharge}$  regressors.]{
        Results for Li (a), Na (b), Mg (c), K (d), Ca (e), Cs (f), Al (g), Rb (h), and Y (i) cathode materials $\Delta E_{discharge}$  regressors.
        i) Parity plot showing the true against predicted cathode stability in discharged state $\Delta E_{discharge}$ for the 4 trained regressors (highlighting the corresponding mean absolute error $MAE_i$), together with the ensemble case, in which each material is represented by the average prediction across the four trained models for training set and ii) for testing set.
        iii) Absolute error distribution over training and testing sets.
    }
    \label{fig:batteries_stability_discharge_parity_1}
\end{figure}
\begin{figure}[H]
    \addtocounter{figure}{-1}
    \centering
    \includegraphics[width=1\linewidth]{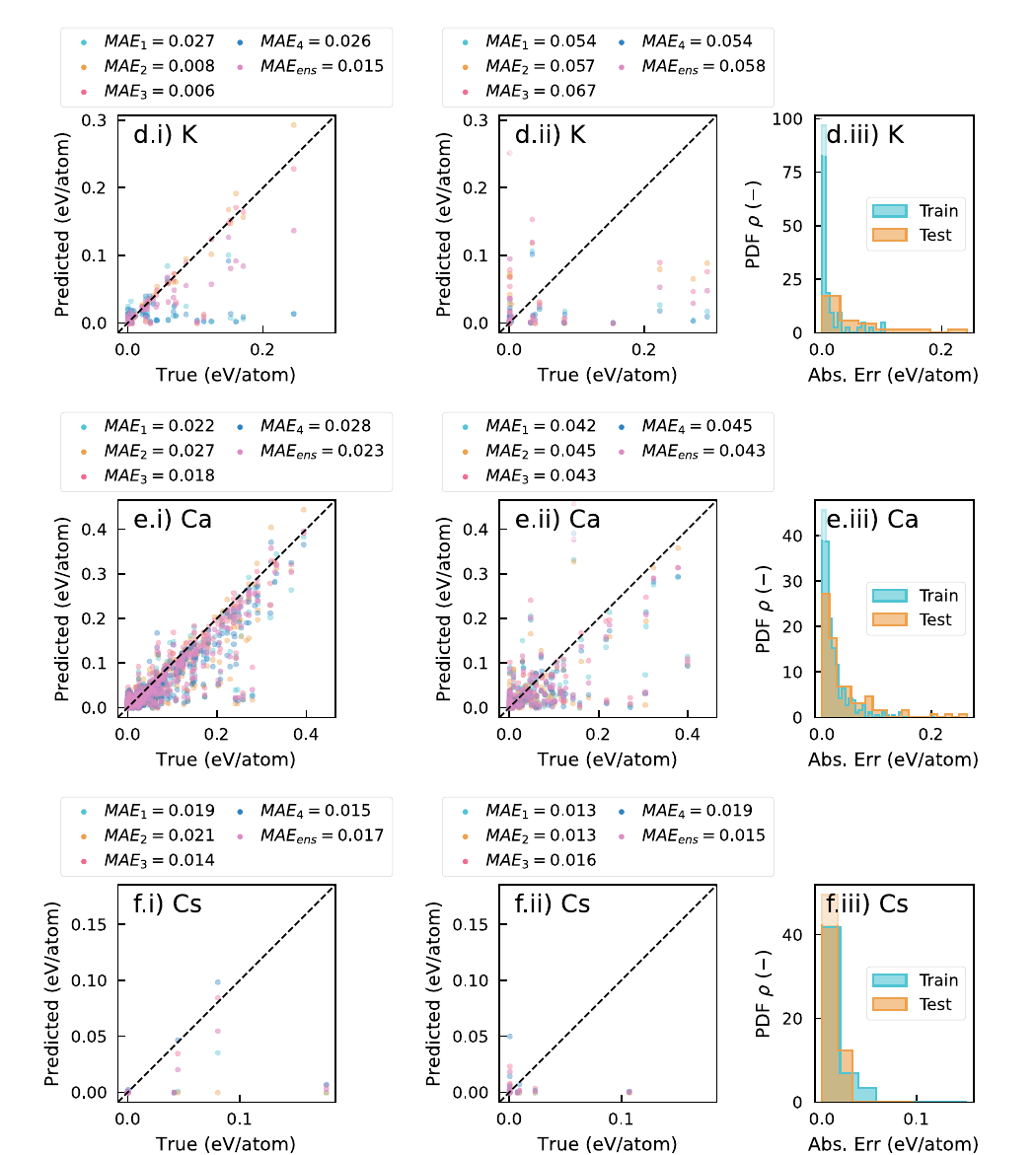}
    \caption[]{
        Results for Li (a), Na (b), Mg (c), K (d), Ca (e), Cs (f), Al (g), Rb (h), and Y (i) cathode materials $\Delta E_{discharge}$  regressors.
        i) Parity plot showing the true against predicted cathode stability in discharged state $\Delta E_{discharge}$ for the 4 trained regressors (highlighting the corresponding mean absolute error $MAE_i$), together with the ensemble case, in which each material is represented by the average prediction across the four trained models for training set and ii) for testing set.
        iii) Absolute error distribution over training and testing sets.
        (Continued)
    }
    \label{fig:batteries_stability_discharge_parity_2}
\end{figure}
\begin{figure}[H]
    \addtocounter{figure}{-1}
    \centering
    \includegraphics[width=1\linewidth]{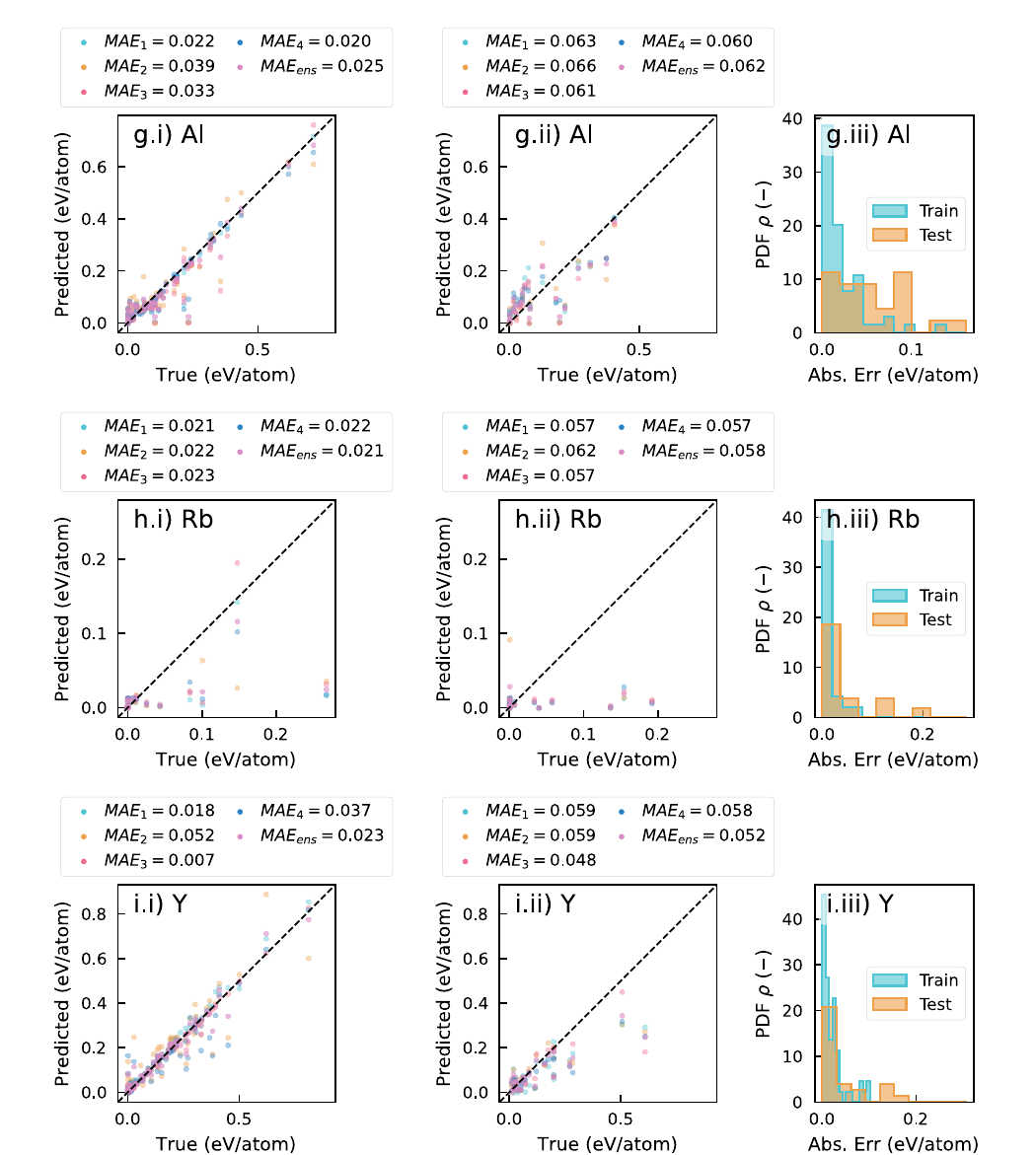}
    \caption[]{
        Results for Li (a), Na (b), Mg (c), K (d), Ca (e), Cs (f), Al (g), Rb (h), and Y (i) cathode materials $\Delta E_{discharge}$  regressors.
        i) Parity plot showing the true against predicted cathode stability in discharged state $\Delta E_{discharge}$ for the 4 trained regressors (highlighting the corresponding mean absolute error $MAE_i$), together with the ensemble case, in which each material is represented by the average prediction across the four trained models for training set and ii) for testing set.
        iii) Absolute error distribution over training and testing sets.
        (Continued)
    }
    \label{fig:batteries_stability_discharge_parity_3}
\end{figure}
\begin{longtable}{c|p{30mm}>{\raggedright\let\newline}cccc|c}
    \caption[Performances of the 4 cathode materials per the 9 working ions $\Delta E_{discharge}$ regressors.]{
        Performances of the 4 cathode materials per the 9 working ions $\Delta E_{discharge}$ regressors (for a total of 36 distinct models), along with the resulting ensemble model, in terms of $R^2$, MAE, RMSE for both training and testing.
        }\label{tab:batteries_stability_discharge_parity} \\
    \toprule
    & & Model 1 & Model 2 & Model 3 & Model 4 & Ens. Model \\
    \midrule
    \endfirsthead
    \caption[]{
        Performances of the 4 cathode materials per the 9 working ions $\Delta E_{discharge}$ regressors (for a total of 36 distinct models), along with the resulting ensemble model, in terms of $R^2$, MAE, RMSE for both training and testing. (Continued)
    } \\
    \toprule
    & & Model 1 & Model 2 & Model 3 & Model 4 & Ens. Model \\
    \midrule
    \endhead
    
    \endfoot
    \multirow{6}*{Li} &
    $R^2$ train $\mathrm{(-)}$ & 0.661 & 0.624 & 0.639 & 0.674 & 0.687 \\*
& $\mathrm{MAE}$ train $\mathrm{(eV)}$ & 0.014 & 0.014 & 0.014 & 0.013 & 0.012 \\*
& $\mathrm{RMSE}$ train $\mathrm{(eV)}$ & 0.022 & 0.023 & 0.023 & 0.021 & 0.021 \\*
& $R^2$ test $\mathrm{(-)}$ & 0.262 & 0.109 & 0.145 & 0.240 & 0.263 \\*
& $\mathrm{MAE}$ test $\mathrm{(eV)}$ & 0.022 & 0.022 & 0.023 & 0.021 & 0.020 \\*
& $\mathrm{RMSE}$ test $\mathrm{(eV)}$ & 0.034 & 0.038 & 0.037 & 0.035 & 0.034 \\
    \nopagebreak
    \midrule
    \multirow{6}*{Na} &
    $R^2$ train $\mathrm{(-)}$ & 0.569 & 0.670 & 0.689 & 0.798 & 0.740 \\*
& $\mathrm{MAE}$ train $\mathrm{(eV)}$ & 0.011 & 0.009 & 0.008 & 0.006 & 0.007 \\*
& $\mathrm{RMSE}$ train $\mathrm{(eV)}$ & 0.019 & 0.016 & 0.016 & 0.013 & 0.015 \\*
& $R^2$ test $\mathrm{(-)}$ & -0.184 & -0.067 & -0.432 & -0.057 & -0.112 \\*
& $\mathrm{MAE}$ test $\mathrm{(eV)}$ & 0.024 & 0.024 & 0.026 & 0.022 & 0.024 \\*
& $\mathrm{RMSE}$ test $\mathrm{(eV)}$ & 0.036 & 0.034 & 0.039 & 0.034 & 0.034 \\
    \nopagebreak
    \midrule
    \multirow{6}*{Mg} & 
    $R^2$ train $\mathrm{(-)}$ & 0.819 & 0.906 & 0.923 & 0.896 & 0.909 \\*
& $\mathrm{MAE}$ train $\mathrm{(eV)}$ & 0.030 & 0.020 & 0.018 & 0.022 & 0.019 \\*
& $\mathrm{RMSE}$ train $\mathrm{(eV)}$ & 0.050 & 0.036 & 0.033 & 0.038 & 0.036 \\*
& $R^2$ test $\mathrm{(-)}$ & 0.660 & 0.611 & 0.662 & 0.659 & 0.669 \\*
& $\mathrm{MAE}$ test $\mathrm{(eV)}$ & 0.051 & 0.057 & 0.052 & 0.052 & 0.051 \\*
& $\mathrm{RMSE}$ test $\mathrm{(eV)}$ & 0.092 & 0.099 & 0.092 & 0.093 & 0.091 \\
    \nopagebreak
    \midrule
    \multirow{6}*{K} & 
    $R^2$ train $\mathrm{(-)}$ & -0.063 & 0.845 & 0.864 & -0.048 & 0.652 \\*
& $\mathrm{MAE}$ train $\mathrm{(eV)}$ & 0.027 & 0.008 & 0.006 & 0.026 & 0.015 \\*
& $\mathrm{RMSE}$ train $\mathrm{(eV)}$ & 0.053 & 0.020 & 0.019 & 0.052 & 0.030 \\*
& $R^2$ test $\mathrm{(-)}$ & -0.242 & 0.083 & -0.345 & -0.282 & -0.115 \\*
& $\mathrm{MAE}$ test $\mathrm{(eV)}$ & 0.054 & 0.057 & 0.067 & 0.054 & 0.058 \\*
& $\mathrm{RMSE}$ test $\mathrm{(eV)}$ & 0.097 & 0.083 & 0.101 & 0.098 & 0.092 \\
    \nopagebreak
    \midrule
    \multirow{6}*{Ca} & 
    $R^2$ train $\mathrm{(-)}$ & 0.778 & 0.692 & 0.829 & 0.699 & 0.772 \\*
& $\mathrm{MAE}$ train $\mathrm{(eV)}$ & 0.022 & 0.027 & 0.018 & 0.028 & 0.023 \\*
& $\mathrm{RMSE}$ train $\mathrm{(eV)}$ & 0.040 & 0.047 & 0.035 & 0.046 & 0.040 \\*
& $R^2$ test $\mathrm{(-)}$ & 0.322 & 0.238 & 0.235 & 0.223 & 0.276 \\*
& $\mathrm{MAE}$ test $\mathrm{(eV)}$ & 0.042 & 0.045 & 0.043 & 0.045 & 0.043 \\*
& $\mathrm{RMSE}$ test $\mathrm{(eV)}$ & 0.070 & 0.074 & 0.074 & 0.075 & 0.072 \\
    \nopagebreak
    \midrule
    \multirow{6}*{Cs} & 
    $R^2$ train $\mathrm{(-)}$ & -0.074 & -0.216 & 0.097 & 0.099 & 0.035 \\*
& $\mathrm{MAE}$ train $\mathrm{(eV)}$ & 0.019 & 0.021 & 0.014 & 0.015 & 0.017 \\*
& $\mathrm{RMSE}$ train $\mathrm{(eV)}$ & 0.048 & 0.051 & 0.044 & 0.044 & 0.045 \\*
& $R^2$ test $\mathrm{(-)}$ & -0.177 & -0.185 & -0.216 & -0.407 & -0.199 \\*
& $\mathrm{MAE}$ test $\mathrm{(eV)}$ & 0.013 & 0.013 & 0.016 & 0.019 & 0.015 \\*
& $\mathrm{RMSE}$ test $\mathrm{(eV)}$ & 0.033 & 0.033 & 0.033 & 0.036 & 0.033 \\
    \nopagebreak
    \midrule
    \multirow{6}*{Al} & 
    $R^2$ train $\mathrm{(-)}$ & 0.914 & 0.834 & 0.863 & 0.939 & 0.916 \\*
& $\mathrm{MAE}$ train $\mathrm{(eV)}$ & 0.022 & 0.039 & 0.033 & 0.020 & 0.025 \\*
& $\mathrm{RMSE}$ train $\mathrm{(eV)}$ & 0.043 & 0.060 & 0.054 & 0.036 & 0.043 \\*
& $R^2$ test $\mathrm{(-)}$ & 0.528 & 0.458 & 0.585 & 0.661 & 0.596 \\*
& $\mathrm{MAE}$ test $\mathrm{(eV)}$ & 0.063 & 0.066 & 0.061 & 0.060 & 0.062 \\*
& $\mathrm{RMSE}$ test $\mathrm{(eV)}$ & 0.082 & 0.088 & 0.077 & 0.069 & 0.076 \\
    \nopagebreak
    \midrule
    \multirow{6}*{Rb} & 
    $R^2$ train $\mathrm{(-)}$ & 0.153 & 0.197 & 0.224 & 0.183 & 0.238 \\*
& $\mathrm{MAE}$ train $\mathrm{(eV)}$ & 0.021 & 0.022 & 0.023 & 0.022 & 0.021 \\*
& $\mathrm{RMSE}$ train $\mathrm{(eV)}$ & 0.056 & 0.055 & 0.054 & 0.055 & 0.053 \\*
& $R^2$ test $\mathrm{(-)}$ & -0.350 & -0.388 & -0.317 & -0.328 & -0.335 \\*
& $\mathrm{MAE}$ test $\mathrm{(eV)}$ & 0.057 & 0.062 & 0.057 & 0.057 & 0.058 \\*
& $\mathrm{RMSE}$ test $\mathrm{(eV)}$ & 0.103 & 0.104 & 0.101 & 0.102 & 0.102 \\
    \nopagebreak
    \midrule
    \multirow{6}*{Y} & 
    $R^2$ train $\mathrm{(-)}$ & 0.976 & 0.765 & 0.994 & 0.849 & 0.952 \\*
& $\mathrm{MAE}$ train $\mathrm{(eV)}$ & 0.018 & 0.052 & 0.007 & 0.037 & 0.023 \\*
& $\mathrm{RMSE}$ train $\mathrm{(eV)}$ & 0.025 & 0.078 & 0.012 & 0.063 & 0.035 \\*
& $R^2$ test $\mathrm{(-)}$ & 0.610 & 0.577 & 0.565 & 0.541 & 0.600 \\*
& $\mathrm{MAE}$ test $\mathrm{(eV)}$ & 0.059 & 0.059 & 0.048 & 0.058 & 0.052 \\*
& $\mathrm{RMSE}$ test $\mathrm{(eV)}$ & 0.091 & 0.095 & 0.096 & 0.099 & 0.092 \\
    \nopagebreak
    \bottomrule
\end{longtable}

\subsection{AI-experts}\label{ssec:batteries_classifiers}

\begin{figure}[H]
    \centering
    \includegraphics[width=1\linewidth]{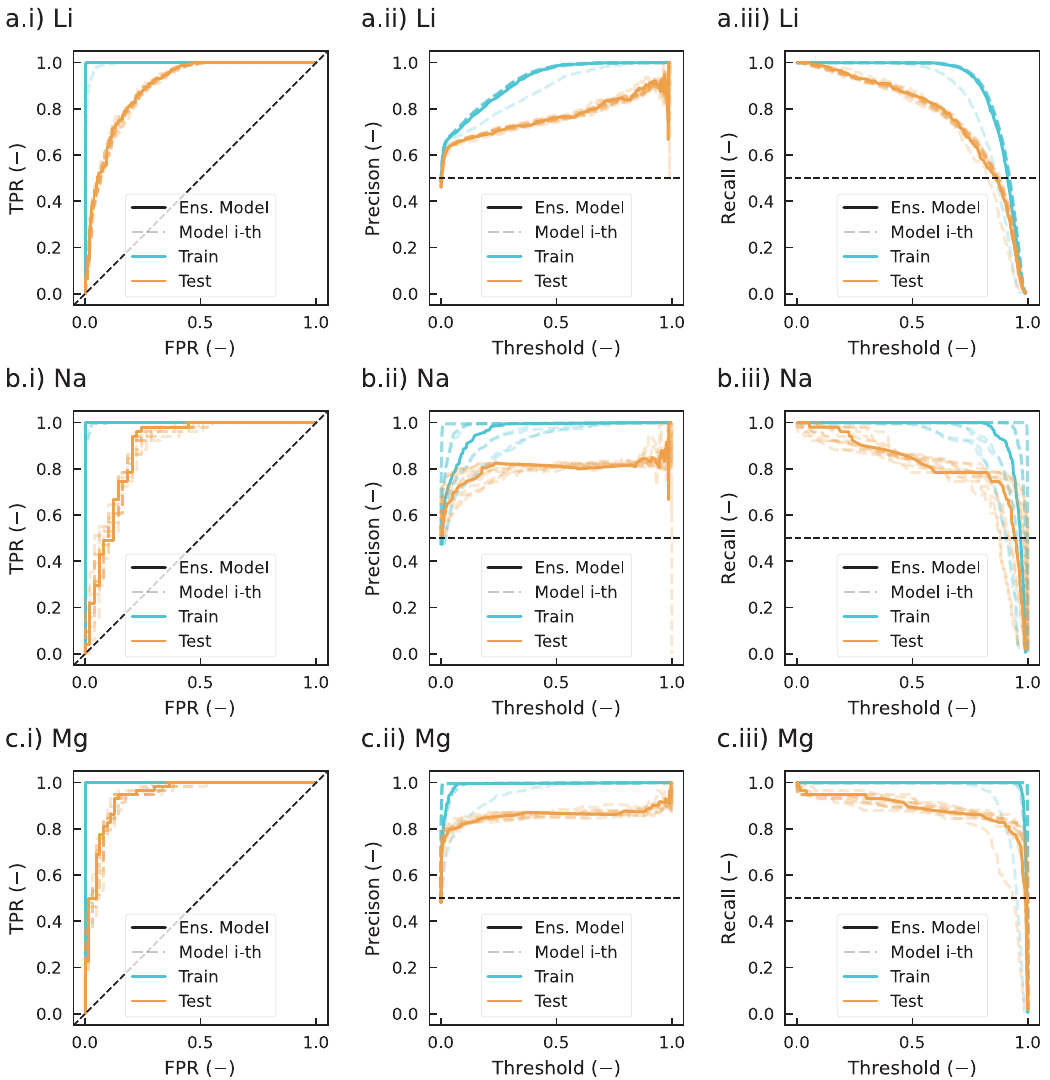}
    \caption[Results for Li, Na, Mg, K, Ca, Cs, Al, Rb, and Y cathode materials AI-experts models.]{
    Results for Li (a), Na (b), Mg (c), K (d), Ca (e), Cs (f), Al (g), Rb (h), and Y (i) cathode materials AI-experts models.
    i) ROC, ii) precision and iii) recall curves over the training/testing sets of all the 10 classifiers, together with the ensemble model curve.
    }
    \label{fig:batteries_classifiers_1}
\end{figure}
\begin{figure}[H]
    \addtocounter{figure}{-1}
    \centering
    \includegraphics[width=1\linewidth]{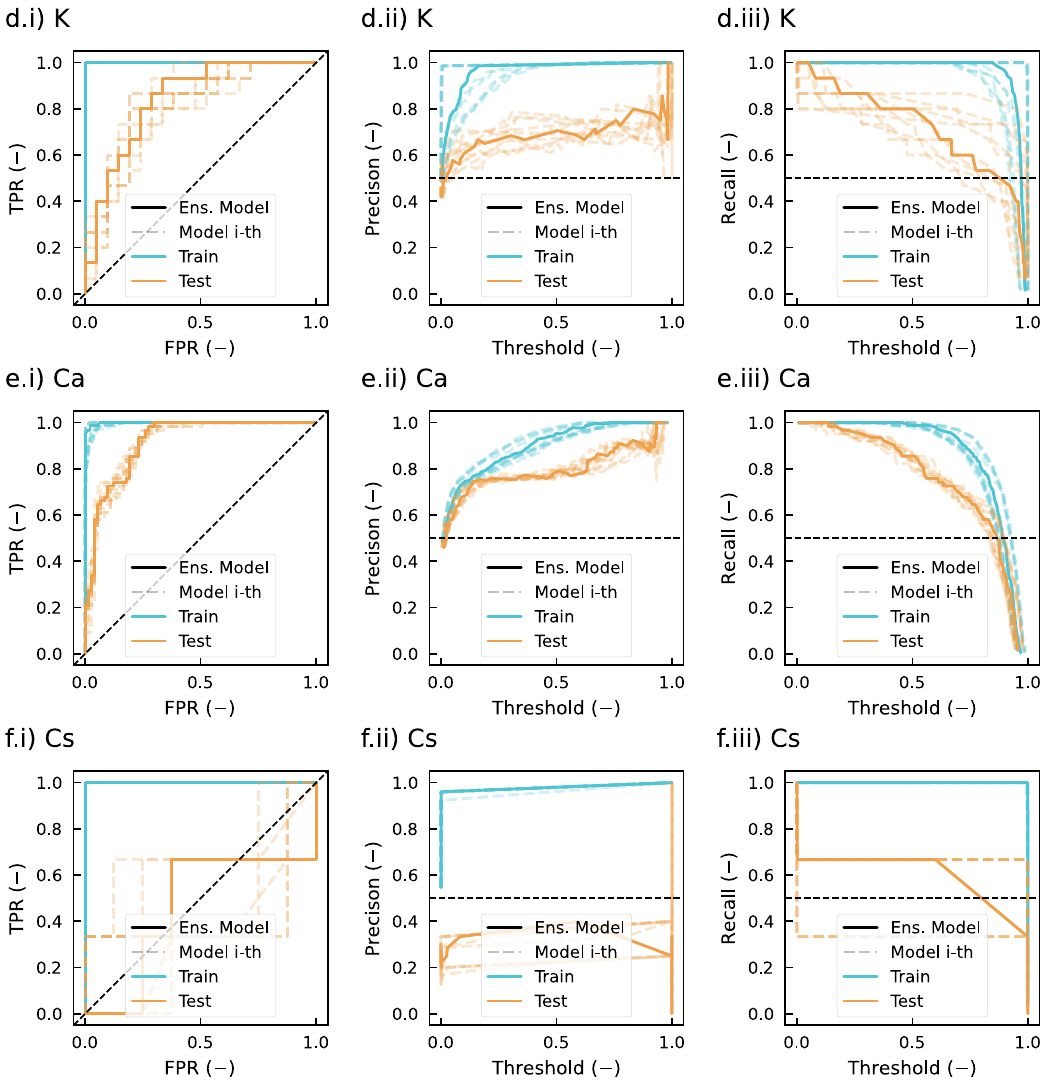}
    \caption[]{
        Results for Li (a), Na (b), Mg (c), K (d), Ca (e), Cs (f), Al (g), Rb (h), and Y (i) cathode materials AI-experts models.
        i) ROC, ii) precision and iii) recall curves over the training/testing sets of all the 10 classifiers, together with the ensemble model curve.
        (Continued)
    }
    \label{fig:batteries_classifiers_2}
\end{figure}
\begin{figure}[H]
    \addtocounter{figure}{-1}
    \centering
    \includegraphics[width=1\linewidth]{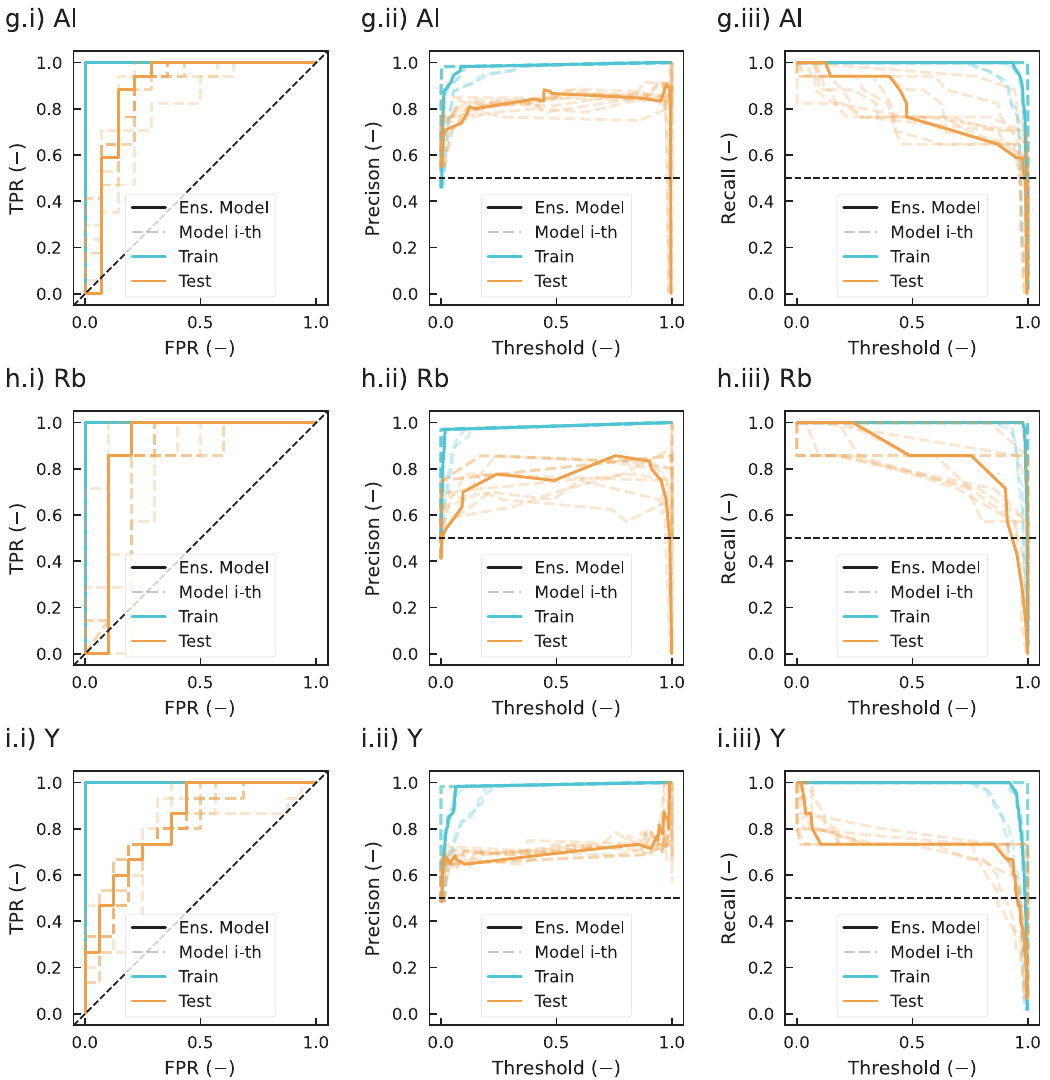}
    \caption[]{
        Results for Li (a), Na (b), Mg (c), K (d), Ca (e), Cs (f), Al (g), Rb (h), and Y (i) cathode materials AI-experts models.
        i) ROC, ii) precision and iii) recall curves over the training/testing sets of all the 10 classifiers, together with the ensemble model curve.
        (Continued)
    }
    \label{fig:batteries_classifiers_3}
\end{figure}
\begin{longtable}{m{3.5mm}|p{15.5mm}>{\raggedright\let\newline}cccccccccc|c}
    \caption[Performances for Li, Na, Mg, K, Ca, Cs, Al, Rb, and Y cathode of the 10 AI-experts models.]{
        Performances for Li, Na, Mg, K, Ca, Cs, Al, Rb, and Y cathode of the 10 AI-experts models, for each of the 10 classifier sand  the resulting ensemble model, in terms of AUC of ROC curve, precision and recall for  testing.
        }\label{tab:batteries_classifiers} \\
    \toprule
    & & M. 1 & M. 2 & M. 3 & M. 4 & M. 5 & M. 6 & M. 7 & M. 8 & M. 9 & M. 10 & Ens. Model \\
    \midrule
    \endfirsthead
    \caption[]{
        Performances for Li, Na, Mg, K, Ca, Cs, Al, Rb, and Y cathode of the 10 AI-experts models, for each of the 10 classifier sand  the resulting ensemble model, in terms of AUC of ROC curve, precision and recall for  testing.
        (Continued)
    } \\
    \toprule
    & & M. 1 & M. 2 & M. 3 & M. 4 & M. 5 & M. 6 & M. 7 & M. 8 & M. 9 & M. 10 & Ens. Model \\
    \midrule
    \endhead
    
    \endfoot
    \multirow{6}*{Li} &
    $\mathrm{AUC}$ train $\mathrm{(-)}$ & 1.000 & 1.000 & 1.000 & 1.000 & 1.000 & 0.994 & 1.000 & 1.000 & 1.000 & 1.000 & 1.000 \\*
& $\mathrm{Precision}$ train $\mathrm{(-)}$ & 0.986 & 0.989 & 0.985 & 0.988 & 0.980 & 0.920 & 0.989 & 0.985 & 0.985 & 0.987 & 0.985 \\*
& $\mathrm{Recall}$ train $\mathrm{(-)}$ & 1.000 & 1.000 & 1.000 & 1.000 & 1.000 & 0.990 & 1.000 & 1.000 & 1.000 & 1.000 & 1.000 \\*
& $\mathrm{AUC}$ test $\mathrm{(-)}$ & 0.896 & 0.891 & 0.893 & 0.896 & 0.897 & 0.900 & 0.893 & 0.893 & 0.891 & 0.896 & 0.897 \\*
& $\mathrm{Precision}$ test $\mathrm{(-)}$ & 0.768 & 0.756 & 0.759 & 0.758 & 0.749 & 0.741 & 0.755 & 0.758 & 0.749 & 0.770 & 0.764 \\*
& $\mathrm{Recall}$ test $\mathrm{(-)}$ & 0.843 & 0.864 & 0.857 & 0.864 & 0.864 & 0.878 & 0.840 & 0.829 & 0.864 & 0.854 & 0.857 \\
    \nopagebreak
    \midrule
    \multirow{6}*{Na} &
    $\mathrm{AUC}$ train $\mathrm{(-)}$ & 1.000 & 1.000 & 1.000 & 1.000 & 1.000 & 1.000 & 0.998 & 0.998 & 1.000 & 1.000 & 1.000 \\*
& $\mathrm{Precision}$ train $\mathrm{(-)}$ & 1.000 & 0.979 & 1.000 & 1.000 & 1.000 & 1.000 & 0.969 & 0.969 & 1.000 & 1.000 & 1.000 \\*
& $\mathrm{Recall}$ train $\mathrm{(-)}$ & 1.000 & 1.000 & 1.000 & 1.000 & 1.000 & 1.000 & 0.995 & 0.989 & 1.000 & 1.000 & 1.000 \\*
& $\mathrm{AUC}$ test $\mathrm{(-)}$ & 0.887 & 0.880 & 0.890 & 0.879 & 0.891 & 0.887 & 0.885 & 0.893 & 0.892 & 0.882 & 0.890 \\*
& $\mathrm{Precision}$ test $\mathrm{(-)}$ & 0.811 & 0.808 & 0.811 & 0.815 & 0.808 & 0.827 & 0.800 & 0.830 & 0.804 & 0.804 & 0.811 \\*
& $\mathrm{Recall}$ test $\mathrm{(-)}$ & 0.843 & 0.824 & 0.843 & 0.863 & 0.824 & 0.843 & 0.784 & 0.863 & 0.804 & 0.804 & 0.843 \\
    \nopagebreak
    \midrule
    \multirow{6}*{Mg} & 
    $\mathrm{AUC}$ train $\mathrm{(-)}$ & 1.000 & 1.000 & 1.000 & 1.000 & 1.000 & 1.000 & 1.000 & 1.000 & 1.000 & 1.000 & 1.000 \\*
& $\mathrm{Precision}$ train $\mathrm{(-)}$ & 1.000 & 0.996 & 1.000 & 1.000 & 1.000 & 1.000 & 1.000 & 1.000 & 1.000 & 1.000 & 1.000 \\*
& $\mathrm{Recall}$ train $\mathrm{(-)}$ & 1.000 & 1.000 & 1.000 & 1.000 & 1.000 & 1.000 & 1.000 & 1.000 & 1.000 & 1.000 & 1.000 \\*
& $\mathrm{AUC}$ test $\mathrm{(-)}$ & 0.952 & 0.949 & 0.950 & 0.935 & 0.938 & 0.947 & 0.948 & 0.937 & 0.933 & 0.940 & 0.946 \\*
& $\mathrm{Precision}$ test $\mathrm{(-)}$ & 0.864 & 0.885 & 0.871 & 0.867 & 0.864 & 0.850 & 0.867 & 0.847 & 0.869 & 0.864 & 0.864 \\*
& $\mathrm{Recall}$ test $\mathrm{(-)}$ & 0.879 & 0.931 & 0.931 & 0.897 & 0.879 & 0.879 & 0.897 & 0.862 & 0.914 & 0.879 & 0.879 \\
    \nopagebreak
    \midrule
    \multirow{6}*{K} & 
    $\mathrm{AUC}$ train $\mathrm{(-)}$ & 1.000 & 1.000 & 1.000 & 1.000 & 1.000 & 1.000 & 1.000 & 1.000 & 1.000 & 1.000 & 1.000 \\*
& $\mathrm{Precision}$ train $\mathrm{(-)}$ & 1.000 & 1.000 & 1.000 & 1.000 & 1.000 & 1.000 & 1.000 & 1.000 & 1.000 & 1.000 & 1.000 \\*
& $\mathrm{Recall}$ train $\mathrm{(-)}$ & 1.000 & 1.000 & 1.000 & 1.000 & 1.000 & 1.000 & 1.000 & 1.000 & 1.000 & 1.000 & 1.000 \\*
& $\mathrm{AUC}$ test $\mathrm{(-)}$ & 0.822 & 0.787 & 0.813 & 0.803 & 0.802 & 0.794 & 0.841 & 0.838 & 0.844 & 0.844 & 0.844 \\*
& $\mathrm{Precision}$ test $\mathrm{(-)}$ & 0.688 & 0.684 & 0.750 & 0.647 & 0.667 & 0.667 & 0.733 & 0.769 & 0.750 & 0.750 & 0.706 \\*
& $\mathrm{Recall}$ test $\mathrm{(-)}$ & 0.733 & 0.867 & 0.800 & 0.733 & 0.667 & 0.800 & 0.733 & 0.667 & 0.800 & 0.600 & 0.800 \\
    \nopagebreak
    \midrule
    \multirow{6}*{Ca} & 
    $\mathrm{AUC}$ train $\mathrm{(-)}$ & 0.995 & 0.999 & 0.995 & 0.994 & 0.993 & 1.000 & 1.000 & 0.995 & 1.000 & 0.994 & 0.999 \\*
& $\mathrm{Precision}$ train $\mathrm{(-)}$ & 0.928 & 0.966 & 0.928 & 0.931 & 0.924 & 0.981 & 0.985 & 0.921 & 0.977 & 0.934 & 0.955 \\*
& $\mathrm{Recall}$ train $\mathrm{(-)}$ & 0.985 & 0.996 & 0.992 & 0.985 & 0.977 & 1.000 & 1.000 & 0.988 & 1.000 & 0.985 & 0.988 \\*
& $\mathrm{AUC}$ test $\mathrm{(-)}$ & 0.916 & 0.905 & 0.920 & 0.910 & 0.913 & 0.912 & 0.913 & 0.923 & 0.910 & 0.920 & 0.917 \\*
& $\mathrm{Precision}$ test $\mathrm{(-)}$ & 0.779 & 0.791 & 0.785 & 0.758 & 0.791 & 0.790 & 0.788 & 0.791 & 0.773 & 0.790 & 0.791 \\*
& $\mathrm{Recall}$ test $\mathrm{(-)}$ & 0.855 & 0.855 & 0.823 & 0.806 & 0.855 & 0.790 & 0.839 & 0.855 & 0.823 & 0.790 & 0.855 \\
    \nopagebreak
    \midrule
    \multirow{6}*{Cs} & 
    $\mathrm{AUC}$ train $\mathrm{(-)}$ & 1.000 & 1.000 & 1.000 & 1.000 & 1.000 & 1.000 & 1.000 & 1.000 & 1.000 & 1.000 & 1.000 \\*
& $\mathrm{Precision}$ train $\mathrm{(-)}$ & 1.000 & 1.000 & 1.000 & 1.000 & 1.000 & 1.000 & 1.000 & 1.000 & 1.000 & 1.000 & 1.000 \\*
& $\mathrm{Recall}$ train $\mathrm{(-)}$ & 1.000 & 1.000 & 1.000 & 1.000 & 1.000 & 1.000 & 1.000 & 1.000 & 1.000 & 1.000 & 1.000 \\*
& $\mathrm{AUC}$ test $\mathrm{(-)}$ & 0.542 & 0.583 & 0.292 & 0.542 & 0.354 & 0.542 & 0.354 & 0.708 & 0.542 & 0.292 & 0.458 \\*
& $\mathrm{Precision}$ test $\mathrm{(-)}$ & 0.400 & 0.400 & 0.250 & 0.400 & 0.250 & 0.400 & 0.250 & 0.333 & 0.400 & 0.250 & 0.400 \\*
& $\mathrm{Recall}$ test $\mathrm{(-)}$ & 0.667 & 0.667 & 0.333 & 0.667 & 0.333 & 0.667 & 0.333 & 0.667 & 0.667 & 0.333 & 0.667 \\
    \nopagebreak
    \midrule
    \multirow{6}*{Al} & 
    $\mathrm{AUC}$ train $\mathrm{(-)}$ & 1.000 & 1.000 & 1.000 & 1.000 & 1.000 & 1.000 & 1.000 & 1.000 & 1.000 & 1.000 & 1.000 \\*
& $\mathrm{Precision}$ train $\mathrm{(-)}$ & 1.000 & 1.000 & 1.000 & 1.000 & 1.000 & 1.000 & 1.000 & 1.000 & 1.000 & 1.000 & 1.000 \\*
& $\mathrm{Recall}$ train $\mathrm{(-)}$ & 1.000 & 1.000 & 1.000 & 1.000 & 1.000 & 1.000 & 1.000 & 1.000 & 1.000 & 1.000 & 1.000 \\*
& $\mathrm{AUC}$ test $\mathrm{(-)}$ & 0.870 & 0.891 & 0.870 & 0.866 & 0.878 & 0.891 & 0.824 & 0.857 & 0.908 & 0.903 & 0.887 \\*
& $\mathrm{Precision}$ test $\mathrm{(-)}$ & 0.786 & 0.867 & 0.867 & 0.846 & 0.867 & 0.867 & 0.750 & 0.842 & 0.889 & 0.882 & 0.857 \\*
& $\mathrm{Recall}$ test $\mathrm{(-)}$ & 0.647 & 0.765 & 0.765 & 0.647 & 0.765 & 0.765 & 0.706 & 0.941 & 0.941 & 0.882 & 0.706 \\
    \nopagebreak
    \midrule
    \multirow{6}*{Rb} & 
    $\mathrm{AUC}$ train $\mathrm{(-)}$ & 1.000 & 1.000 & 1.000 & 1.000 & 1.000 & 1.000 & 1.000 & 1.000 & 1.000 & 1.000 & 1.000 \\*
& $\mathrm{Precision}$ train $\mathrm{(-)}$ & 1.000 & 1.000 & 1.000 & 1.000 & 1.000 & 1.000 & 1.000 & 1.000 & 1.000 & 1.000 & 1.000 \\*
& $\mathrm{Recall}$ train $\mathrm{(-)}$ & 1.000 & 1.000 & 1.000 & 1.000 & 1.000 & 1.000 & 1.000 & 1.000 & 1.000 & 1.000 & 1.000 \\*
& $\mathrm{AUC}$ test $\mathrm{(-)}$ & 0.871 & 0.836 & 0.914 & 0.971 & 0.857 & 0.800 & 0.871 & 0.786 & 0.836 & 0.829 & 0.886 \\*
& $\mathrm{Precision}$ test $\mathrm{(-)}$ & 0.833 & 0.857 & 0.778 & 0.778 & 0.833 & 0.750 & 0.833 & 0.625 & 0.857 & 0.750 & 0.857 \\*
& $\mathrm{Recall}$ test $\mathrm{(-)}$ & 0.714 & 0.857 & 1.000 & 1.000 & 0.714 & 0.857 & 0.714 & 0.714 & 0.857 & 0.857 & 0.857 \\
    \nopagebreak
    \midrule
    \multirow{6}*{Y} & 
    $\mathrm{AUC}$ train $\mathrm{(-)}$ & 1.000 & 1.000 & 1.000 & 1.000 & 1.000 & 1.000 & 1.000 & 1.000 & 1.000 & 1.000 & 1.000 \\*
& $\mathrm{Precision}$ train $\mathrm{(-)}$ & 1.000 & 1.000 & 1.000 & 1.000 & 1.000 & 1.000 & 1.000 & 1.000 & 1.000 & 1.000 & 1.000 \\*
& $\mathrm{Recall}$ train $\mathrm{(-)}$ & 1.000 & 1.000 & 1.000 & 1.000 & 1.000 & 1.000 & 1.000 & 1.000 & 1.000 & 1.000 & 1.000 \\*
& $\mathrm{AUC}$ test $\mathrm{(-)}$ & 0.825 & 0.821 & 0.808 & 0.821 & 0.825 & 0.721 & 0.846 & 0.808 & 0.787 & 0.833 & 0.833 \\*
& $\mathrm{Precision}$ test $\mathrm{(-)}$ & 0.733 & 0.733 & 0.688 & 0.733 & 0.688 & 0.733 & 0.733 & 0.688 & 0.714 & 0.733 & 0.733 \\*
& $\mathrm{Recall}$ test $\mathrm{(-)}$ & 0.733 & 0.733 & 0.733 & 0.733 & 0.733 & 0.733 & 0.733 & 0.733 & 0.667 & 0.733 & 0.733 \\
    \nopagebreak
    \bottomrule
\end{longtable}






